\documentclass[longauth]{aaEC}

\usepackage{graphicx}
\usepackage{natbib}
\usepackage{scalerel}

\usepackage[table]{xcolor}

\usepackage{color}
\usepackage{caption}
\usepackage{subcaption}
\usepackage{array}

\usepackage{threeparttable}

\usepackage{tikz}
\usetikzlibrary{shapes.geometric, arrows}
\usepackage{txfonts}
\usepackage{siunitx}
\usepackage[pdfencoding=auto,psdextra]{hyperref}
\hypersetup{
    colorlinks=true,
    linkcolor=blue,
    filecolor=magenta,      
    urlcolor=blue,
    citecolor=blue
}
\usepackage[utf8]{inputenc}
\usepackage{euclid}
\usepackage[switch, modulo]{lineno}

\tikzstyle{input} = [rectangle, rounded corners, 
minimum width=3cm, 
minimum height=1cm,
text centered, 
draw=black,
fill=orange!30]

\tikzstyle{item} = [trapezium, 
trapezium stretches=true, 
trapezium left angle=70, 
trapezium right angle=110, 
minimum width=3cm, 
minimum height=1cm, text centered, 
draw=black, fill=blue!30]

\tikzstyle{sed} = [rectangle, 
minimum width=3.5cm, 
minimum height=1cm, 
text centered, 
draw=black, 
fill=red!30]

\tikzstyle{decision} = [diamond, 
minimum width=3cm, 
minimum height=1cm, 
text centered, 
draw=black, 
fill=yellow!30]
\tikzstyle{arrow} = [thick,->,>=stealth]

\newcommand{\head}[2]{\multicolumn{1}{>{\centering\arraybackslash}p{#1}}{#2}}

\newcommand{\ebv}{{$E(B-V)$}}
\newcommand*{\spr}{{\tt SPRITZ}}
\newcommand{\qsfit}{{\tt QSFIT}}

\makeatletter
\renewcommand*\aa@pageof{, page \thepage{} of \pageref*{LastPage}}
\makeatother

\def\NUMDetectableIeEWS{3.1\,$\times$\,10$^{7}$\,}
\def\NUMDetectableYeEWS{2.2\,$\times$\,10$^{7}$\,}
\def\NUMDetectableJeEWS{3.0\,$\times$\,10$^{7}$\,}
\def\NUMDetectableHeEWS{3.5\,$\times$\,10$^{7}$\,}
\def\NUMDetectableOneFiltEWS{4.0\,$\times$\,10$^{7}$\,}
\def\NUMDetectableAllFiltEWS{2.1\,$\times$\,10$^{7}$\,}

\def\SDDetectableIeEWS{2.2\,$\times$\,10$^{3}$\,}
\def\SDDetectableYeEWS{1.5\,$\times$\,10$^{3}$\,}
\def\SDDetectableJeEWS{2.1\,$\times$\,10$^{3}$\,}
\def\SDDetectableHeEWS{2.4\,$\times$\,10$^{3}$\,}
\def\SDDetectableOneFiltEWS{2.8\,$\times$\,10$^{3}$\,}
\def\SDDetectableAllFiltEWS{1.4\,$\times$\,10$^{3}$\,}

\def\NUMDetectableIeEDS{1.9\,$\times$\,10$^{5}$\,}
\def\NUMDetectableYeEDS{1.6\,$\times$\,10$^{5}$\,}
\def\NUMDetectableJeEDS{2.0\,$\times$\,10$^{5}$\,}
\def\NUMDetectableHeEDS{2.3\,$\times$\,10$^{5}$\,}
\def\NUMDetectableOneFiltEDS{2.4\,$\times$\,10$^{5}$\,}
\def\NUMDetectableAllFiltEDS{1.6\,$\times$\,10$^{5}$\,}

\def\SDDetectableIeEDS{3.8\,$\times$\,10$^{3}$\,}
\def\SDDetectableYeEDS{3.2\,$\times$\,10$^{3}$\,}
\def\SDDetectableJeEDS{4.0\,$\times$\,10$^{3}$\,}
\def\SDDetectableHeEDS{4.5\,$\times$\,10$^{3}$\,}
\def\SDDetectableOneFiltEDS{4.7\,$\times$\,10$^{3}$\,}
\def\SDDetectableAllFiltEDS{3.1\,$\times$\,10$^{3}$\,}

\def\NUMSelectedEuclidOnlyTypeOneEWS{4.8\,$\times$\,10$^{6}$\,}
\def\SDSelectedEuclidOnlyTypeOneEWS{331\,}
\def\CSelectedEuclidOnlyTypeOneEWS{0.23}
\def\CTypeOneSelectedEuclidOnlyTypeOneEWS{0.52}
\def\NUMSelectedEuclidLSSTTypeOneEWS{5.7\,$\times$\,10$^{6}$\,}
\def\SDSelectedEuclidLSSTTypeOneEWS{393\,}
\def\CSelectedEuclidLSSTTypeOneEWS{0.45}
\def\CTypeOneSelectedEuclidLSSTTypeOneEWS{0.75}
\def\NUMSelectedEuclidLSSTAllEWS{6.0\,$\times$\,10$^{6}$\,}
\def\SDSelectedEuclidLSSTAllEWS{413\,}
\def\CSelectedEuclidLSSTAllEWS{0.51}
\def\CTypeOneSelectedEuclidLSSTAllEWS{0.65}
\def\CTypeTwoSelectedEuclidLSSTAllEWS{0.33}

\def\NUMSelectedEuclidOnlyTypeOneEDS{1.7\,$\times$\,10$^{4}$\,}
\def\SDSelectedEuclidOnlyTypeOneEDS{346\,}
\def\CSelectedEuclidOnlyTypeOneEDS{0.11}
\def\CTypeOneSelectedEuclidOnlyTypeOneEDS{0.40}
\def\NUMSelectedEuclidLSSTTypeOneEDS{2.0\,$\times$\,10$^{4}$\,}
\def\SDSelectedEuclidLSSTTypeOneEDS{392\,}
\def\CSelectedEuclidLSSTTypeOneEDS{0.45}
\def\CTypeOneSelectedEuclidLSSTTypeOneEDS{0.76}
\def\NUMSelectedEuclidLSSTAllEDS{2.9\,$\times$\,10$^{4}$\,}
\def\SDSelectedEuclidLSSTAllEDS{579\,}
\def\CSelectedEuclidLSSTAllEDS{0.32}
\def\CTypeOneSelectedEuclidLSSTAllEDS{0.51}
\def\CTypeTwoSelectedEuclidLSSTAllEDS{0.18}

\defcitealias{scaramella2022}{EC: Scaramella et al. (2022)}
\begin{document}

\title{\Euclid preparation}
\subtitle{Observational expectations for redshift $z<7$ active galactic nuclei in the Euclid Wide and Deep surveys\thanks{This paper is published on behalf of the Euclid Consortium}}    
\titlerunning{\Euclid Preparation. Expectations for $z<7$ AGN in the \Euclid surveys}
\authorrunning{Euclid Collaboration: M.~Selwood et al.}

\newcommand{\orcid}[1]{} 
\author{Euclid Collaboration: M.~Selwood\orcid{0009-0002-3235-0825}\thanks{\email{matthew.selwood@bristol.ac.uk}}\inst{\ref{aff1}}
\and S.~Fotopoulou\orcid{0000-0002-9686-254X}\inst{\ref{aff1}}
\and M.~N.~Bremer\inst{\ref{aff1}}
\and L.~Bisigello\orcid{0000-0003-0492-4924}\inst{\ref{aff2},\ref{aff3}}
\and H.~Landt\orcid{0000-0001-8391-6900}\inst{\ref{aff4}}
\and E.~Ba\~nados\orcid{0000-0002-2931-7824}\inst{\ref{aff5}}
\and G.~Zamorani\orcid{0000-0002-2318-301X}\inst{\ref{aff6}}
\and F.~Shankar\orcid{0000-0001-8973-5051}\inst{\ref{aff7}}
\and D.~Stern\orcid{0000-0003-2686-9241}\inst{\ref{aff8}}
\and E.~Lusso\orcid{0000-0003-0083-1157}\inst{\ref{aff9},\ref{aff10}}
\and L.~Spinoglio\orcid{0000-0001-8840-1551}\inst{\ref{aff11}}
\and V.~Allevato\orcid{0000-0001-7232-5152}\inst{\ref{aff12}}
\and F.~Ricci\orcid{0000-0001-5742-5980}\inst{\ref{aff13},\ref{aff14}}
\and A.~Feltre\orcid{0000-0001-6865-2871}\inst{\ref{aff9}}
\and F.~Mannucci\orcid{0000-0002-4803-2381}\inst{\ref{aff9}}
\and M.~Salvato\orcid{0000-0001-7116-9303}\inst{\ref{aff15}}
\and R.~A.~A.~Bowler\orcid{0000-0003-3917-1678}\inst{\ref{aff16}}
\and M.~Mignoli\orcid{0000-0002-9087-2835}\inst{\ref{aff6}}
\and D.~Vergani\orcid{0000-0003-0898-2216}\inst{\ref{aff6}}
\and F.~La~Franca\orcid{0000-0002-1239-2721}\inst{\ref{aff13}}
\and A.~Amara\inst{\ref{aff17}}
\and S.~Andreon\orcid{0000-0002-2041-8784}\inst{\ref{aff18}}
\and N.~Auricchio\orcid{0000-0003-4444-8651}\inst{\ref{aff6}}
\and M.~Baldi\orcid{0000-0003-4145-1943}\inst{\ref{aff19},\ref{aff6},\ref{aff20}}
\and S.~Bardelli\orcid{0000-0002-8900-0298}\inst{\ref{aff6}}
\and R.~Bender\orcid{0000-0001-7179-0626}\inst{\ref{aff15},\ref{aff21}}
\and C.~Bodendorf\inst{\ref{aff15}}
\and D.~Bonino\orcid{0000-0002-3336-9977}\inst{\ref{aff22}}
\and E.~Branchini\orcid{0000-0002-0808-6908}\inst{\ref{aff23},\ref{aff24},\ref{aff18}}
\and M.~Brescia\orcid{0000-0001-9506-5680}\inst{\ref{aff25},\ref{aff12},\ref{aff26}}
\and J.~Brinchmann\orcid{0000-0003-4359-8797}\inst{\ref{aff27}}
\and S.~Camera\orcid{0000-0003-3399-3574}\inst{\ref{aff28},\ref{aff29},\ref{aff22}}
\and V.~Capobianco\orcid{0000-0002-3309-7692}\inst{\ref{aff22}}
\and C.~Carbone\orcid{0000-0003-0125-3563}\inst{\ref{aff30}}
\and J.~Carretero\orcid{0000-0002-3130-0204}\inst{\ref{aff31},\ref{aff32}}
\and S.~Casas\orcid{0000-0002-4751-5138}\inst{\ref{aff33}}
\and M.~Castellano\orcid{0000-0001-9875-8263}\inst{\ref{aff14}}
\and S.~Cavuoti\orcid{0000-0002-3787-4196}\inst{\ref{aff12},\ref{aff26}}
\and A.~Cimatti\inst{\ref{aff34}}
\and G.~Congedo\orcid{0000-0003-2508-0046}\inst{\ref{aff35}}
\and C.~J.~Conselice\orcid{0000-0003-1949-7638}\inst{\ref{aff16}}
\and L.~Conversi\orcid{0000-0002-6710-8476}\inst{\ref{aff36},\ref{aff37}}
\and Y.~Copin\orcid{0000-0002-5317-7518}\inst{\ref{aff38}}
\and F.~Courbin\orcid{0000-0003-0758-6510}\inst{\ref{aff39}}
\and H.~M.~Courtois\orcid{0000-0003-0509-1776}\inst{\ref{aff40}}
\and M.~Cropper\orcid{0000-0003-4571-9468}\inst{\ref{aff41}}
\and A.~Da~Silva\orcid{0000-0002-6385-1609}\inst{\ref{aff42},\ref{aff43}}
\and H.~Degaudenzi\orcid{0000-0002-5887-6799}\inst{\ref{aff44}}
\and A.~M.~Di~Giorgio\orcid{0000-0002-4767-2360}\inst{\ref{aff11}}
\and J.~Dinis\inst{\ref{aff42},\ref{aff43}}
\and F.~Dubath\orcid{0000-0002-6533-2810}\inst{\ref{aff44}}
\and X.~Dupac\inst{\ref{aff37}}
\and S.~Dusini\orcid{0000-0002-1128-0664}\inst{\ref{aff45}}
\and M.~Farina\orcid{0000-0002-3089-7846}\inst{\ref{aff11}}
\and S.~Farrens\orcid{0000-0002-9594-9387}\inst{\ref{aff46}}
\and S.~Ferriol\inst{\ref{aff38}}
\and M.~Frailis\orcid{0000-0002-7400-2135}\inst{\ref{aff47}}
\and E.~Franceschi\orcid{0000-0002-0585-6591}\inst{\ref{aff6}}
\and S.~Galeotta\orcid{0000-0002-3748-5115}\inst{\ref{aff47}}
\and B.~Gillis\orcid{0000-0002-4478-1270}\inst{\ref{aff35}}
\and C.~Giocoli\orcid{0000-0002-9590-7961}\inst{\ref{aff6},\ref{aff48}}
\and A.~Grazian\orcid{0000-0002-5688-0663}\inst{\ref{aff49}}
\and F.~Grupp\inst{\ref{aff15},\ref{aff21}}
\and L.~Guzzo\orcid{0000-0001-8264-5192}\inst{\ref{aff50},\ref{aff18}}
\and S.~V.~H.~Haugan\orcid{0000-0001-9648-7260}\inst{\ref{aff51}}
\and H.~Hoekstra\orcid{0000-0002-0641-3231}\inst{\ref{aff52}}
\and M.~S.~Holliman\inst{\ref{aff53}}
\and W.~Holmes\inst{\ref{aff8}}
\and I.~Hook\orcid{0000-0002-2960-978X}\inst{\ref{aff54}}
\and F.~Hormuth\inst{\ref{aff55}}
\and A.~Hornstrup\orcid{0000-0002-3363-0936}\inst{\ref{aff56},\ref{aff57}}
\and P.~Hudelot\inst{\ref{aff58}}
\and K.~Jahnke\orcid{0000-0003-3804-2137}\inst{\ref{aff5}}
\and E.~Keih\"anen\orcid{0000-0003-1804-7715}\inst{\ref{aff59}}
\and S.~Kermiche\orcid{0000-0002-0302-5735}\inst{\ref{aff60}}
\and A.~Kiessling\orcid{0000-0002-2590-1273}\inst{\ref{aff8}}
\and B.~Kubik\orcid{0009-0006-5823-4880}\inst{\ref{aff38}}
\and M.~K\"ummel\orcid{0000-0003-2791-2117}\inst{\ref{aff21}}
\and M.~Kunz\orcid{0000-0002-3052-7394}\inst{\ref{aff61}}
\and H.~Kurki-Suonio\orcid{0000-0002-4618-3063}\inst{\ref{aff62},\ref{aff63}}
\and R.~Laureijs\inst{\ref{aff64}}
\and S.~Ligori\orcid{0000-0003-4172-4606}\inst{\ref{aff22}}
\and P.~B.~Lilje\orcid{0000-0003-4324-7794}\inst{\ref{aff51}}
\and V.~Lindholm\orcid{0000-0003-2317-5471}\inst{\ref{aff62},\ref{aff63}}
\and I.~Lloro\inst{\ref{aff65}}
\and D.~Maino\inst{\ref{aff50},\ref{aff30},\ref{aff66}}
\and E.~Maiorano\orcid{0000-0003-2593-4355}\inst{\ref{aff6}}
\and O.~Mansutti\orcid{0000-0001-5758-4658}\inst{\ref{aff47}}
\and O.~Marggraf\orcid{0000-0001-7242-3852}\inst{\ref{aff67}}
\and K.~Markovic\orcid{0000-0001-6764-073X}\inst{\ref{aff8}}
\and N.~Martinet\orcid{0000-0003-2786-7790}\inst{\ref{aff68}}
\and F.~Marulli\orcid{0000-0002-8850-0303}\inst{\ref{aff69},\ref{aff6},\ref{aff20}}
\and R.~Massey\orcid{0000-0002-6085-3780}\inst{\ref{aff4}}
\and E.~Medinaceli\orcid{0000-0002-4040-7783}\inst{\ref{aff6}}
\and S.~Mei\orcid{0000-0002-2849-559X}\inst{\ref{aff70}}
\and M.~Melchior\inst{\ref{aff71}}
\and Y.~Mellier\inst{\ref{aff72},\ref{aff58}}
\and M.~Meneghetti\orcid{0000-0003-1225-7084}\inst{\ref{aff6},\ref{aff20}}
\and E.~Merlin\orcid{0000-0001-6870-8900}\inst{\ref{aff14}}
\and G.~Meylan\inst{\ref{aff39}}
\and M.~Moresco\orcid{0000-0002-7616-7136}\inst{\ref{aff69},\ref{aff6}}
\and L.~Moscardini\orcid{0000-0002-3473-6716}\inst{\ref{aff69},\ref{aff6},\ref{aff20}}
\and E.~Munari\orcid{0000-0002-1751-5946}\inst{\ref{aff47},\ref{aff73}}
\and S.-M.~Niemi\inst{\ref{aff64}}
\and J.~W.~Nightingale\orcid{0000-0002-8987-7401}\inst{\ref{aff74},\ref{aff75}}
\and C.~Padilla\orcid{0000-0001-7951-0166}\inst{\ref{aff76}}
\and S.~Paltani\orcid{0000-0002-8108-9179}\inst{\ref{aff44}}
\and F.~Pasian\orcid{0000-0002-4869-3227}\inst{\ref{aff47}}
\and K.~Pedersen\inst{\ref{aff77}}
\and W.~J.~Percival\orcid{0000-0002-0644-5727}\inst{\ref{aff78},\ref{aff79},\ref{aff80}}
\and V.~Pettorino\inst{\ref{aff64}}
\and G.~Polenta\orcid{0000-0003-4067-9196}\inst{\ref{aff81}}
\and M.~Poncet\inst{\ref{aff82}}
\and L.~A.~Popa\inst{\ref{aff83}}
\and L.~Pozzetti\orcid{0000-0001-7085-0412}\inst{\ref{aff6}}
\and F.~Raison\orcid{0000-0002-7819-6918}\inst{\ref{aff15}}
\and R.~Rebolo\inst{\ref{aff84},\ref{aff85}}
\and A.~Renzi\orcid{0000-0001-9856-1970}\inst{\ref{aff3},\ref{aff45}}
\and J.~Rhodes\inst{\ref{aff8}}
\and G.~Riccio\inst{\ref{aff12}}
\and Hans-Walter~Rix\orcid{0000-0003-4996-9069}\inst{\ref{aff5}}
\and E.~Romelli\orcid{0000-0003-3069-9222}\inst{\ref{aff47}}
\and M.~Roncarelli\orcid{0000-0001-9587-7822}\inst{\ref{aff6}}
\and E.~Rossetti\orcid{0000-0003-0238-4047}\inst{\ref{aff19}}
\and R.~Saglia\orcid{0000-0003-0378-7032}\inst{\ref{aff21},\ref{aff15}}
\and D.~Sapone\orcid{0000-0001-7089-4503}\inst{\ref{aff86}}
\and B.~Sartoris\orcid{0000-0003-1337-5269}\inst{\ref{aff21},\ref{aff47}}
\and R.~Scaramella\orcid{0000-0003-2229-193X}\inst{\ref{aff14},\ref{aff87}}
\and M.~Schirmer\orcid{0000-0003-2568-9994}\inst{\ref{aff5}}
\and P.~Schneider\orcid{0000-0001-8561-2679}\inst{\ref{aff67}}
\and T.~Schrabback\orcid{0000-0002-6987-7834}\inst{\ref{aff88}}
\and A.~Secroun\orcid{0000-0003-0505-3710}\inst{\ref{aff60}}
\and G.~Seidel\orcid{0000-0003-2907-353X}\inst{\ref{aff5}}
\and S.~Serrano\orcid{0000-0002-0211-2861}\inst{\ref{aff89},\ref{aff90},\ref{aff91}}
\and C.~Sirignano\orcid{0000-0002-0995-7146}\inst{\ref{aff3},\ref{aff45}}
\and G.~Sirri\orcid{0000-0003-2626-2853}\inst{\ref{aff20}}
\and L.~Stanco\orcid{0000-0002-9706-5104}\inst{\ref{aff45}}
\and C.~Surace\orcid{0000-0003-2592-0113}\inst{\ref{aff68}}
\and P.~Tallada-Cresp\'{i}\orcid{0000-0002-1336-8328}\inst{\ref{aff31},\ref{aff32}}
\and D.~Tavagnacco\orcid{0000-0001-7475-9894}\inst{\ref{aff47}}
\and A.~N.~Taylor\inst{\ref{aff35}}
\and H.~I.~Teplitz\orcid{0000-0002-7064-5424}\inst{\ref{aff92}}
\and I.~Tereno\inst{\ref{aff42},\ref{aff93}}
\and R.~Toledo-Moreo\orcid{0000-0002-2997-4859}\inst{\ref{aff94}}
\and F.~Torradeflot\orcid{0000-0003-1160-1517}\inst{\ref{aff32},\ref{aff31}}
\and I.~Tutusaus\orcid{0000-0002-3199-0399}\inst{\ref{aff95}}
\and L.~Valenziano\orcid{0000-0002-1170-0104}\inst{\ref{aff6},\ref{aff96}}
\and T.~Vassallo\orcid{0000-0001-6512-6358}\inst{\ref{aff21},\ref{aff47}}
\and A.~Veropalumbo\orcid{0000-0003-2387-1194}\inst{\ref{aff18},\ref{aff24}}
\and Y.~Wang\orcid{0000-0002-4749-2984}\inst{\ref{aff92}}
\and J.~Weller\orcid{0000-0002-8282-2010}\inst{\ref{aff21},\ref{aff15}}
\and E.~Zucca\orcid{0000-0002-5845-8132}\inst{\ref{aff6}}
\and A.~Biviano\orcid{0000-0002-0857-0732}\inst{\ref{aff47},\ref{aff73}}
\and M.~Bolzonella\orcid{0000-0003-3278-4607}\inst{\ref{aff6}}
\and E.~Bozzo\orcid{0000-0002-8201-1525}\inst{\ref{aff44}}
\and C.~Burigana\orcid{0000-0002-3005-5796}\inst{\ref{aff2},\ref{aff96}}
\and C.~Colodro-Conde\inst{\ref{aff84}}
\and G.~De~Lucia\orcid{0000-0002-6220-9104}\inst{\ref{aff47}}
\and D.~Di~Ferdinando\inst{\ref{aff20}}
\and J.~A.~Escartin~Vigo\inst{\ref{aff15}}
\and R.~Farinelli\inst{\ref{aff6}}
\and K.~George\orcid{0000-0002-1734-8455}\inst{\ref{aff21}}
\and J.~Gracia-Carpio\inst{\ref{aff15}}
\and M.~Martinelli\orcid{0000-0002-6943-7732}\inst{\ref{aff14},\ref{aff87}}
\and N.~Mauri\orcid{0000-0001-8196-1548}\inst{\ref{aff34},\ref{aff20}}
\and C.~Neissner\orcid{0000-0001-8524-4968}\inst{\ref{aff76},\ref{aff32}}
\and Z.~Sakr\orcid{0000-0002-4823-3757}\inst{\ref{aff97},\ref{aff95},\ref{aff98}}
\and V.~Scottez\inst{\ref{aff72},\ref{aff99}}
\and M.~Tenti\orcid{0000-0002-4254-5901}\inst{\ref{aff20}}
\and M.~Viel\orcid{0000-0002-2642-5707}\inst{\ref{aff73},\ref{aff47},\ref{aff100},\ref{aff101},\ref{aff102}}
\and M.~Wiesmann\orcid{0009-0000-8199-5860}\inst{\ref{aff51}}
\and Y.~Akrami\orcid{0000-0002-2407-7956}\inst{\ref{aff103},\ref{aff104}}
\and S.~Anselmi\orcid{0000-0002-3579-9583}\inst{\ref{aff45},\ref{aff3},\ref{aff105}}
\and C.~Baccigalupi\orcid{0000-0002-8211-1630}\inst{\ref{aff100},\ref{aff47},\ref{aff101},\ref{aff73}}
\and M.~Ballardini\orcid{0000-0003-4481-3559}\inst{\ref{aff106},\ref{aff6},\ref{aff107}}
\and M.~Bethermin\orcid{0000-0002-3915-2015}\inst{\ref{aff108},\ref{aff68}}
\and A.~Blanchard\orcid{0000-0001-8555-9003}\inst{\ref{aff95}}
\and L.~Blot\orcid{0000-0002-9622-7167}\inst{\ref{aff109},\ref{aff105}}
\and S.~Borgani\orcid{0000-0001-6151-6439}\inst{\ref{aff110},\ref{aff73},\ref{aff47},\ref{aff101}}
\and S.~Bruton\orcid{0000-0002-6503-5218}\inst{\ref{aff111}}
\and R.~Cabanac\orcid{0000-0001-6679-2600}\inst{\ref{aff95}}
\and A.~Calabro\orcid{0000-0003-2536-1614}\inst{\ref{aff14}}
\and G.~Canas-Herrera\orcid{0000-0003-2796-2149}\inst{\ref{aff64},\ref{aff112}}
\and A.~Cappi\inst{\ref{aff6},\ref{aff113}}
\and C.~S.~Carvalho\inst{\ref{aff93}}
\and G.~Castignani\orcid{0000-0001-6831-0687}\inst{\ref{aff6}}
\and T.~Castro\orcid{0000-0002-6292-3228}\inst{\ref{aff47},\ref{aff101},\ref{aff73},\ref{aff102}}
\and K.~C.~Chambers\orcid{0000-0001-6965-7789}\inst{\ref{aff114}}
\and S.~Contarini\orcid{0000-0002-9843-723X}\inst{\ref{aff15},\ref{aff69}}
\and T.~Contini\orcid{0000-0003-0275-938X}\inst{\ref{aff95}}
\and A.~R.~Cooray\orcid{0000-0002-3892-0190}\inst{\ref{aff115}}
\and O.~Cucciati\orcid{0000-0002-9336-7551}\inst{\ref{aff6}}
\and S.~Davini\orcid{0000-0003-3269-1718}\inst{\ref{aff24}}
\and B.~De~Caro\inst{\ref{aff45},\ref{aff3}}
\and G.~Desprez\inst{\ref{aff116}}
\and A.~D\'iaz-S\'anchez\orcid{0000-0003-0748-4768}\inst{\ref{aff117}}
\and S.~Di~Domizio\orcid{0000-0003-2863-5895}\inst{\ref{aff23},\ref{aff24}}
\and H.~Dole\orcid{0000-0002-9767-3839}\inst{\ref{aff118}}
\and S.~Escoffier\orcid{0000-0002-2847-7498}\inst{\ref{aff60}}
\and A.~G.~Ferrari\orcid{0009-0005-5266-4110}\inst{\ref{aff34},\ref{aff20}}
\and I.~Ferrero\orcid{0000-0002-1295-1132}\inst{\ref{aff51}}
\and F.~Finelli\orcid{0000-0002-6694-3269}\inst{\ref{aff6},\ref{aff96}}
\and A.~Fontana\orcid{0000-0003-3820-2823}\inst{\ref{aff14}}
\and F.~Fornari\orcid{0000-0003-2979-6738}\inst{\ref{aff96}}
\and L.~Gabarra\orcid{0000-0002-8486-8856}\inst{\ref{aff119}}
\and K.~Ganga\orcid{0000-0001-8159-8208}\inst{\ref{aff70}}
\and J.~Garc\'ia-Bellido\orcid{0000-0002-9370-8360}\inst{\ref{aff103}}
\and V.~Gautard\inst{\ref{aff120}}
\and E.~Gaztanaga\orcid{0000-0001-9632-0815}\inst{\ref{aff90},\ref{aff89},\ref{aff121}}
\and F.~Giacomini\orcid{0000-0002-3129-2814}\inst{\ref{aff20}}
\and G.~Gozaliasl\orcid{0000-0002-0236-919X}\inst{\ref{aff122},\ref{aff62}}
\and A.~Hall\orcid{0000-0002-3139-8651}\inst{\ref{aff35}}
\and H.~Hildebrandt\orcid{0000-0002-9814-3338}\inst{\ref{aff123}}
\and J.~Hjorth\orcid{0000-0002-4571-2306}\inst{\ref{aff124}}
\and J.~J.~E.~Kajava\orcid{0000-0002-3010-8333}\inst{\ref{aff125},\ref{aff126}}
\and V.~Kansal\orcid{0000-0002-4008-6078}\inst{\ref{aff127},\ref{aff128}}
\and D.~Karagiannis\orcid{0000-0002-4927-0816}\inst{\ref{aff129},\ref{aff130}}
\and C.~C.~Kirkpatrick\inst{\ref{aff59}}
\and L.~Legrand\orcid{0000-0003-0610-5252}\inst{\ref{aff131}}
\and G.~Libet\inst{\ref{aff82}}
\and A.~Loureiro\orcid{0000-0002-4371-0876}\inst{\ref{aff132},\ref{aff133}}
\and J.~Macias-Perez\orcid{0000-0002-5385-2763}\inst{\ref{aff134}}
\and G.~Maggio\orcid{0000-0003-4020-4836}\inst{\ref{aff47}}
\and M.~Magliocchetti\orcid{0000-0001-9158-4838}\inst{\ref{aff11}}
\and R.~Maoli\orcid{0000-0002-6065-3025}\inst{\ref{aff135},\ref{aff14}}
\and C.~J.~A.~P.~Martins\orcid{0000-0002-4886-9261}\inst{\ref{aff136},\ref{aff27}}
\and S.~Matthew\inst{\ref{aff35}}
\and L.~Maurin\orcid{0000-0002-8406-0857}\inst{\ref{aff118}}
\and R.~B.~Metcalf\orcid{0000-0003-3167-2574}\inst{\ref{aff69},\ref{aff6}}
\and P.~Monaco\orcid{0000-0003-2083-7564}\inst{\ref{aff110},\ref{aff47},\ref{aff101},\ref{aff73}}
\and C.~Moretti\orcid{0000-0003-3314-8936}\inst{\ref{aff100},\ref{aff102},\ref{aff47},\ref{aff73},\ref{aff101}}
\and G.~Morgante\inst{\ref{aff6}}
\and S.~Nadathur\orcid{0000-0001-9070-3102}\inst{\ref{aff121}}
\and L.~Nicastro\orcid{0000-0001-8534-6788}\inst{\ref{aff6}}
\and Nicholas~A.~Walton\orcid{0000-0003-3983-8778}\inst{\ref{aff137}}
\and L.~Patrizii\inst{\ref{aff20}}
\and A.~Pezzotta\orcid{0000-0003-0726-2268}\inst{\ref{aff15}}
\and M.~P\"ontinen\orcid{0000-0001-5442-2530}\inst{\ref{aff62}}
\and V.~Popa\inst{\ref{aff83}}
\and C.~Porciani\orcid{0000-0002-7797-2508}\inst{\ref{aff67}}
\and D.~Potter\orcid{0000-0002-0757-5195}\inst{\ref{aff138}}
\and I.~Risso\orcid{0000-0003-2525-7761}\inst{\ref{aff139}}
\and P.-F.~Rocci\inst{\ref{aff118}}
\and M.~Sahl\'en\orcid{0000-0003-0973-4804}\inst{\ref{aff140}}
\and A.~G.~S\'anchez\orcid{0000-0003-1198-831X}\inst{\ref{aff15}}
\and A.~Schneider\orcid{0000-0001-7055-8104}\inst{\ref{aff138}}
\and E.~Sefusatti\orcid{0000-0003-0473-1567}\inst{\ref{aff47},\ref{aff73},\ref{aff101}}
\and M.~Sereno\orcid{0000-0003-0302-0325}\inst{\ref{aff6},\ref{aff20}}
\and P.~Simon\inst{\ref{aff67}}
\and A.~Spurio~Mancini\orcid{0000-0001-5698-0990}\inst{\ref{aff141},\ref{aff41}}
\and J.~Steinwagner\inst{\ref{aff15}}
\and G.~Testera\inst{\ref{aff24}}
\and R.~Teyssier\orcid{0000-0001-7689-0933}\inst{\ref{aff142}}
\and S.~Toft\orcid{0000-0003-3631-7176}\inst{\ref{aff57},\ref{aff143},\ref{aff144}}
\and S.~Tosi\orcid{0000-0002-7275-9193}\inst{\ref{aff23},\ref{aff24},\ref{aff18}}
\and A.~Troja\orcid{0000-0003-0239-4595}\inst{\ref{aff3},\ref{aff45}}
\and M.~Tucci\inst{\ref{aff44}}
\and C.~Valieri\inst{\ref{aff20}}
\and J.~Valiviita\orcid{0000-0001-6225-3693}\inst{\ref{aff62},\ref{aff63}}
\and G.~Verza\orcid{0000-0002-1886-8348}\inst{\ref{aff145},\ref{aff146}}
\and J.~R.~Weaver\orcid{0000-0003-1614-196X}\inst{\ref{aff147}}
\and I.~A.~Zinchenko\inst{\ref{aff21}}}
										   
\institute{School of Physics, HH Wills Physics Laboratory, University of Bristol, Tyndall Avenue, Bristol, BS8 1TL, UK\label{aff1}
\and
INAF, Istituto di Radioastronomia, Via Piero Gobetti 101, 40129 Bologna, Italy\label{aff2}
\and
Dipartimento di Fisica e Astronomia "G. Galilei", Universit\`a di Padova, Via Marzolo 8, 35131 Padova, Italy\label{aff3}
\and
Department of Physics, Centre for Extragalactic Astronomy, Durham University, South Road, DH1 3LE, UK\label{aff4}
\and
Max-Planck-Institut f\"ur Astronomie, K\"onigstuhl 17, 69117 Heidelberg, Germany\label{aff5}
\and
INAF-Osservatorio di Astrofisica e Scienza dello Spazio di Bologna, Via Piero Gobetti 93/3, 40129 Bologna, Italy\label{aff6}
\and
Department of Physics and Astronomy, University of Southampton, Southampton, SO17 1BJ, UK\label{aff7}
\and
Jet Propulsion Laboratory, California Institute of Technology, 4800 Oak Grove Drive, Pasadena, CA, 91109, USA\label{aff8}
\and
INAF-Osservatorio Astrofisico di Arcetri, Largo E. Fermi 5, 50125, Firenze, Italy\label{aff9}
\and
Dipartimento di Fisica e Astronomia, Universit\`{a} di Firenze, via G. Sansone 1, 50019 Sesto Fiorentino, Firenze, Italy\label{aff10}
\and
INAF-Istituto di Astrofisica e Planetologia Spaziali, via del Fosso del Cavaliere, 100, 00100 Roma, Italy\label{aff11}
\and
INAF-Osservatorio Astronomico di Capodimonte, Via Moiariello 16, 80131 Napoli, Italy\label{aff12}
\and
Department of Mathematics and Physics, Roma Tre University, Via della Vasca Navale 84, 00146 Rome, Italy\label{aff13}
\and
INAF-Osservatorio Astronomico di Roma, Via Frascati 33, 00078 Monteporzio Catone, Italy\label{aff14}
\and
Max Planck Institute for Extraterrestrial Physics, Giessenbachstr. 1, 85748 Garching, Germany\label{aff15}
\and
Jodrell Bank Centre for Astrophysics, Department of Physics and Astronomy, University of Manchester, Oxford Road, Manchester M13 9PL, UK\label{aff16}
\and
School of Mathematics and Physics, University of Surrey, Guildford, Surrey, GU2 7XH, UK\label{aff17}
\and
INAF-Osservatorio Astronomico di Brera, Via Brera 28, 20122 Milano, Italy\label{aff18}
\and
Dipartimento di Fisica e Astronomia, Universit\`a di Bologna, Via Gobetti 93/2, 40129 Bologna, Italy\label{aff19}
\and
INFN-Sezione di Bologna, Viale Berti Pichat 6/2, 40127 Bologna, Italy\label{aff20}
\and
Universit\"ats-Sternwarte M\"unchen, Fakult\"at f\"ur Physik, Ludwig-Maximilians-Universit\"at M\"unchen, Scheinerstrasse 1, 81679 M\"unchen, Germany\label{aff21}
\and
INAF-Osservatorio Astrofisico di Torino, Via Osservatorio 20, 10025 Pino Torinese (TO), Italy\label{aff22}
\and
Dipartimento di Fisica, Universit\`a di Genova, Via Dodecaneso 33, 16146, Genova, Italy\label{aff23}
\and
INFN-Sezione di Genova, Via Dodecaneso 33, 16146, Genova, Italy\label{aff24}
\and
Department of Physics "E. Pancini", University Federico II, Via Cinthia 6, 80126, Napoli, Italy\label{aff25}
\and
INFN section of Naples, Via Cinthia 6, 80126, Napoli, Italy\label{aff26}
\and
Instituto de Astrof\'isica e Ci\^encias do Espa\c{c}o, Universidade do Porto, CAUP, Rua das Estrelas, PT4150-762 Porto, Portugal\label{aff27}
\and
Dipartimento di Fisica, Universit\`a degli Studi di Torino, Via P. Giuria 1, 10125 Torino, Italy\label{aff28}
\and
INFN-Sezione di Torino, Via P. Giuria 1, 10125 Torino, Italy\label{aff29}
\and
INAF-IASF Milano, Via Alfonso Corti 12, 20133 Milano, Italy\label{aff30}
\and
Centro de Investigaciones Energ\'eticas, Medioambientales y Tecnol\'ogicas (CIEMAT), Avenida Complutense 40, 28040 Madrid, Spain\label{aff31}
\and
Port d'Informaci\'{o} Cient\'{i}fica, Campus UAB, C. Albareda s/n, 08193 Bellaterra (Barcelona), Spain\label{aff32}
\and
Institute for Theoretical Particle Physics and Cosmology (TTK), RWTH Aachen University, 52056 Aachen, Germany\label{aff33}
\and
Dipartimento di Fisica e Astronomia "Augusto Righi" - Alma Mater Studiorum Universit\`a di Bologna, Viale Berti Pichat 6/2, 40127 Bologna, Italy\label{aff34}
\and
Institute for Astronomy, University of Edinburgh, Royal Observatory, Blackford Hill, Edinburgh EH9 3HJ, UK\label{aff35}
\and
European Space Agency/ESRIN, Largo Galileo Galilei 1, 00044 Frascati, Roma, Italy\label{aff36}
\and
ESAC/ESA, Camino Bajo del Castillo, s/n., Urb. Villafranca del Castillo, 28692 Villanueva de la Ca\~nada, Madrid, Spain\label{aff37}
\and
Universit\'e Claude Bernard Lyon 1, CNRS/IN2P3, IP2I Lyon, UMR 5822, Villeurbanne, F-69100, France\label{aff38}
\and
Institute of Physics, Laboratory of Astrophysics, Ecole Polytechnique F\'ed\'erale de Lausanne (EPFL), Observatoire de Sauverny, 1290 Versoix, Switzerland\label{aff39}
\and
UCB Lyon 1, CNRS/IN2P3, IUF, IP2I Lyon, 4 rue Enrico Fermi, 69622 Villeurbanne, France\label{aff40}
\and
Mullard Space Science Laboratory, University College London, Holmbury St Mary, Dorking, Surrey RH5 6NT, UK\label{aff41}
\and
Departamento de F\'isica, Faculdade de Ci\^encias, Universidade de Lisboa, Edif\'icio C8, Campo Grande, PT1749-016 Lisboa, Portugal\label{aff42}
\and
Instituto de Astrof\'isica e Ci\^encias do Espa\c{c}o, Faculdade de Ci\^encias, Universidade de Lisboa, Campo Grande, 1749-016 Lisboa, Portugal\label{aff43}
\and
Department of Astronomy, University of Geneva, ch. d'Ecogia 16, 1290 Versoix, Switzerland\label{aff44}
\and
INFN-Padova, Via Marzolo 8, 35131 Padova, Italy\label{aff45}
\and
Universit\'e Paris-Saclay, Universit\'e Paris Cit\'e, CEA, CNRS, AIM, 91191, Gif-sur-Yvette, France\label{aff46}
\and
INAF-Osservatorio Astronomico di Trieste, Via G. B. Tiepolo 11, 34143 Trieste, Italy\label{aff47}
\and
Istituto Nazionale di Fisica Nucleare, Sezione di Bologna, Via Irnerio 46, 40126 Bologna, Italy\label{aff48}
\and
INAF-Osservatorio Astronomico di Padova, Via dell'Osservatorio 5, 35122 Padova, Italy\label{aff49}
\and
Dipartimento di Fisica "Aldo Pontremoli", Universit\`a degli Studi di Milano, Via Celoria 16, 20133 Milano, Italy\label{aff50}
\and
Institute of Theoretical Astrophysics, University of Oslo, P.O. Box 1029 Blindern, 0315 Oslo, Norway\label{aff51}
\and
Leiden Observatory, Leiden University, Einsteinweg 55, 2333 CC Leiden, The Netherlands\label{aff52}
\and
Higgs Centre for Theoretical Physics, School of Physics and Astronomy, The University of Edinburgh, Edinburgh EH9 3FD, UK\label{aff53}
\and
Department of Physics, Lancaster University, Lancaster, LA1 4YB, UK\label{aff54}
\and
Felix Hormuth Engineering, Goethestr. 17, 69181 Leimen, Germany\label{aff55}
\and
Technical University of Denmark, Elektrovej 327, 2800 Kgs. Lyngby, Denmark\label{aff56}
\and
Cosmic Dawn Center (DAWN), Denmark\label{aff57}
\and
Institut d'Astrophysique de Paris, UMR 7095, CNRS, and Sorbonne Universit\'e, 98 bis boulevard Arago, 75014 Paris, France\label{aff58}
\and
Department of Physics and Helsinki Institute of Physics, Gustaf H\"allstr\"omin katu 2, 00014 University of Helsinki, Finland\label{aff59}
\and
Aix-Marseille Universit\'e, CNRS/IN2P3, CPPM, Marseille, France\label{aff60}
\and
Universit\'e de Gen\`eve, D\'epartement de Physique Th\'eorique and Centre for Astroparticle Physics, 24 quai Ernest-Ansermet, CH-1211 Gen\`eve 4, Switzerland\label{aff61}
\and
Department of Physics, P.O. Box 64, 00014 University of Helsinki, Finland\label{aff62}
\and
Helsinki Institute of Physics, Gustaf H{\"a}llstr{\"o}min katu 2, University of Helsinki, Helsinki, Finland\label{aff63}
\and
European Space Agency/ESTEC, Keplerlaan 1, 2201 AZ Noordwijk, The Netherlands\label{aff64}
\and
NOVA optical infrared instrumentation group at ASTRON, Oude Hoogeveensedijk 4, 7991PD, Dwingeloo, The Netherlands\label{aff65}
\and
INFN-Sezione di Milano, Via Celoria 16, 20133 Milano, Italy\label{aff66}
\and
Universit\"at Bonn, Argelander-Institut f\"ur Astronomie, Auf dem H\"ugel 71, 53121 Bonn, Germany\label{aff67}
\and
Aix-Marseille Universit\'e, CNRS, CNES, LAM, Marseille, France\label{aff68}
\and
Dipartimento di Fisica e Astronomia "Augusto Righi" - Alma Mater Studiorum Universit\`a di Bologna, via Piero Gobetti 93/2, 40129 Bologna, Italy\label{aff69}
\and
Universit\'e Paris Cit\'e, CNRS, Astroparticule et Cosmologie, 75013 Paris, France\label{aff70}
\and
University of Applied Sciences and Arts of Northwestern Switzerland, School of Engineering, 5210 Windisch, Switzerland\label{aff71}
\and
Institut d'Astrophysique de Paris, 98bis Boulevard Arago, 75014, Paris, France\label{aff72}
\and
IFPU, Institute for Fundamental Physics of the Universe, via Beirut 2, 34151 Trieste, Italy\label{aff73}
\and
School of Mathematics, Statistics and Physics, Newcastle University, Herschel Building, Newcastle-upon-Tyne, NE1 7RU, UK\label{aff74}
\and
Department of Physics, Institute for Computational Cosmology, Durham University, South Road, DH1 3LE, UK\label{aff75}
\and
Institut de F\'{i}sica d'Altes Energies (IFAE), The Barcelona Institute of Science and Technology, Campus UAB, 08193 Bellaterra (Barcelona), Spain\label{aff76}
\and
Department of Physics and Astronomy, University of Aarhus, Ny Munkegade 120, DK-8000 Aarhus C, Denmark\label{aff77}
\and
Waterloo Centre for Astrophysics, University of Waterloo, Waterloo, Ontario N2L 3G1, Canada\label{aff78}
\and
Department of Physics and Astronomy, University of Waterloo, Waterloo, Ontario N2L 3G1, Canada\label{aff79}
\and
Perimeter Institute for Theoretical Physics, Waterloo, Ontario N2L 2Y5, Canada\label{aff80}
\and
Space Science Data Center, Italian Space Agency, via del Politecnico snc, 00133 Roma, Italy\label{aff81}
\and
Centre National d'Etudes Spatiales -- Centre spatial de Toulouse, 18 avenue Edouard Belin, 31401 Toulouse Cedex 9, France\label{aff82}
\and
Institute of Space Science, Str. Atomistilor, nr. 409 M\u{a}gurele, Ilfov, 077125, Romania\label{aff83}
\and
Instituto de Astrof\'isica de Canarias, Calle V\'ia L\'actea s/n, 38204, San Crist\'obal de La Laguna, Tenerife, Spain\label{aff84}
\and
Departamento de Astrof\'isica, Universidad de La Laguna, 38206, La Laguna, Tenerife, Spain\label{aff85}
\and
Departamento de F\'isica, FCFM, Universidad de Chile, Blanco Encalada 2008, Santiago, Chile\label{aff86}
\and
INFN-Sezione di Roma, Piazzale Aldo Moro, 2 - c/o Dipartimento di Fisica, Edificio G. Marconi, 00185 Roma, Italy\label{aff87}
\and
Universit\"at Innsbruck, Institut f\"ur Astro- und Teilchenphysik, Technikerstr. 25/8, 6020 Innsbruck, Austria\label{aff88}
\and
Institut d'Estudis Espacials de Catalunya (IEEC),  Edifici RDIT, Campus UPC, 08860 Castelldefels, Barcelona, Spain\label{aff89}
\and
Institute of Space Sciences (ICE, CSIC), Campus UAB, Carrer de Can Magrans, s/n, 08193 Barcelona, Spain\label{aff90}
\and
Satlantis, University Science Park, Sede Bld 48940, Leioa-Bilbao, Spain\label{aff91}
\and
Infrared Processing and Analysis Center, California Institute of Technology, Pasadena, CA 91125, USA\label{aff92}
\and
Instituto de Astrof\'isica e Ci\^encias do Espa\c{c}o, Faculdade de Ci\^encias, Universidade de Lisboa, Tapada da Ajuda, 1349-018 Lisboa, Portugal\label{aff93}
\and
Universidad Polit\'ecnica de Cartagena, Departamento de Electr\'onica y Tecnolog\'ia de Computadoras,  Plaza del Hospital 1, 30202 Cartagena, Spain\label{aff94}
\and
Institut de Recherche en Astrophysique et Plan\'etologie (IRAP), Universit\'e de Toulouse, CNRS, UPS, CNES, 14 Av. Edouard Belin, 31400 Toulouse, France\label{aff95}
\and
INFN-Bologna, Via Irnerio 46, 40126 Bologna, Italy\label{aff96}
\and
Institut f\"ur Theoretische Physik, University of Heidelberg, Philosophenweg 16, 69120 Heidelberg, Germany\label{aff97}
\and
Universit\'e St Joseph; Faculty of Sciences, Beirut, Lebanon\label{aff98}
\and
Junia, EPA department, 41 Bd Vauban, 59800 Lille, France\label{aff99}
\and
SISSA, International School for Advanced Studies, Via Bonomea 265, 34136 Trieste TS, Italy\label{aff100}
\and
INFN, Sezione di Trieste, Via Valerio 2, 34127 Trieste TS, Italy\label{aff101}
\and
ICSC - Centro Nazionale di Ricerca in High Performance Computing, Big Data e Quantum Computing, Via Magnanelli 2, Bologna, Italy\label{aff102}
\and
Instituto de F\'isica Te\'orica UAM-CSIC, Campus de Cantoblanco, 28049 Madrid, Spain\label{aff103}
\and
CERCA/ISO, Department of Physics, Case Western Reserve University, 10900 Euclid Avenue, Cleveland, OH 44106, USA\label{aff104}
\and
Laboratoire Univers et Th\'eorie, Observatoire de Paris, Universit\'e PSL, Universit\'e Paris Cit\'e, CNRS, 92190 Meudon, France\label{aff105}
\and
Dipartimento di Fisica e Scienze della Terra, Universit\`a degli Studi di Ferrara, Via Giuseppe Saragat 1, 44122 Ferrara, Italy\label{aff106}
\and
Istituto Nazionale di Fisica Nucleare, Sezione di Ferrara, Via Giuseppe Saragat 1, 44122 Ferrara, Italy\label{aff107}
\and
Universit\'e de Strasbourg, CNRS, Observatoire astronomique de Strasbourg, UMR 7550, 67000 Strasbourg, France\label{aff108}
\and
Kavli Institute for the Physics and Mathematics of the Universe (WPI), University of Tokyo, Kashiwa, Chiba 277-8583, Japan\label{aff109}
\and
Dipartimento di Fisica - Sezione di Astronomia, Universit\`a di Trieste, Via Tiepolo 11, 34131 Trieste, Italy\label{aff110}
\and
Minnesota Institute for Astrophysics, University of Minnesota, 116 Church St SE, Minneapolis, MN 55455, USA\label{aff111}
\and
Institute Lorentz, Leiden University, Niels Bohrweg 2, 2333 CA Leiden, The Netherlands\label{aff112}
\and
Universit\'e C\^{o}te d'Azur, Observatoire de la C\^{o}te d'Azur, CNRS, Laboratoire Lagrange, Bd de l'Observatoire, CS 34229, 06304 Nice cedex 4, France\label{aff113}
\and
Institute for Astronomy, University of Hawaii, 2680 Woodlawn Drive, Honolulu, HI 96822, USA\label{aff114}
\and
Department of Physics \& Astronomy, University of California Irvine, Irvine CA 92697, USA\label{aff115}
\and
Department of Astronomy \& Physics and Institute for Computational Astrophysics, Saint Mary's University, 923 Robie Street, Halifax, Nova Scotia, B3H 3C3, Canada\label{aff116}
\and
Departamento F\'isica Aplicada, Universidad Polit\'ecnica de Cartagena, Campus Muralla del Mar, 30202 Cartagena, Murcia, Spain\label{aff117}
\and
Universit\'e Paris-Saclay, CNRS, Institut d'astrophysique spatiale, 91405, Orsay, France\label{aff118}
\and
Department of Physics, Oxford University, Keble Road, Oxford OX1 3RH, UK\label{aff119}
\and
CEA Saclay, DFR/IRFU, Service d'Astrophysique, Bat. 709, 91191 Gif-sur-Yvette, France\label{aff120}
\and
Institute of Cosmology and Gravitation, University of Portsmouth, Portsmouth PO1 3FX, UK\label{aff121}
\and
Department of Computer Science, Aalto University, PO Box 15400, Espoo, FI-00 076, Finland\label{aff122}
\and
Ruhr University Bochum, Faculty of Physics and Astronomy, Astronomical Institute (AIRUB), German Centre for Cosmological Lensing (GCCL), 44780 Bochum, Germany\label{aff123}
\and
DARK, Niels Bohr Institute, University of Copenhagen, Jagtvej 155, 2200 Copenhagen, Denmark\label{aff124}
\and
Department of Physics and Astronomy, Vesilinnantie 5, 20014 University of Turku, Finland\label{aff125}
\and
Serco for European Space Agency (ESA), Camino bajo del Castillo, s/n, Urbanizacion Villafranca del Castillo, Villanueva de la Ca\~nada, 28692 Madrid, Spain\label{aff126}
\and
ARC Centre of Excellence for Dark Matter Particle Physics, Melbourne, Australia\label{aff127}
\and
Centre for Astrophysics \& Supercomputing, Swinburne University of Technology, Victoria 3122, Australia\label{aff128}
\and
School of Physics and Astronomy, Queen Mary University of London, Mile End Road, London E1 4NS, UK\label{aff129}
\and
Department of Physics and Astronomy, University of the Western Cape, Bellville, Cape Town, 7535, South Africa\label{aff130}
\and
ICTP South American Institute for Fundamental Research, Instituto de F\'{\i}sica Te\'orica, Universidade Estadual Paulista, S\~ao Paulo, Brazil\label{aff131}
\and
Oskar Klein Centre for Cosmoparticle Physics, Department of Physics, Stockholm University, Stockholm, SE-106 91, Sweden\label{aff132}
\and
Astrophysics Group, Blackett Laboratory, Imperial College London, London SW7 2AZ, UK\label{aff133}
\and
Univ. Grenoble Alpes, CNRS, Grenoble INP, LPSC-IN2P3, 53, Avenue des Martyrs, 38000, Grenoble, France\label{aff134}
\and
Dipartimento di Fisica, Sapienza Universit\`a di Roma, Piazzale Aldo Moro 2, 00185 Roma, Italy\label{aff135}
\and
Centro de Astrof\'{\i}sica da Universidade do Porto, Rua das Estrelas, 4150-762 Porto, Portugal\label{aff136}
\and
Institute of Astronomy, University of Cambridge, Madingley Road, Cambridge CB3 0HA, UK\label{aff137}
\and
Department of Astrophysics, University of Zurich, Winterthurerstrasse 190, 8057 Zurich, Switzerland\label{aff138}
\and
Dipartimento di Fisica, Universit\`a degli studi di Genova, and INFN-Sezione di Genova, via Dodecaneso 33, 16146, Genova, Italy\label{aff139}
\and
Theoretical astrophysics, Department of Physics and Astronomy, Uppsala University, Box 515, 751 20 Uppsala, Sweden\label{aff140}
\and
Department of Physics, Royal Holloway, University of London, TW20 0EX, UK\label{aff141}
\and
Department of Astrophysical Sciences, Peyton Hall, Princeton University, Princeton, NJ 08544, USA\label{aff142}
\and
Cosmic Dawn Center (DAWN)\label{aff143}
\and
Niels Bohr Institute, University of Copenhagen, Jagtvej 128, 2200 Copenhagen, Denmark\label{aff144}
\and
Center for Cosmology and Particle Physics, Department of Physics, New York University, New York, NY 10003, USA\label{aff145}
\and
Center for Computational Astrophysics, Flatiron Institute, 162 5th Avenue, 10010, New York, NY, USA\label{aff146}
\and
Department of Astronomy, University of Massachusetts, Amherst, MA 01003, USA\label{aff147}}    

\date{\today}

\abstract{
We forecast the expected population of active galactic nuclei (AGN) observable in the Euclid Wide Survey (EWS) and Euclid Deep Survey (EDS). Starting from an X-ray luminosity function (XLF) we generate volume-limited samples of the AGN expected in the \Euclid survey footprints. Each AGN is assigned an SED appropriate for its X-ray luminosity and redshift, with perturbations sampled from empirical distributions. The photometric detectability of each AGN is assessed via mock observation of the assigned SED. 
We estimate 40 million AGN will be detectable in at least one \Euclid band in the EWS and 0.24 million in the EDS, corresponding to surface densities of \SDDetectableOneFiltEWS\,deg$^{-2}$ and \SDDetectableOneFiltEDS\,deg$^{-2}$. 
Employing \Euclid-only colour selection criteria on our simulated data we select a sample of \NUMSelectedEuclidOnlyTypeOneEWS (\SDSelectedEuclidOnlyTypeOneEWS\,deg$^{-2}$) AGN in the EWS and \NUMSelectedEuclidOnlyTypeOneEDS (\SDSelectedEuclidOnlyTypeOneEDS\,deg$^{-2}$) in the EDS, amounting to 10\% and 8\% of the AGN detectable in the EWS and EDS. 
Including ancillary Rubin/LSST bands improves the completeness and purity of AGN selection. These data roughly double the total number of selected AGN to comprise 21\% and 15\% of the \Euclid detectable AGN in the EWS and EDS. 
The total expected sample of colour-selected AGN contains 6.0\,$\times$\,10$^{6}$ (74\%) unobscured AGN and 2.1\,$\times$\,10$^{6}$ (26\%) obscured AGN, covering $0.02 \leq z \lesssim 5.2$ and $43 \leq \log_{10} (L_{\rm bol} / {\rm erg}\, {\rm s}^{-1}) \leq 47$.
With this simple colour selection, expected surface densities are already comparable to the yield of modern X-ray and mid-infrared surveys of similar area.
The relative uncertainty on our expectation for \Euclid detectable AGN is 6.7\% for the EWS and 12.5\% for the EDS, driven by the uncertainty of the XLF. 
}


\maketitle

\section{Introduction}
Active galactic nuclei (AGN) mark a phase of luminous accretion of matter onto a central supermassive black hole (SMBH). AGN therefore indicate periods of growth for the massive compact objects, thought to ubiquitously occupy the centers of massive galaxies \citep{magorrian1998}. A co-evolution between AGN and their host-galaxies has long been invoked through a series of observational relations linking their respective physical properties \citep[e.g.,][]{ferrarese&merritt2000,gultekin2009}. The interaction of AGN and their host-galaxies, or AGN feedback \citep[][and references therein]{fabian2012}, is a crucial mechanism to reproduce observed galaxy properties \citep{granato2004,shankar2006, marulli2008} and quench star formation in massive galaxies \citep{gabor2010, dubois2013, piotrowska2022}. In cosmological simulations AGN feedback has been identified as a necessary ingredient to replicate present day distributions of galaxies \citep[e.g.,][]{bower2006,croton2006,sijacki2007,schaye2015,rosito2021,ward2022}.  

The statistical evolution of galaxy and AGN populations over cosmic time can be described by a luminosity function (LF). AGN LFs have been of central importance since quasars (i.e. high-luminosity unobscured AGN) were first discovered as distinct cosmological objects, as early as 1968 \citep{schmidt1968}. The six decade-spanning field of work has seen AGN LFs constructed in the radio \citep[e.g.,][]{dunlop1990}, submillimeter \citep[e.g.,][]{vaccari2010}, infrared \citep[IR; e.g.,][]{gruppioni2013,lacy2015}, UV/optical \citep[e.g.,][]{boyle1987,pei1995,kulkarni2019,adams2023}, X-ray \citep[e.g.,][]{miyaji2000,ueda2003,lafranca2005,ueda2014,aird2015,buchner2015,fotopoulou2016xlf}, and bolometric \citep[e.g.,][]{hopkins2007,shen2020} domains, spanning a wide range of redshifts.

By deriving and comparing LFs from differently selected samples of galaxies or AGN, their formation histories can be compared and inferred \citep[e.g.,][]{dai2009, gruppioni2013, lacy2015}. For example, the evolution of nuclear obscuration in AGN can be probed by measuring and comparing the LFs of obscured and unobscured AGN. Such studies found that high-luminosity obscured AGN peak in space density at a higher redshift than their unobscured counterparts, hinting at an evolutionary scenario \citep[e.g.,][]{lacy2005,hopkins2008,glickman2018}.

Almost all studies have shown that the AGN LF has a strong evolution with redshift, which is an ubiquitous feature across different wavebands \citep[e.g.,][]{boyle2000, ueda2003, richards2006lf, hasinger2005, hopkins2007, gruppioni2013,buchner2015,fotopoulou2016xlf, shen2020}. This evolution is not a simple change of normalisation, but also affects the slope of the parameterising function. For example, X-ray and some optical, radio, and IR studies report a flattening of the faint-end slope of the AGN LF with increasing redshift. This implies the peak space-density of low-luminosity AGN is at a lower redshift than that of bright quasars, indicating a ``cosmic downsizing" of AGN \citep[e.g.,][]{cowie1996, hasinger2005, aird2015}. AGN feedback is often invoked as the mechanism behind this phenomenon, shutting down the supply of in-falling cold gas to the central SMBH via galactic-scale dust ejection with powerful outflows or by the gradual heating of the host-galaxy's dark matter halo \citep[e.g.,][and references therein]{fabian2012}.

The European Space Agency (ESA) \Euclid space telescope \citep{laureijs2011,mellier2024}, successfully launched in July 2023, is the premier dark energy mission of ESA. \Euclid will observe $\sim$14\,500\,deg$^{2}$ of the extra-Galactic sky over a six-year period, providing high spatial resolution imaging sampled at $0\farcs 1$\,pixel$^{-1}$ in the optical \citep{cropper2024} and $\sim 0\farcs 3$\,pixel$^{-1}$ in the near-infrared \citep[NIR;][]{jahnke2024} for billions of astrophysical sources \citep{scaramella2022}. Despite having a primary mission focused on cosmology, the rich data set generated by the \Euclid surveys will drive significant progress in many areas of astronomy. To effectively exploit \Euclid data for AGN legacy science, we must first understand the available sample size and characteristics of the AGN detectable with \Euclid photometry.

This work aims to forecast the expected number of $z < 7$ unobscured and obscured AGN observable with \Euclid photometry in the EWS and EDS. We first determine the sample size and properties of AGN \textit{detectable} with \Euclid photometry, before focusing on the sample of AGN we can \textit{select} with \Euclid and Rubin/LSST \citep{ivezic2019} photometric criteria. For a comprehensive analysis of $7 \leq z \leq 9$ AGN in \Euclid we refer the reader to \citet{barnett2019}. We compare the output of our simulations at $z \sim 7$ with those of \citet{barnett2019} in Appendix \ref{app:barnett}. 

In Sect. \ref{sec:lumfuncs} we introduce the XLF utilized in this work. Section \ref{sec:method} outlines our AGN sample generation method from input data to resulting photometry. Our main results for the number of \textit{detected} AGN in the \Euclid surveys are presented in Sect. \ref{sec:results}. The drivers and impact of uncertainty within our framework, the expected sample sizes of AGN \textit{selected} using \Euclid photometric colour criteria, and comparisons to the yield of AGN from surveys in different wavebands are discussed in Sect. \ref{sec:discussion}.

Throughout this work we refer to AGN as unobscured or obscured based on their optical properties, i.e. \textit{unobscured} AGN denote Type 1 AGN with broad permitted emission lines (full-width at half-maximum; FWHM $\gtrsim 2000$\,km\,s$^{-1}$) and \textit{obscured} AGN signifies Type 2 AGN with no observable broad line components.
We assume a $\Lambda$CDM cosmology with $H_{0}$ = 70\,\kmsMpc, $\Omega_{\rm m}$ = 0.3 and $\Omega_{\rm \Lambda}$ = 0.7. All magnitudes are in the AB system \citep{oke&gunn1983} unless stated otherwise.


\section{X-ray AGN luminosity function}\label{sec:lumfuncs}

The differential LF, $\diff\phi(L, z)/\diff L$, is defined as the number of AGN per unit comoving volume per unit luminosity interval

\begin{equation}\label{data/LF/diffLF}
\frac{\diff\phi(L,z)}{\diff L}  = \frac{\diff^{2}N}{\diff V_{\rm c} \, \diff L}(L,z),
\end{equation}

\noindent where $N$ is the number of AGN with luminosity $L$ in the comoving volume $V_{\rm c}$ at redshift $z$ and $\phi$ is the comoving number density ($\diff N / \diff V_{\rm c}$). 
The expected number of AGN, $\langle N \rangle$, in luminosity and comoving volume interval $\Delta L\,\Delta V_{\rm c}(\Delta z)$ can be extracted from a differential LF as defined in Eq. (\ref{data/LF/diffLF}) via the calculation

\begin{equation}\label{eqn:lf_exp}
    \langle N \rangle = \int_{L_{\rm min}}^{L_{\rm max}} \int_{z_{\rm min}}^{z_{\rm max}(L)} \frac{\diff\phi(L,z)}{\diff L} \,\frac{\diff V_{\rm c}}{\diff z\,\diff\Omega}(z)\;\Omega(L,z)\,\diff z\,\diff L,
\end{equation}

\noindent where $L_{\rm min}$ and $L_{\rm max}$ are the minimum and maximum of the luminosity interval $\Delta L$ respectively, $z_{\rm min}$ is the minimum of the redshift interval $\Delta z$, $z_{\rm max}(L)$ is either the maximum redshift at which an object of luminosity $L$ can still be detected or the maximum of the redshift interval, $\diff V_{\rm c} / \diff z\,\diff\Omega$ is the differential comoving volume element of the Universe at redshift $z$, and $\Omega(L, z)$ is the sky coverage available for an object of luminosity $L$ and redshift $z$. The sky coverage available for an object is calculated from the flux-area curve of the survey(s) for which the expectation value of the number of objects is being derived. 

\begin{figure*}[!ht]
   \centering
   \includegraphics[width=\hsize]{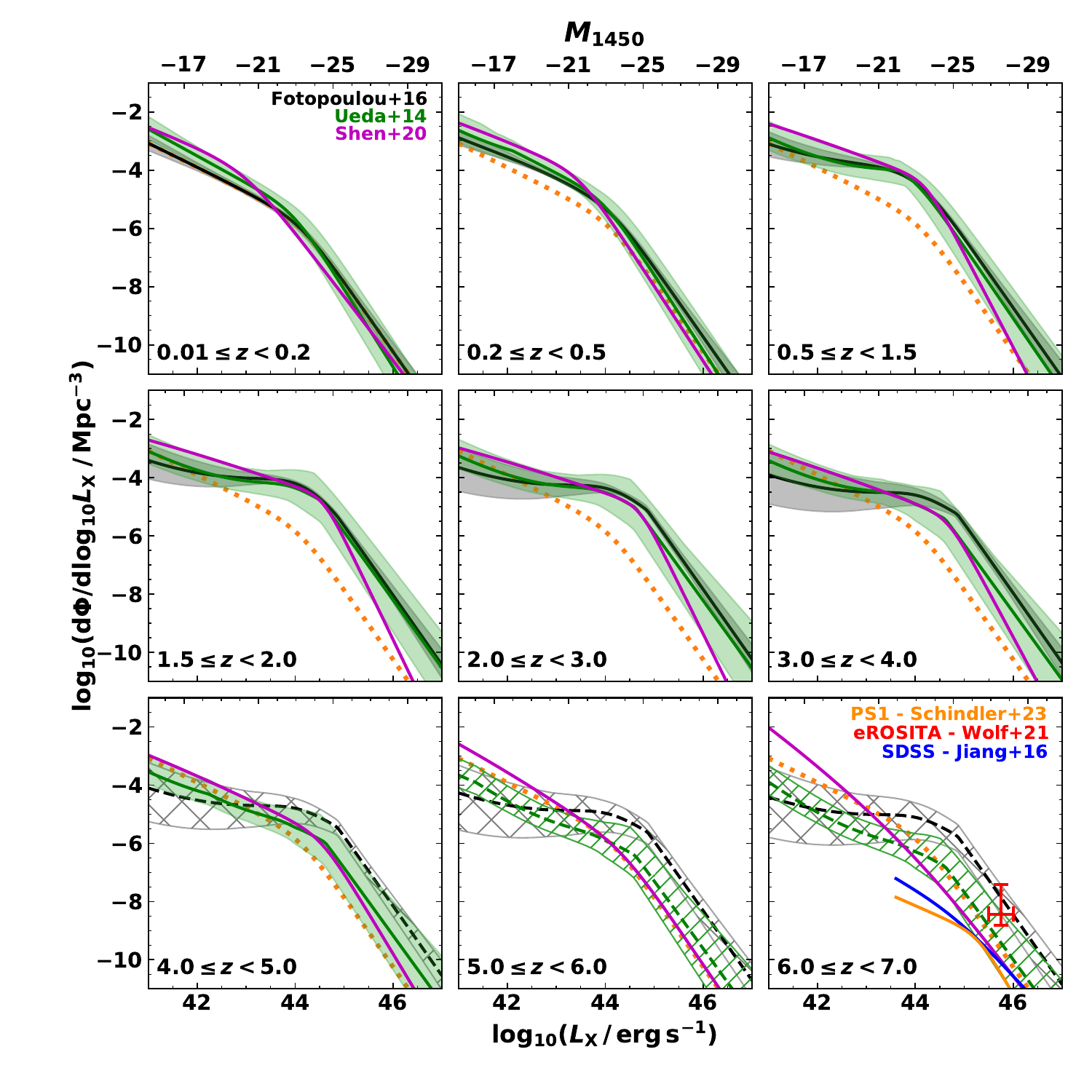}
      \caption{Comparison of different AGN LFs homogenised to the 2--10\,keV X-ray band. Corresponding absolute UV magnitudes at 1450\,\AA{}, $M_{1450}$, are displayed on the upper axes. In each panel the LFs are realised for the central redshift value. The hard X-ray LF of \citet{fotopoulou2016xlf} employed in this work is shown in black. The grey shaded regions depict the 1$\sigma$ uncertainty. For reference, we plot the \citet{fotopoulou2016xlf} XLF evaluated at $z=0.1$ as orange dotted lines in each panel. The hard XLF of \citet{ueda2014} is shown in green, with the green shaded regions corresponding to the 1$\sigma$ uncertainty generated with sampling from the published parameter uncertainties. The magenta lines portray the bolometric quasar LF of \citet{shen2020}, converted to the X-ray domain. In the final panel the \citet{jiang2016} $z>6$ SDSS quasar LF is represented by the blue curve. The solid orange curve gives the \citet{schindler2023} $z\sim6$ quasar LF derived from Pan-STARRS1 and SHELLQs observations. The red uncertainty interval represents eROSITA high-redshift constraints on the XLF \citep{wolf2021}. In all cases dashed curves and hatched uncertainty intervals indicate extrapolation.}
         \label{fig:lf_comparison}
\end{figure*}

This work concerns the determination of the total population of $z < 7$ AGN detectable with \textit{Euclid}. We therefore require an analysis that represents both unobscured and obscured AGN as they are observed in the Universe. Given these considerations, we adopted the observed 5--10\,keV XLF constructed in \cite{fotopoulou2016xlf}. The 5--10\,keV band avoids absorption of the X-ray spectrum up to obscuring hydrogen column densities of $N_{\rm H} \sim$ 10$^{23}$\,cm$^{-2}$. The observed 5--10\,keV fluxes of AGN with $N_{\rm H} =$ 10$^{23}$\,cm$^{-2}$ are $>90\%$ of the intrinsic flux for $z > 1$, and $>80\%$ at lower redshifts. This means the 5--10\,keV band selects an unbiased population of Compton-thin ($N_{\rm H} \lesssim$ 10$^{24}$\,cm$^{-2}$) AGN \citep{dellaceca2008}. 
Even though the 5--10\,keV XLF is constrained by observations up to $z \sim 4$, we show in Fig. \ref{fig:lf_comparison} that our extrapolations are consistent with the latest XLF constraints at much higher redshifts \citep[$z \sim 6$ e.g.,][]{wolf2021,barlowhall2023}.
AGN LFs constructed in IR bands, which also incorporate both unobscured and obscured AGN, are not well constrained beyond $z \sim 3$ \citep[e.g.,][]{gruppioni2013,lacy2015}, therefore requiring greater extrapolation than the XLF with no high-redshift constraints to compare to. Whilst UV/optical AGN LFs are constrained by observations up to $z \sim 7$ \cite[e.g.,][]{wang2019,matsuoka2023}, this waveband probes only the unobscured (quasar) population. Thus, strong assumptions are required to incorporate obscured AGN. By using an extrapolated XLF the total AGN space density, which is consistent with the latest observational constraints, is preserved and treated self-consistently. 

Furthermore, the 5--10\,keV band is straightforwardly reconciled with the 2--10\,keV band assuming a power-law spectrum with a representative photon index. This is advantageous because there are a wide range of established models and empirical analyses connecting 2--10\,keV emission of AGN with bolometric luminosity \citep{shen2020,duras2020}, UV/optical emission \citep{lusso2010}, obscuration properties \citep{merloni2014}, and broad-band SED shapes \citep{salvato2009,fotopoulou2016,shen2020}. We leverage these models in our work to produce robust multiwavelength photometry that represent the best of our current knowledge in connecting X-ray luminosity and redshift of AGN with observed multiwavelength emission from empirical data.

Whilst the 5--10\,keV band is highly complete for Compton-thin AGN, the effect of absorption is more substantial for AGN with $N_{\rm H} =$ 10$^{24}$\,cm$^{-2}$, especially at low redshifts. At $z > 2$ more than 80\% of the intrinsic flux can still be observed. Our choice of the 5--10\,keV XLF is therefore incomplete for Compton-thick AGN ($N_{\rm H} \gtrsim$ 10$^{24}$\,cm$^{-2}$) in the local Universe. The fraction of missed Compton-thick AGN in hard X-ray observations is uncertain, with estimates ranging from 20--40\%, depending on redshift \citep{civano2015,laloux2023,pouliasis2024} and up to 80\%, depending on the selection \citep{fiore2008}. Modelling of the cosmic X-ray background, averaging across redshifts and luminosities, suggests that Compton-thick AGN should be at least as abundant as Compton-thin AGN probed by 2--10\,keV XLFs \citep[][]{gilli2007}. The nature of Compton-thick objects can in principle be confirmed by subsequent observations at higher X-ray energies (e.g., 14--195\,keV) and/or at IR wavelengths, which correct the usual 2--10\,keV X-ray emission \citep[e.g.,][]{spinoglio2022}. 
The number of confirmed hard X-ray identified Compton-thick AGN in the $z \leq 1.5$ Universe is only in the tens of sources \citep[e.g.,][]{ajello2012,civano2015,marchesi2018}, which if missed by our analysis would make a negligible difference to our results.  
Comparing IR and hard X-ray selected samples of AGN at $0.2 < z < 1.2$, \citet{mendez2013} derive an upper limit suggesting that $\sim 10\%$ of IR-selected AGN are Compton-thick.
Compton-thick objects, at least those not optically obscured, will be detected using \Euclid photometry and spectroscopy.
We therefore assess that the numbers of detectable AGN presented in this work may be a conservative estimate due to the selection effects biased against heavily obscured and Compton-thick AGN in the X-ray regime. It is plausible that Compton-thick AGN will add up to an additional $\sim 10\%$ to our \Euclid detected AGN estimates at $z < 2$.

In \cite{fotopoulou2016xlf}, XLF parameters were estimated using a sample of $1\,115$ X-ray selected AGN with $0.01 < z < 4.0$ and $41 < \logten (L_{\rm X} / {\rm erg}\, {\rm s}^{-1}) < 46$. The sample was compiled of AGN from a mixture of wide area, medium area, and pencil-beam X-ray fields: The Monitor of All-sky X-ray Image extra-Galactic survey \citep[MAXI;][]{ueda2011}, The XMM-\textit{Newton} Hard Bright Serendipitous Survey \citep[HBSS;][]{dellaceca2004}, XMM-COSMOS \citep{cappelluti2009}, XMM-Lockman Hole \citep{brunner2008}, XMM-\textit{Chandra} Deep Field South \citep[XMM-CDFS;][]{ranalli2013}, \textit{Chandra}-COSMOS \citep{elvis2009}, AEGIS-X Deep \citep[AEGIS-XD;][]{nandra2015}, and \textit{Chandra}-\textit{Chandra} Deep Field South \citep{xue2011}. Using Bayesian model selection, a luminosity-dependent density evolution (LDDE) model was shown to best describe the data. In LDDE models the number density of AGN changes over cosmic time with low-luminosity and high-luminosity AGN evolving on different timescales. This evolution is implemented through the luminosity dependence of the critical redshift, $z_{\rm c}$. 

The LDDE model presented in \citet{fotopoulou2016xlf} uses the formalism introduced by \citet{ueda2003} and is described by the following

\begin{equation}\label{LDDE_main}
    \frac{\diff \phi(L, z)}{\diff \logten L}  = \frac{\diff\phi(L,\,z=0)}{\diff\logten L} \, \epsilon(L,\,z),
\end{equation}

\noindent where $\diff\phi(L,\,z)/\diff\logten L$ represents the differential LF, $\diff\phi(L,\,z=0)/\diff\logten L$ describes the local ($z\sim0$) LF and $\epsilon(L,\,z)$ denotes the LF evolution factor. The local X-ray LF is well described by a broken power-law distribution

\begin{equation}\label{LDDE_local}
   \frac{\diff\phi(L,\,z=0)}{\diff\logten L} = \frac{A}{\left(\frac{L}{L_{0}}\right)^{\gamma_{1}} + \left( \frac{L}{L_{0}}\right)^{\gamma_{2}}} ,
\end{equation}

\noindent where $A$ is the LF normalization, $L_{0}$ is the luminosity at which the break occurs and $\gamma_{1}$, $\gamma_{2}$ are the slopes of the power-law above and below $L_{0}$. The evolution factor has the form

\begin{equation}\label{LDDE_evolutionfactor}
    \epsilon(L,\,z) = \frac{(1 + z_{\rm c})^{p_{1}} + (1 + z_{\rm c})^{p_{2}}}{\left(\frac{1 + z}{1 + z_{\rm c}}\right)^{-p_{1}} + \left(\frac{1 + z}{1 + z_{\rm c}}\right)^{-p_{2}}},
\end{equation}

\noindent where $p_{1}$, $p_{2}$ are slopes of the evolution factor broken power law and $z_{\rm c}$ is the luminosity-dependent critical redshift, expressed by

\begin{equation}\label{LDDE_zcrit}
    z_{\rm c}(L) =  \left\{ \begin{array}{rcl}
z^{*}_{\rm c} & \mbox{for} & L \geq L_{\alpha} \\
z^{*}_{\rm c}\,\left(\frac{L}{L_{\alpha}}\right)^{\alpha}  & \mbox{for} & L < L_{\alpha} \\
\end{array}\right.,
\end{equation}

\noindent where $z^{*}_{\rm c}$ is the high-luminosity critical redshift. The $\alpha$ exponent and $L_{\alpha}$ luminosity are parameters calculated in the fit of the XLF. For the main simulation, we used the mode of the parameter posteriors presented in \citet{fotopoulou2016xlf} given in Table \ref{tab:XLFparams}. The impact of the parameter uncertainties as well as extrapolation of the XLF is examined in Sect. \ref{sec:uncertainties}.

Throughout this manuscript we primarily consider X-ray luminosity and fluxes in the 2--10\,keV domain in erg\,s$^{-1}$ and erg\,s$^{-1}$\,cm$^{-2}$, respectively. When used as input to the XLF we convert to the 5--10\,keV domain by assuming an AGN X-ray spectrum following a power-law distribution, $F(E) \propto E^{-\Gamma}$ where $\Gamma$ is the photon index. We take $\Gamma$ = 1.9, the midpoint of the range of $\,\Gamma$ = 1.8--2.0 found for samples of radio-quiet AGN \citep[e.g.,][]{nandra&pounds1994,reeves&turner2000,piconcelli2005,page2005,young2009}. 

We compare the XLF invoked in this work with a selection of AGN LFs in Fig. \ref{fig:lf_comparison}. Over the full redshift range considered in this work we compared to the well established $\logten (N_{\rm H} / {\rm cm}^{-2}) \leq 24$ hard XLF of \citet{ueda2014} and the bolometric quasar LF of \citet{shen2020}. In the high-redshift ($z \gtrsim 6$) regime we consider the $z>6$ Sloan Digital Sky Survey \citep[SDSS;][]{york2000} quasar LF of \citet{jiang2016}, which was used to derive $z > 7$ AGN expectations for \Euclid in \citet{barnett2019}, the recent \citet{schindler2023} $z \sim 6$ quasar LF derived from Pan-STARRS1 \citep[PS1;][]{banados2016,banados2023} and Subaru High-z
Exploration of Low-Luminosity Quasars \citep[SHELLQs;][]{matsuoka2018} survey observations, as well as constraints placed on the XLF at $z>6$ by \citet{wolf2021} using the extended ROentgen Survey with an Imaging Telescope Array \citep[eROSITA;][]{merloni2012}. 

To convert the \citet{shen2020} bolometric LF to the 2--10\,keV domain, we applied the X-ray bolometric correction derived and utilized in the same work. Explicitly, the \citet{shen2020} X-ray bolometric correction is parameterised as 
\begin{equation}
    \frac{L_{\rm bol}}{L_{\rm \text{2--10}\,keV}} = c_{1}\left(  \frac{L_{\rm bol}}{10^{10}\,L_{\odot}} \right)^{k_{1}} + c_{2}\left(  \frac{L_{\rm bol}}{10^{10}\,L_{\odot}} \right)^{k_{2}},
\end{equation}
\noindent where ($c_{1}$, $k_{1}$, $c_{2}$, $k_{2}$) = ($3.759$, $-0.361$, $9.830$, $-0.0063$). 

To convert the high-redshift quasar UV LFs of \citet{jiang2016} and \citet{schindler2019} to the 2--10\,keV domain we followed the method outlined in \citet{ricci2017}. Briefly, an assumed X-ray power law with photon index $\Gamma = 1.9$ was used to acquire the monochromatic luminosity at 2\,keV, $L_{\rm 2\,keV}$. This was converted into a monochromatic luminosity at 2500\,\AA{}, $L_{2500}$, through the equation 
\begin{equation}
    \logten{L_{2500}} = (1.050 \pm 0.036)\logten{L_{\rm 2\,keV}} + (2.246 \pm 1.003),
\end{equation}
\noindent an inversion of the relation derived in \citet{lusso2010}. Both monochromatic luminosities are in ${\rm erg\,s}^{-1}\,{\rm cm}^{-2}\,{\rm Hz}^{-1}$. Next, a UV power-law SED $L_{\nu} \propto \nu^{\alpha_{\nu}}$ \citep[e.g.,][]{giallongo2015} with $\alpha_{\nu} = -0.44$ for $1200\,{\text \AA} < \lambda < 5000\,{\text \AA}$ \citep{natali1998,vandenberk2001} and $\alpha_{\nu} = -1.57$ for $228\,{\text \AA} < \lambda < 1200\,{\text \AA}$ \citep{telfer2002} was adopted to obtain the UV monochromatic luminosity at 1450\,\AA{}, $L_{1450}$. Finally we calculated the absolute magnitude at 1450\,\AA, $M_{1450}$, using
\begin{equation}
    L_{1450} = 4 \pi d^{2} 10^{-0.4M_{1450}} f_{0},
\end{equation}
\noindent where $d = 10\,{\rm pc} = 3.0857\times10^{19}\,{\rm cm}$ and $f_{0} = 3.65\times10^{-20}\,{\rm erg\,s}^{-1}\,{\rm cm}^{-2}\,{\rm Hz}^{-1}$ is the zero-point. 

Across the redshift range probed here, the \citet{fotopoulou2016xlf} and \citet{ueda2014} XLFs are consistent within 1$\sigma$ uncertainties. Both of these XLFs are also consistent within 1$\sigma$ with the recent $z < 4$ XLF determination of \citet{peca2023}. At $z \geq 4$, where constraining observational data are sparse and we must extrapolate the XLFs, there is some tension on the normalisation. The \citet{ueda2014} and \citet{shen2020} parametrisations show a steeper decline in space density compared to that of \citet{fotopoulou2016xlf}. The UV/optically derived quasar LFs of \citet{jiang2016} and \citet{schindler2023} agree with this decline. The recent X-ray derived result of \citet{wolf2021} however, advocates for a higher space density at $z>6$, demonstrating consistency with our extrapolation of the \citet{fotopoulou2016xlf} XLF. Through similar means, \citet{barlowhall2023} also derive constraints on the high-redshift XLF that are consistent with \citet{wolf2021}. Our extrapolation of the \citet{fotopoulou2016xlf} XLF is therefore in agreement with the most recent XLF constraints across the entire redshift range considered in this work.

We verified that the space density of our simulated \Euclid-detectable unobscured AGN are consistent with empirical UV/optical quasar samples at $1 \leq z \leq 6$. In the $6 \leq z \leq 7$ regime there is a space-density excess of up to an order of magnitude in our simulated sample at $-26 \lesssim M_{1450} \lesssim -23$. Recent results from observations with The James Webb Space Telescope \citep[JWST;][]{gardner2006} point towards a steeper than expected AGN LF at low luminosities in the $z \gtrsim 3$ regime \citep[e.g.,][]{harikane2023,kocevski2023,maiolino2023}, with some results suggesting that the space density of AGN is up to an order of magnitude greater than extrapolations of quasar UV LFs \citep{mattee2023}. Additionally, there are hints that the UV/optical AGN LF gives an underestimated space density at $z\sim4$ due to the incompleteness of canonically used colour selections \citep[e.g.,][]{boutsia2018}. We also observe an excess of \Euclid-detectable unobscured AGN in our simulation compared to empirical UV/optical quasars at $z < 1$. It is known that XLFs give a higher space density of AGN at low redshifts compared to UV/optical quasar LFs \citep[compare to e.g.,][]{kulkarni2019}. As explored in \citet{ricci2017}, much of this excess is due to the obscured AGN incorporated in the XLF. Some of these AGN, particularly those with high-luminosities, were assigned as unobscured by the \citet{merloni2014} probabilistic model in our SED assignment prescription (Sect. \ref{sec:sed_models}), driving this comparative excess.

\begin{table}
\caption{Parameter values employed for the XLF of \citet{fotopoulou2016xlf}. These values denote the mode of the posterior draws sampled in the Bayesian analysis of the LDDE XLF.}
\label{tab:XLFparams}      
\centering                          
\begin{tabular}{l r}        
\hline                 
Parameter & Value \\    
\hline                        
    $\logten (L_{0} / {\rm erg}\, {\rm s}^{-1})$ & 43.77  \\      
    $\gamma_{1}$ & 0.87  \\
    $\gamma_{2}$ & 2.40  \\
    $p_{1}$ & 5.89  \\
    $p_{2}$ & $-$2.30  \\
    $z^{*}_{\rm c}$ & 2.12  \\
    $\logten (L_{\alpha}/ {\rm erg}\, {\rm s}^{-1})$ & 44.51  \\
    $\alpha$ & 0.24 \\
    $\logten (A/ {\rm Mpc}^{-3})$ & $-$5.97  \\
\hline                                   
\end{tabular}
\end{table}

\section{Method}\label{sec:method}
\begin{figure}
\centering
\begin{tikzpicture}[node distance=2cm]

\node (xlf) [input, align=center] {XLF \\ \citet{fotopoulou2016xlf}};
\node (euclid) [input, right of=xlf, xshift=2.2cm, align=center] {\Euclid Area (EWS / EDS) \\ \citetalias{scaramella2022}};

\node (integration) [input, below of=xlf, xshift=2cm, align=center] {Integrate XLF \\ $43 \leq \logten (L_{\rm bol} / {\rm erg}\, {\rm s}^{-1}) \leq 50$, $\delta \logten L_{\rm bol} = 0.1$ \\
$0.01 \leq z \leq 7$, $\delta z = 0.01$ };

\node (sample) [item, below of=integration, align=center] {AGN 
volume-limited Sample};

\node (optclass) [decision, below of=sample, yshift=-1cm, align=center] {Optically Obscured? \\ \citet{merloni2014}};

\node (qsoclass) [sed, below of=optclass, xshift=-2cm, yshift=-1cm, align=center] {Mean Quasar SED \\ \citet{shen2020}};
\node (agnclass) [sed, below of=optclass, xshift=2cm, yshift=-1cm, align=center] {AGN SED Class ($L_{\rm X}$, $z$) \\ \citet{fotopoulou2016}};

\node (ebv) [sed, below of=qsoclass, xshift=2cm, align=center] {XXL \ebv{} Distributions \\ \citet{fotopoulou2016}};

\node (igmext) [sed, below of=ebv, align=center] {IGM Extinction \\ \citet{madau1995}};

\node (mockobs) [input, below of=igmext, align=center] {Mock Observations \\
(\Euclid, Rubin/LSST, DECam, \textit{WISE}, \\\textit{GALEX}, \textit{Spitzer}, 2MASS, VISTA)
};

\node (catalog) [item, below of=mockobs, align=center] {Final AGN Photometric Catalog \\
$N_{\rm AGN,\,EWS} = 1.2\times10^{8}$\\
$N_{\rm AGN,\,EDS} = 4.0\times10^{5}$
};

\node (detection) [input, below of=catalog, align=center] {
Assess \Euclid Detectable AGN (5$\sigma$)\\
$N_{\rm AGN,\,EWS} =\ $\NUMDetectableOneFiltEWS\\
$N_{\rm AGN,\,EDS} =\ $\NUMDetectableOneFiltEDS
};

\node (selection) [input, below of=detection, yshift=-0.2cm, align=center] {
Assess \Euclid Selected AGN Sample\\
Bisigello et al. (in prep.)\\
$N_{\rm AGN,\,EWS} = 8.1\times10^{6}$\\
$N_{\rm AGN,\,EDS} = 3.5\times10^{4}$
};

\draw [arrow] (xlf) -- (integration);
\draw [arrow] (euclid) -- (integration);
\draw [arrow] (integration) -- (sample);
\draw [arrow] (sample) -- (optclass);
\draw [arrow] (optclass) -- node[anchor=east] {Unobscured\,} (qsoclass);
\draw [arrow] (optclass) -- node[anchor=west] {\,Obscured} (agnclass);
\draw [arrow] (qsoclass) -- (ebv);
\draw [arrow] (agnclass) -- (ebv);
\draw [arrow] (ebv) -- (igmext);
\draw [arrow] (igmext) -- (mockobs);
\draw [arrow] (mockobs) -- (catalog);
\draw [arrow] (catalog) -- (detection);
\draw [arrow] (detection) -- (selection);

\end{tikzpicture}
\caption{Sketch outlining the method adopted in this work to attain observational expectations for AGN in the \Euclid surveys.}
\label{fig:flowchart}
\end{figure}
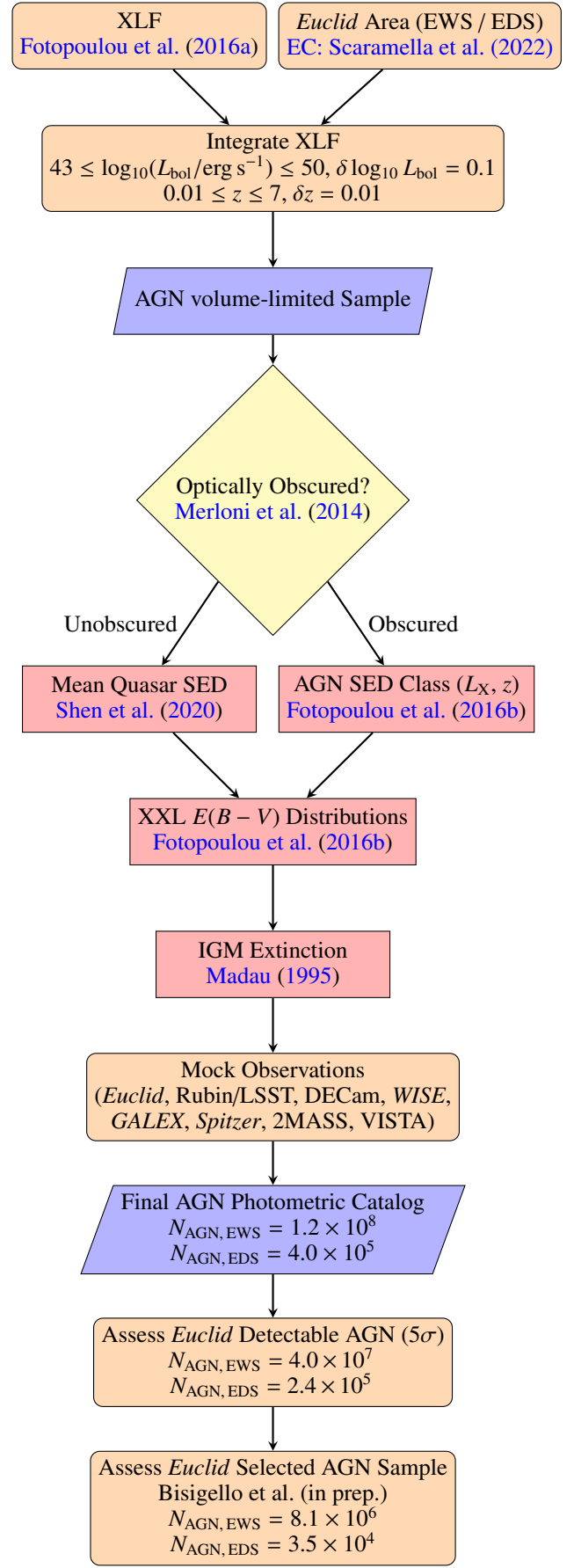

We now describe the main methodology, going from input XLF to output AGN photometry. We explain the adopted \Euclid survey parameters and modelling in Sect. \ref{sec:euclidAC}. In Sect. \ref{sec:samplegeneration} we explain the calculation to generate the volume-limited samples of AGN. The SED assignment model is presented in Sect. \ref{sec:sed_models}. Finally, Sect. \ref{sec:detectabilityassessment} describes how we derive the final photometric measurements and assess \Euclid detectability of each AGN in our sample.
We present a flowchart outlining the adopted methodology in Fig. \ref{fig:flowchart}.

\subsection{Euclid surveys}\label{sec:euclidAC}

The \Euclid space telescope possesses two photometric instruments; the Visible Imager \citep[VIS,][]{cropper2024} and the Near Infrared Spectrograph and Photometer \citep[NISP,][]{jahnke2024}. The VIS instrument carries a single broadband optical filter, \IE{} (5300--9200\,\AA) which covers the wavelength range of the traditional \textit{riz} bands. NISP possesses three near-infrared (NIR) photometric filters \citep{schimer2022}: \YE{} (9500--12\,120\,\AA), \JE{} (11\,680--15\,670\,\AA), and \HE{} (15\,220--20\,210\,\AA).

Over its six year nominal lifetime, \Euclid will perform two core surveys. The EWS is a program observing $\sim$14\,500\,deg$^{2}$ of the extra-Galactic sky with \Euclid's four photometric filters \citep{scaramella2022}. The EDS will provide observations two magnitudes deeper than the EWS for several distinct fields totalling 50\,deg$^{2}$. The EDS not only helps with calibrations of the EWS data but also extends the scientific scope of \Euclid to fainter galaxies and AGN. Throughout this work we adopt the expected EWS photometric depths presented in \citet{scaramella2022} for a 5$\sigma$ point source detection, summarised in Table \ref{tab:euc_filters}.

The high Galactic latitude of the extragalactic observations made with \Euclid means that there will be minimal Galactic extinction, with only 7--8\% of the \Euclid sky exceeding a Galactic \ebv{} of 0.1 \citep{galametz2017}. Due to this and the assumption that observed magnitudes will be corrected accordingly, there is no need to account for the effects of Galactic extinction on observed magnitudes throughout this work.


\begin{table}
\caption{\Euclid photometric filters with their corresponding effective wavelength ($\lambda_{\rm eff}$) in angstrom and observational depths (5$\sigma$ point source) for the EWS and EDS based on values reported in \citet{scaramella2022}.}            
\label{tab:euc_filters}      
\centering                          
\begin{tabular}{lccc}        
\hline                 
Band & $\lambda_{\rm eff}$ (\AA) & EWS Limiting Mag & EDS Limiting Mag \\    
\hline                        
\rule{0pt}{2ex}
    \IE & 7200 & 26.2 & 28.2  \\       
    \YE & 10\,810 & 24.3 & 26.3 \\
    \JE & 13\,670 & 24.5 & 26.5  \\
    \HE & 17\,710 & 24.4 & 26.4  \\
\hline                                   
\end{tabular}
\end{table}

The EWS and EDS areas of 14\,500\,deg$^{2}$ (4.42\,sr) and 50\,deg$^{2}$ \citep[0.015\,sr,][]{scaramella2022} were adopted as the available area applied to every source within our 2--10\,keV flux limits for our sample generation calculation. We imposed an upper flux limit of F$_{\rm \text{2--10}\,keV,\,upper}$ = 10$^{-11}$\,erg\,s$^{-1}$\,cm$^{-2}$ to prevent sources with un-physically large incident fluxes being recovered by the XLF integration for our volume-limited sample. The upper flux limit is the observed flux of the brightest source in the \textit{ROSAT} All Sky Catalog \citep[RASS;][]{boller2016}. 

We also applied a lower flux limit of F$_{\rm \text{2--10}\,keV,\,lower}$ = 10$^{-19}$\,erg\,s$^{-1}$\,cm$^{-2}$. This value was empirically determined to probe beyond \Euclid's optical magnitude limits with an appropriate margin. The lower flux limit was determined as the incident flux when the \citet{shen2020} mean quasar SED was scaled to the minimum X-ray luminosity ($\logten [L_{\rm \text{2--10}\,keV} / {\rm erg}\, {\rm s}^{-1}] = 41.8$) and maximum redshift ($z = 7$) probed in this work. Determining an area curve minimum flux limit ensures that the important parameter space of AGN close to and beyond the \Euclid detectability threshold are considered in our work whilst saving on computation by not simulating a large amount of sources that cannot be detected with \Euclid.

\subsection{AGN sample generation}\label{sec:samplegeneration}
We integrated the XLF to create a volume-limited sample of statistically expected X-ray AGN present in the EWS and EDS footprints. We performed this calculation following Eq. (\ref{eqn:lf_exp}) with the EWS and EDS area curves described in Sect. \ref{sec:euclidAC}. In each instance we integrated the XLF over $0.01 \leq z \leq 7$ with a constant redshift interval of $\delta z = 0.01$. 

We defined the X-ray luminosity integration range to correspond to a bolometric luminosity interval of $43 \leq \logten (L_{\rm bol} / {\rm erg}\, {\rm s}^{-1}) \leq 50$. This is the bolometric luminosity interval corresponding to the approximate range for which observed AGN are present in the bolometric LF data of \citet{shen2020}. We calculated the corresponding 2--10\,keV luminosity range by adopting the 2--10\,keV AGN bolometric correction of \citet{duras2020}. We therefore integrate our XLF in the 2--10\,keV luminosity range $41.8 \leq \logten (L_{\rm \text{2--10}\,keV} / {\rm erg}\, {\rm s}^{-1}) \leq 46.3$, with a constant interval of $\delta \logten{L_{\rm \text{2--10}\,keV}} = 0.1$. Adopting the conversion introduced in Sect. \ref{sec:lumfuncs}, this $L_{\rm \text{2--10}\,keV}$ integration range corresponds to the $M_{1450}$ range $-29.0 \leq M_{1450} \leq -17.2$.

Our choice of \Euclid area curve lower flux limit in Sect. \ref{sec:euclidAC} ensures the completeness of our samples at the lowest fluxes. The flux observed from a source with our highest redshift ($z=7$) at our lowest probed 2--10\,keV luminosity ($\logten [L_{\rm \text{2--10}\,keV} / {\rm erg}\, {\rm s}^{-1}]=41.8$) is 1.11\,$\times$\,10$^{-18}$\,erg\,s$^{-1}$\,cm$^{-2}$, approximately an order of magnitude above the lower flux limit.

The result of our integration is a set of distinct ($\delta\logten{L_{\rm \text{2--10}\,keV}}$, $\delta z$) bins with the expected number of AGN, $\langle N \rangle$, within the corresponding X-ray luminosity and redshift range. We neglected all integration bins where $\langle N \rangle < 1$ as it does not make sense to treat parameter ranges where we statistically expect fewer than one AGN. These cases arise due to the statistical nature of the operation. For our final volume-limited sample, we re-sampled our integration bins with $\langle N \rangle > 1$. Accordingly, bin $i$ corresponding to an X-ray luminosity, redshift interval of ($\delta\logten{L_{{\rm \text{2--10}\,keV,}\,i}}$, $\delta z_{i}$) and expectation value $\langle N \rangle = N_{i}$ first has $N_{i}$ rounded to an integer, and is then re-sampled to become $N_{i}$ AGN in our data set. Each separate AGN is assigned an X-ray luminosity and redshift value sampled uniformly from the parent integration bin parameter range ($\delta\logten{L_{{\rm \text{2--10}\,keV,}\,i}}$, $\delta z_{i}$). 
   
\subsection{SED allocation model}\label{sec:sed_models}

AGN have variations in spectral shape which can correlate with the luminosity and redshift. The shape of the SED is a result of the interplay between the black hole accretion rate (BHAR) of the AGN and the star-formation rate (SFR) of the host galaxy. In cases with high obscuration of the AGN, low BHAR or high SFR, observed AGN SEDs can have significant contributions from its host galaxy which results in composite sources.

Aiming to encapsulate the diversity of observed AGN SEDs and their variations, a number of characteristic SEDs are adopted in this work. We incorporate probabilistic spectral variations (e.g., dust extinction, optical to X-ray slope) sampled from empirical distributions to recreate the heterogeneity of observed AGN fluxes and colours. Our model ensures that AGN in our sample with similar luminosities and redshifts have realistic multi-wavelength variation in their SEDs, creating a range of photometric detectability when normalised at the same wavelength.

\subsubsection{AGN optical class assignment}\label{sec:opticaltype}

As our XLF incorporates unobscured and obscured AGN in an unbiased fashion up to $N_{\rm H} \sim$ 10$^{23}$\,cm$^{-2}$, we introduced a probabilistic model to assign a particular AGN as either unobscured or obscured, optically. For this purpose, we leveraged the optically obscured AGN fraction evolution of \citet{merloni2014}. 

In \citet{merloni2014}, the optically obscured AGN fraction ($f_{\rm obsc}$) as a function of redshift and $L_{\rm \text{2--10\,keV}}$ was derived from a sample of 1310 X-ray selected AGN from the XMM-COSMOS field in the redshift range $0.3 \leq z \leq 3.5$. Within their sample AGN were classified as optically unobscured or obscured based on their optical/NIR properties. AGN were considered unobscured if there were broad emission lines (FWHM > 2000\,km\,s$^{-1}$) in their optical spectra and obscured otherwise. If no optical spectra were available for a given AGN they were instead classified via the best-fit template class obtained by SED fitting. The optically obscured fraction evolution is parameterised as

\begin{equation}\label{eqn:merloni2014}
    f_{\rm obsc} = B (1 + z)^{\delta},
\end{equation}

\noindent where $B$ and $\delta$ are luminosity-dependent parameters. The 2--10\,keV luminosities and redshift values in our work exceed the ranges considered in \citet{merloni2014}, we therefore extrapolated their function to higher and lower luminosities, and to higher redshifts. This has the effect of low X-ray luminosity AGN giving obscured AGN fractions greater than unity at high redshifts. We therefore truncate the maximum obscured fraction at 100\%. This modification takes effect in the low-luminosity case at $z \sim 3$ and implies that objects with redshifts greater than this and $\logten (L_{\rm \text{2--10}\,keV} / {\rm erg}\, {\rm s}^{-1}) \leq 43.5$ are obscured AGN and therefore effectively excluded from detection with \Euclid. The resulting implementation of the model obeys 

\begin{equation}\label{eqn:merloni2014_mod}
    f_{\rm obsc} = \min[B (1 + z)^{\delta},\, 1].
\end{equation}

\begin{figure}
\centering
\includegraphics[width=\linewidth]{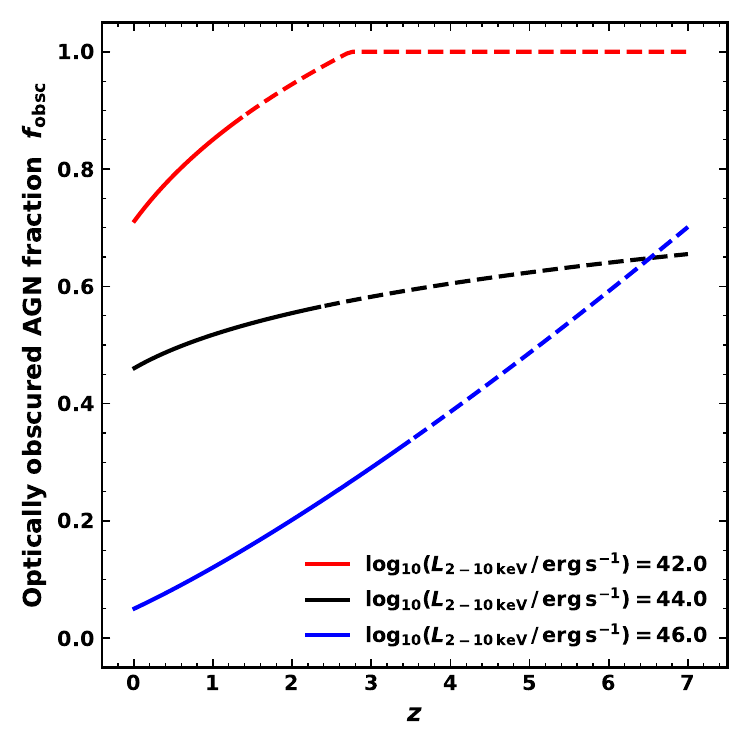}
  \caption{Evolution of the \citet{merloni2014} optically obscured AGN fraction as a function of redshift for different $L_{\rm \text{2--10}\,keV}$ values probed. Dashed lines depict extrapolation.}
     \label{fig:absorption_evo}
\end{figure}

\noindent The luminosity dependent values of the parameters $B$ and $\delta$ are presented in Table \ref{tab:merloni2014params}. The redshift evolution of the optically obscured fraction of AGN for three $L_{\rm 2-10keV}$ values, spanning the range considered in this work are depicted in Fig. \ref{fig:absorption_evo}.

\begin{table}
\caption{Luminosity-dependent parameter values used in this work for the \citet{merloni2014} optically obscured AGN fraction (Eq. \ref{eqn:merloni2014}).}            
\label{tab:merloni2014params}      
\centering                          
\begin{tabular}{l c c}        
\hline                 
$\logten (L_{\rm \text{2--10}\,keV} / {\rm erg}\, {\rm s}^{-1})$ & $B$ & $\delta$ \\    
\hline                        
    $\leq 43.5$ & 0.71 & 0.26 \\
    43.5--44.3 & 0.46 & 0.17 \\
    $\geq 44.3$ & 0.05 & 1.27 \\
\hline                                   
\end{tabular}
\end{table}

In practice, for each AGN considered we calculated $f_{\rm obsc}$ corresponding to its ($L_{\rm \text{2--10}\,keV}$, $z$) co-ordinate and used this as the probability that the AGN is optically obscured. We drew a random number from a uniform distribution between 0 and 1. If the number was below the corresponding obscured AGN fraction the AGN was assigned as obscured and vice versa for unobscured. This is a binary choice between AGN optical types which in reality is a continuum of different obscuration fractions. The binary optical AGN classification is for simplicity. Empirically sampled \ebv{} for intrinsic reddening of our AGN SEDs is introduced in Sect. \ref{sec:xrayopt_extinction}.

\subsubsection{Unobscured AGN}\label{sec:xrayt1sed}

\begin{figure*}
   \centering
   \includegraphics[width=\hsize]{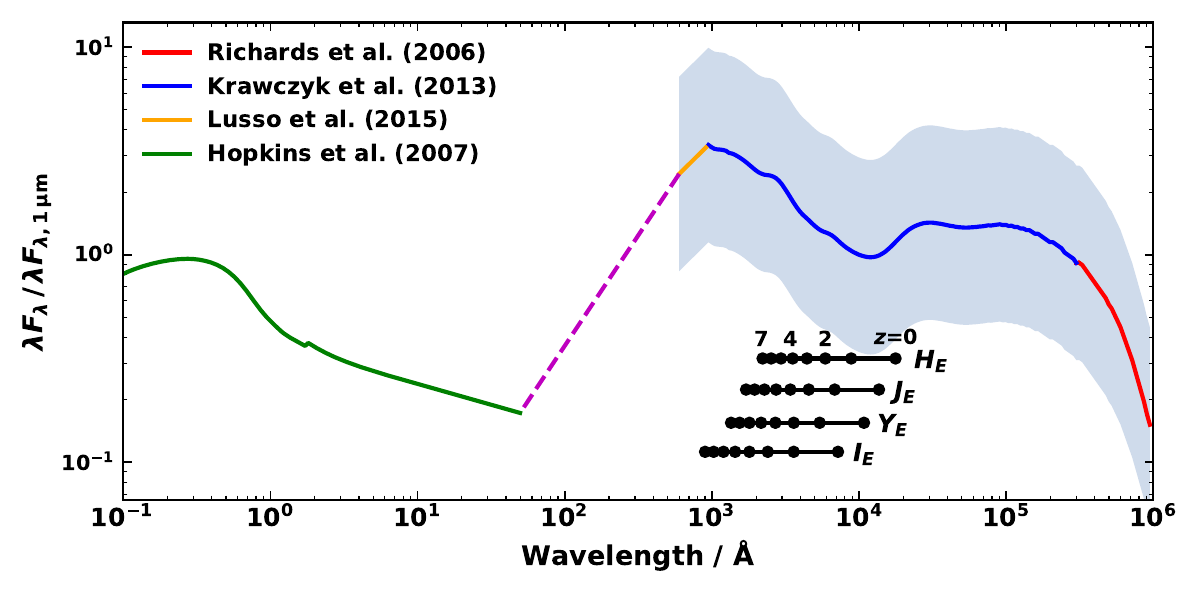}
      \caption{The broad-band mean quasar SED introduced in \citet{shen2020} which we assign to unobscured AGN, normalised at 1\,\micron. This SED is a combination of multiple templates from the literature: IR template of \citet[][red]{richards2006b}, optical/UV SED template of \citet[][blue]{krawczyk2013}, EUV power-law model based on the spectral index of \citet[][orange]{lusso2015}, which is directly connected (magenta) to the X-ray template of \citet[][green]{hopkins2007}. The template in this figure is a rest-frame realization of the SED with no IGM absorption, generated with the mean $\alpha_{\rm ox}$ of \citet{lusso2010}, $\langle\alpha_{\rm ox}\rangle = -1.37$. The blue shaded region indicates the range in possible realisations of the optical portion of the SED considering $\alpha_{\rm ox}$ values with the reported dispersion of 0.18. The effective wavelength redshift evolution for each \Euclid filter over the redshift range probed in this work are depicted as black lines.
              }
         \label{fig:shen2020sed}
\end{figure*}

To model unobscured AGN we adopted the broad-band mean quasar SED assembled in \citet{shen2020} that spans from ultra-hard X-rays to far-infrared (FIR). This SED represents the average continuum emission of unobscured AGN neglecting emission line contributions. The SED was utilized in \citet{shen2020} for their bolometric AGN LF derivation, calculating bolometric corrections and reconciling unobscured AGN emission between different wavebands. The full SED model, across all wavelengths is shown in Fig. \ref{fig:shen2020sed}.

In the optical/UV, the SED template of \citet{krawczyk2013} was adopted. This mean template was derived from 96\,716 luminous broad-lined quasars that do not show signs of dust reddening. The incorporated quasars are at $0.064 < z < 5.46$. The template covers the wavelength range 912\,\AA{}\,--\,30\,\micron. 
In the IR regime, the \citet{krawczyk2013} SED template is extended to 100\,\micron{} using the \citet{richards2006b} SED. This IR SED includes dust emission, removing the need for an additional dust emission model. 
On the short wavelength side, the optical/UV SED is extended into the extreme UV ($\lambda <$ 912\,\AA) with a power-law model, $f_{\nu} \propto \nu^{\,\alpha_{\nu}}$, with spectral index $\alpha_{\nu} = -1.70$ \citep{lusso2015}. This model extends from 600\,\AA{} to 912\,\AA, where the flux is directly connected to an X-ray SED template. 
The X-ray SED adopted in this work is that used in \citet{hopkins2007}. Extending shortwards of 0.5\,keV, this template follows a power law with intrinsic photon index $\Gamma=1.8$ and an exponential cut-off at 500\,keV. Following \citet{ueda2003}, a reflection component is included with a reflection solid angle of 2$\pi$, inclination $\cos{i}=0.5$ and Solar abundances.

To scale the X-ray and optical portions of the SED, an optical-to-X-ray luminosity relation must be utilized. The literature reports a non-linear correlation between the X-ray and optical luminosities for unobscured AGN \citep{steffen2006,just2007,lusso2010}. In particular this relates the luminosities at 2\,keV and 2500\,\AA{} with the expression

\begin{equation}
    \logten{L_{\nu}(2\,{\rm keV})} = \beta \logten{L_{\nu}(2500\,{\text \AA{}})} + C ,
\end{equation}

\noindent where $L_{\nu}$ is the luminosity density in units of erg\,s$^{-1}$\,Hz$^{-1}$. This relation can be parameterised by the optical-to-X-ray spectral index, $\alpha_{\rm ox}$, defined as

\begin{equation}
    \alpha_{\rm ox} = 0.384\logten\left(\frac{{L_{\nu}(2\,{\rm keV})}}{L_{\nu}(2500\,{\text \AA})}\right).
\end{equation}

\cite{lusso2010} reported a mean $\alpha_{\rm ox}$ value $\langle\alpha_{\rm ox}\rangle \approx -1.37 \pm 0.01$ with a dispersion of 0.18, for a sample of 545 X-ray selected unobscured AGN from the XMM-COSMOS survey observed at $0.04 < z < 4.25$ with $40.6 \leq \logten (L_{\rm \text{2--10}\,keV} / {\rm erg}\, {\rm s}^{-1}) \leq 45.3$. The sample used in \citet{lusso2010} has similar properties to the AGN expected to be probed in this work. Therefore, for each unobscured AGN in this work we draw a value of $\alpha_{\rm ox}$ from a normal distribution centered on 1.37 with a standard deviation of 0.18. A single $\alpha_{\rm ox}$ value sampled with empirical scatter can be adopted in this case because no significant correlation is observed between $\alpha_{\rm ox}$ and redshift or $L_{\rm \text{2--10}\,keV}$ for X-ray selected samples \citep{lusso2010}. Drawing $\alpha_{\rm ox}$ values from an empirically derived distribution perturbs the normalization of the optical-IR portion of the SED with respect to the X-ray luminosity, allowing us to introduce natural variations in AGN flux within this single SED model for unobscured AGN.

\subsubsection{Obscured AGN}\label{sec:type2agnSED}

X-ray selected obscured AGN display a considerable range of spectral shapes. Hard X-rays allow AGN to be selected even with a substantial portion of their optical AGN emission obscured by dust. Spectral shapes therefore range from Seyfert 2-like SEDs with high AGN contribution to that of quiescent and star-forming galaxies (SFG) when host-galaxy contamination is high. 

To assign appropriate obscured AGN SEDs depending on both X-ray luminosity and redshift we leveraged the SED fitting results of \citet{fotopoulou2016}. SED fitting was performed with UV--MIR photometry of the 1000 brightest X-ray sources in the XXL survey \citep{pierre2016}, which includes 972 unobscured and obscured AGN in the redshift range $0.01 < z < 4$. The XXL survey covers an area of 50\,deg$^{2}$ with a medium X-ray flux limit of $\sim 5\,\times\,10^{-15}\,{\rm erg}\,{\rm s}^{-1}\,{\rm cm}^{-2}$ in the 0.5--2\,keV band, providing a middle ground between deep and all-sky surveys. Properties reported for AGN in this field should thus provide a realistic representation of those expected to be encountered in the EWS and EDS. The extensive multiwavelength followup programme in the XXL fields has provided a means to connect X-ray AGN emission with empirical broad-band SEDs in a self-consistent manner. The SED templates used in the fitting are drawn from those used in the \citet{ilbert2009} SED fitting analysis of COSMOS sources and \citet{salvato2009} SED fitting of XMM-COSMOS sources. The templates in \citet{fotopoulou2016} are categorised into QSO, AGN, Starburst (SB), SFG, and Passive categories. The sources were assigned in each category prior to the SED fitting, using a Random Forest classifier as described in \citet{fotopoulou&paltani2018}. The SED templates used in this work have therefore been demonstrated to well approximate the empirical colour-distributions of X-ray selected AGN. In all cases, rather than the assignment of a particular SED meaning that the AGN has solely the properties of its assigned class, it means that observed photometry of the AGN is best described by the allocated template.

\begin{figure}
\centering
\includegraphics[width=\hsize]{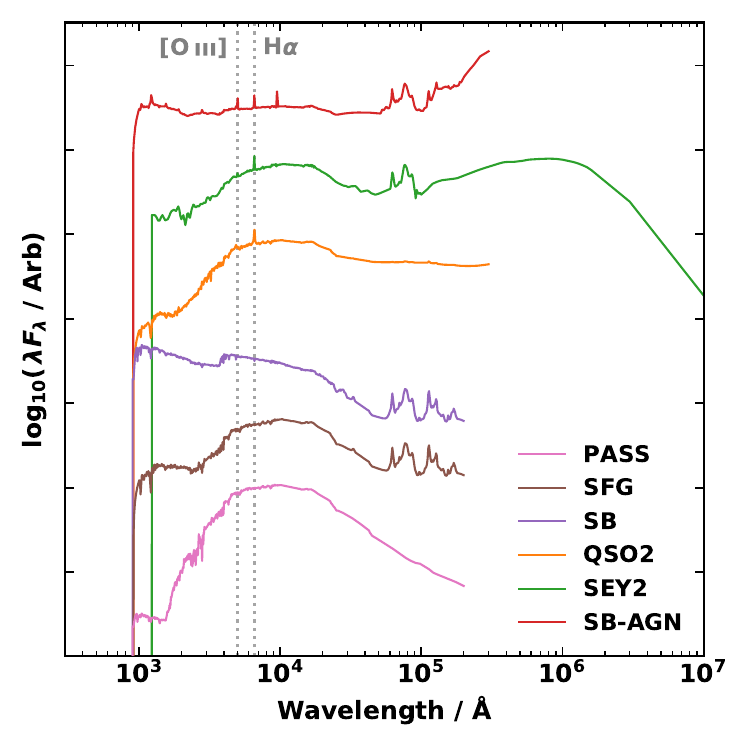}
  \caption{Representative SEDs utilized for each obscured AGN SED class (`PASS': pink, `SFG': brown, `SB': purple, `QSO2': orange, `SEY2': green, `SB-AGN': red) in this work. Based on the SED shapes in \citet{fotopoulou2016}. Grey dotted lines highlight the rest-frame wavelength of the [\ion{O}{iii}]$\lambda$5007\,\AA{} and H$\alpha$ emission lines. Templates are assigned based on the relative probability of each template class in high- and low-luminosity groups at a given redshift.}
     \label{fig:xrayt2_seds}
\end{figure}

For the purposes of assigning obscured AGN spectra we dropped all sources which have a best-fit template in the QSO category, leaving a sample of 652 AGN. Within the SED category denominated `AGN' in \cite{fotopoulou2016} there are three characteristic SED shapes; Seyfert 2-like, QSO2-like, and SB-AGN composite-like. We therefore split the AGN category into three further classes based on the characteristic shapes, resulting in six SED classes considered for obscured AGN in this work; (1) PASS, (2) SFG, (3) SB, (4) QSO2, (5) SEY2, and (6) SB-AGN. 
These six obscured AGN classes provide a range of different AGN-galaxy contributions and stellar populations as observed for X-ray AGN. We selected a singular SED from the set of SEDs belonging to each defined class which was used to represent its class throughout this work. The selected representative SED for each class is presented in Fig. \ref{fig:xrayt2_seds}.

All the obscured AGN SEDs adopted here are empirical and therefore include nebular emission lines. We note the presence of particularly prominent emission lines in the SED templates for our AGN-dominated classes (QSO2, SEY2, and SB-AGN). All three of these representative SEDs include the H$\alpha$ emission line. The QSO2 template additionally features the H$\beta$ line. The SEY2 template includes [\ion{O}{iii}]$\lambda$5007\,\AA and the template for SB-AGN incorporates a range of prominent emission lines from \ion{Ly}{$\alpha$} at 1216\,\AA{} to [\ion{S}{iii}]$\lambda$9533\,\AA. 

To include an X-ray luminosity dependence in our obscured AGN SED assignment we split the sources in each of our six SED classes into high and low X-ray luminosity groups, creating 12 SED groups in total. We defined high X-ray luminosity sources as those with $\logten (L_{\rm \text{2--10}\,keV} / {\rm erg}\, {\rm s}^{-1}) \geq 44$ and low X-ray luminosity sources as those with $\logten (L_{\rm \text{2--10}\,keV} / {\rm erg}\, {\rm s}^{-1}) < 44$. We fit log-normal probability distributions to the redshift-space distribution of our 12 SED groups. The resulting normalised probability distributions for each class were then extrapolated to $z = 7$. The normalised and extrapolated redshift-space probability distributions are presented in Appendix \ref{app:sedclassdists}. For a given source with a ($\logten{L_{\rm \text{2--10}\,keV}}$, $z$) co-ordinate we used our log-normal distribution fits in the appropriate X-ray luminosity group to find the probability of the source occurring in each SED class at its given redshift. These six probabilities were scaled to provide a summed total probability of unity. We used these scaled probabilities to draw an SED class for each obscured AGN. 

The adopted methodology provides a template appropriate for the X-ray luminosity and redshift of the obscured AGN. SED templates are more probable if they were assigned more frequently in the XXL SED fitting analysis at a given redshift, therefore reflecting the available empirical data. In both the high- and low-luminosity groups the AGN-dominated SED classes (QSO2, SEY2, and SB-AGN) are most probable at $z > 1.5$. The SB SED class has a similar probability as the AGN-dominated classes in the high-luminosity group in this redshift regime. For both luminosity groups at $z < 1.5$ there are no dominantly probable SED classes but a balanced chance of any of the classes being assigned to an AGN. 

None of the SED templates assigned to obscured AGN in this work extend to X-ray wavelengths. We therefore adopted bolometric corrections to appropriately scale the SEDs considering their respective $L_{\rm \text{2--10}\,keV}$ values. We first transformed from 2--10\,keV luminosity to the bolometric domain with the bolometric correction of \citet{duras2020}. Following this, we leveraged the zero-intercept 2\,\micron{} bolometric correction of \citet{runnoe2012b} to transform from a bolometric luminosity to the $\lambda L_{\lambda}$(2\,\micron) luminosity, which we used to scale our SEDs. 
We opted to scale our SEDs at 2\,\micron{} in the rest-frame as the corresponding portion of the SED is free from major effects of dust reddening, non-continuum emission features and polycyclic aromatic hydrocarbon (PAH) emission. Furthermore, because our obscured AGN SEDs are derived from X-ray observations, three of our classes (PASS, SFG, SB) do not contain significant AGN contributions to their optical--MIR AGN emission (Fig. \ref{fig:xrayt2_seds}). Due to this, scaling these SEDs at wavelengths longer than $\sim$2\,\micron{} results in erroneous photometry. 

The \cite{runnoe2012b} 2\,\micron{} bolometric correction was originally derived from a sample of UV-bright unobscured AGN, which is not the context in which we are applying it here. We argue however that at rest-frame NIR wavelengths the average SEDs of obscured and unobscured AGN are remarkably similar \citep[e.g.,][see Fig. 3 in \citealt{hickox2017}]{alonsoherrero2006,polletta2007}. In this regime there is a minimum in AGN emission where the accretion disk power-law continuum emission in unobscured AGN falls off and the black body emission of the torus begins, which is ubiquitous to all AGN. NIR bolometric corrections for unobscured or obscured AGN are seldom investigated in the literature, largely due to complications in deconvolving the AGN and host-galaxy contribution to spectra at these wavelengths \citep{elvis1994, richards2006b, runnoe2012b}. Due to the compatibility of obscured and unobscured AGN spectra at 2\,\micron{} and the sparsity of alternative options we elected to use the \cite{runnoe2012b} 2\,\micron{} bolometric correction in this work whilst accepting that it is a limitation of our method. 

We apply both the X-ray and 2\,\micron{} bolometric corrections for the bulk of this work employing the nominal parameter values reported in their respective works (i.e., neglecting parameter uncertainty). We adopted this procedure so that we can fully constrain the impact of the bolometric correction dispersion on our resulting AGN number estimates (see Sect. \ref{sec:uncertainties}). 

\subsubsection{Intrinsic and IGM extinction}\label{sec:xrayopt_extinction}

Extinction is the combination of the absorption and scattering of photons by intervening dust and gas along the line of sight. We treat two different sources of extinction in AGN SEDs; intrinsic extinction and intergalactic medium (IGM) extinction. Intrinsic extinction is caused by dust and gas originating at the redshift of the source, often in the host-galaxy. This form of extinction, often referred to as reddening, is most severe in the UV/optical but acts on wavelengths up to the IR regime, causing variations in observed SEDs and therefore observed colours. IGM extinction considers the absorption and scattering of photons by intervening gas and dust in the line of sight to astrophysical sources at cosmological distances. In particular, photons with wavelengths shorter than the rest-frame \ion{Ly}{$\alpha$} transition (1216\,\AA{}) are attenuated by intergalactic \ion{H}{i}.

To apply intrinsic reddening with realistic \ebv{} values for AGN SEDs, we again leveraged the results of the XXL SED fitting analysis of \citet{fotopoulou2016}. Their analysis was conducted on 972 X-ray selected AGN in the XXL survey. Intrinsic X-ray properties of the AGN were derived directly from their X-ray spectra, while optical and longer wavelength properties related to the AGN host-galaxies were ascertained via broad-band SED fitting of the multiwavelength photometry of each source. Relevant to this section, the SED fitting allowed the best-fit SED template for each AGN to be determined along with estimations of the intrinsic \ebv.

\begin{figure}[h]
\centering
\includegraphics[width=\hsize]{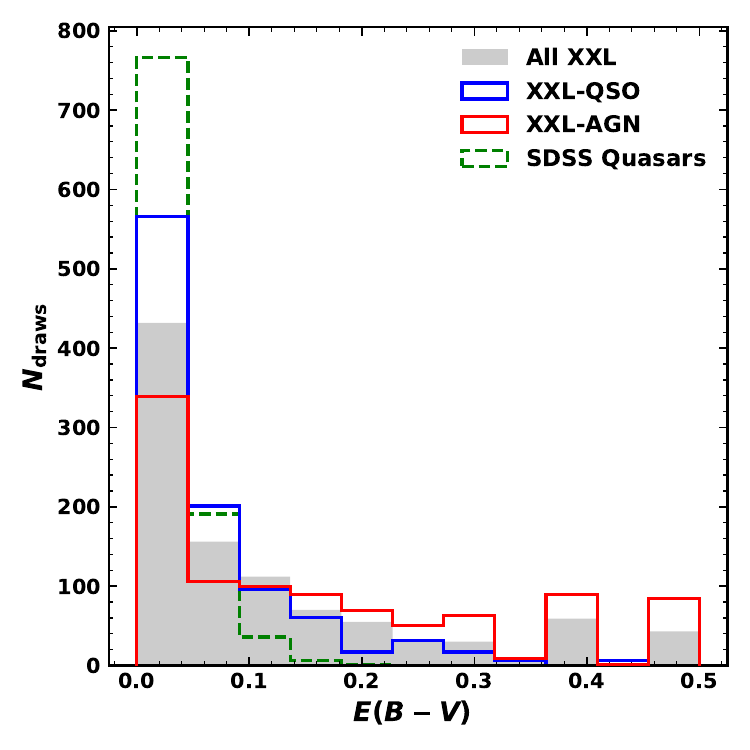}
  \caption{Comparison of 1000 draws from the unobscured and obscured AGN \ebv{} distributions used in this work. The distributions are based on the XXL AGN \ebv{} distribution derived in \citet[][grey shaded]{fotopoulou2016}. The XXL-QSO distribution (blue) is the distribution found for XXL AGN assigned the QSO class template and is used for unobscured AGN in this work. The XXL-AGN distribution (red) is constructed for all XXL sources not assigned the QSO class and is used for obscured AGN in this work. For comparison, we also include the \ebv{} distribution found for SDSS quasars \citep[green,][]{hopkins2004}.
          }
     \label{fig:xray_ebvdist}
\end{figure}

We divided the SED fitting results into two classes; (1) XXL-QSO and (2) XXL-AGN, defined as all sources with a best-fitting SED template belonging to the QSO class or not, respectively. We constructed \ebv{} probability distributions for the XXL-QSO and XXL-AGN classes based on source frequency at each discrete \ebv{} value used in the analysis of \citet{fotopoulou2016}. The discrete colour excess values lie in the range $0.0 \leq E(B-V) \leq 0.5$ in steps of 0.05. The results of 1000 draws of each XXL AGN \ebv{} distribution created for this work is depicted in Fig. \ref{fig:xray_ebvdist}. For reference, we also plot 1000 draws of the distribution of all XXL sources analysed in \citet{fotopoulou2016} and 1000 draws of the \ebv{} distribution derived from SDSS quasars \citep{hopkins2004}. 

Our distributions show the XXL-QSO class have generally lower \ebv{} values than the XXL-AGN class. XXL-AGN sources account for the higher colour excess values observed in the all-sources distribution. This is expected as QSOs are by definition less obscured than AGN due to the requirement of a clear viewing angle towards the central engine to observe a QSO-like spectrum. Comparing the SDSS quasar \ebv{} distribution to that of the XXL-QSO class, it is clear that X-ray selected AGN probe higher \ebv{} values than their optically selected counterparts. This is mainly due to optical selection effects. It is therefore beneficial to adopt an XXL-based \ebv{} distribution for our analysis as it provides a less biased distribution of \ebv{} values.

We found no significant trend between \ebv{} values and $L_{\rm \text{2--10}\,keV}$, redshift or SED fitting template class in the XXL data. Therefore, in our analysis we used our XXL-QSO distribution to sample \ebv{} values for dust extinction in unobscured AGN SEDs and the XXL-AGN distribution for obscured AGN SEDs. Intrinsic extinction was applied to our SED models assuming a Small Magellanic Cloud (SMC) extinction law \citep{prevot1984}. The SMC extinction law has been demonstrated to be appropriate for use with AGN \citep[e.g.,][]{hopkins2004, salvato2009} for all but the most extremely reddened objects \citep{zafar2015}.

We applied IGM extinction to all our SEDs following the redshift-dependent prescription presented in \citet{madau1995}. Although updated versions IGM extinction models are available \citep[e.g.,][]{meiksin2006,inoue2014,thomas2017} we assert that the adopted method is sufficient for the analyses described in this study. Only the \Euclid \IE{} band is affected by IGM extinction at the highest redshifts probed in this work ($z\gtrsim6.5$), with the largest effect being for ancillary bands considered from Rubin/LSST. We are not assessing photometric dropouts and therefore it is beyond the scope of this work to include a more detailed IGM extinction model.

\subsection{Assessing AGN detectability}\label{sec:detectabilityassessment}

We assigned every source in our EWS and EDS AGN samples an appropriate broad-band SED according to the models described in Sect. \ref{sec:sed_models}. We scaled and transformed each SED to retrieve an incident flux SED in line with the appointed luminosity and redshift co-ordinate of the AGN. The detectability of each AGN was assessed by performing a synthetic photometric observation of its assigned SED using \Euclid's filter transmission curves. To perform this mock observation we used the \texttt{sedpy} Python package \citep{johnson2019}. 
We additionally observed each SED with a selection of ancillary bands from other facilities, which we outline in Appendix \ref{app:photometry}.

When considering \Euclid photometric bands, we perturbed the resulting fluxes based on the associated uncertainty for each \Euclid filter following the prescription of Bisigello et al. (in prep.). In this formulation fluxes are perturbed considering a normal distribution where the standard deviation is equal to the expected photometric uncertainty for each AGN. The photometric uncertainty in each \Euclid filter is described by the sum in quadrature of the photon noise (${\sigma}_{\rm noise}$) and the background flux density error (${\sigma}_{\rm bkg}$). The photon noise is defined as 

\begin{equation}
    {\sigma}_{\rm noise} = \frac{f_{5\sigma}}{5}\,\frac{r}{r_{\rm ref}},
\end{equation}

\noindent where $f_{5\sigma}$ is the 5$\sigma$ survey depth in each filter (Table \ref{tab:euc_filters}), $r$ is the projected radius of each source, and $r_{\rm ref}$ is the median effective radius of a source with flux densities equal to $f_{5\sigma}$. As we do not assign our AGN a physical size or physical properties (e.g., stellar mass) that can be used to derive a physical size we assume that $r=r_{\rm ref}=\ang{;;0.23}$. This corresponds to the radius of the photometric aperture used to measure the flux of a point source, 1.25 times the FWHM of the \Euclid point spread function \citep[PSF;][]{scaramella2022}. This assumption neglects the dependence of ${\sigma}_{\rm noise}$ on source radius. Our modelling therefore results in an underestimation of the photometric uncertainty for low-redshift extended objects and an overestimation for high-redshift truly point-like objects. The background flux error component is derived as 

\begin{equation}
    {\sigma}_{\rm bkg} = {\sigma}_{\rm noise}\,\sqrt{\frac{f}{f_{\rm sky}\,\pi\,r^{2}}},
\end{equation}

\noindent where $f$ is the flux of the object and $f_{\rm sky}$ represents the reference sky surface background corresponding to 22.33, 22.10, 22.11, and 22.28\,mag\,arcsec$^{-2}$ for \IE{}, \YE{}, \JE{}, and \HE{}, respectively.
Resultant observed magnitudes generated with perturbed fluxes were used to determine if each AGN will be detectable with \Euclid. The photometry prescription used here assumes that all flux is captured by photometric measurements. Our derived AGN colours are validated in Appendix \ref{app:colourvalidation}. 

We stress that an AGN deemed to be \textit{detectable} with \Euclid photometry in this work does not necessarily mean it will be identified as an AGN. Rather, it is expected to be detected above 5$\sigma$ in the photometric filters of the corresponding \Euclid survey. We consider the \textit{selection} of \Euclid detectable AGN via photometry in Sect. \ref{sec:euclidselection}.

\section{Results}\label{sec:results}

\begin{table*}
\caption{The expected number of AGN detectable with \Euclid photometry in the EWS. Numbers are reported on a per-filter basis as well as the number of AGN detectable in at least one \Euclid filter, (\IE\,|\,\YE\,|\,\JE\,|\,\HE), and the number of AGN detectable in all \Euclid filters, (\IE\,$\land$\,\YE\,$\land$\,\JE\,$\land$\,\HE). Corresponding surface densities of \Euclid detectable AGN are presented as well as splits by obscured/unobscured classification and redshift range.}             
\label{tab:results_wide}      
\centering                          
\begin{tabular}{l c c c c c c}        
\hline            
Band & \head{2cm}{Detectable AGN} & \head{2cm}{Surface Density (deg$^{-2}$)} & \head{2cm}{Unobscured AGN} & \head{2cm}{Obscured AGN} & $0.1 \leq z \leq 4$ & $4 < z \leq 7$\\    
\hline                        
& & & & & & \\[-8pt]
    \IE & \NUMDetectableIeEWS & \SDDetectableIeEWS & $1.2\times10^{7}$ & $1.9\times10^{7}$ & $3.0\times10^{7}$ & $1.1\times10^{6}$ \\
    \YE & \NUMDetectableYeEWS & \SDDetectableYeEWS & $8.7\times10^{6}$ & $1.3\times10^{7}$ & $2.1\times10^{7}$ & $8.4\times10^{5}$ \\  
    \JE & \NUMDetectableJeEWS & \SDDetectableJeEWS & $9.7\times10^{6}$ & $2.0\times10^{7}$ & $2.9\times10^{7}$ & $1.0\times10^{6}$ \\  
    \HE & \NUMDetectableHeEWS & \SDDetectableHeEWS & $9.9\times10^{6}$ & $2.5\times10^{7}$ & $3.4\times10^{7}$ & $1.1\times10^{6}$ \\  
    (\IE\,|\,\YE\,|\,\JE\,|\,\HE) & \NUMDetectableOneFiltEWS & \SDDetectableOneFiltEWS & $1.2\times10^{7}$ & $2.8\times10^{7}$ & $3.9\times10^{7}$ & $1.4\times10^{6}$ \\
    (\IE\,$\land$\,\YE\,$\land$\,\JE\,$\land$\,\HE) & \NUMDetectableAllFiltEWS & \SDDetectableAllFiltEWS & $8.4\times10^{6}$ & $1.2\times10^{7}$ & $2.0\times10^{7}$ & $7.2\times10^{5}$ \\
\hline

\end{tabular}
\end{table*}

\begin{table*}
\caption{The expected number of AGN detectable with \Euclid photometry in the EDS. Numbers are reported on a per-filter basis as well as the number of AGN detectable in at least one \Euclid filter, (\IE\,|\,\YE\,|\,\JE\,|\,\HE), and the number of AGN detectable in all \Euclid filters, (\IE\,$\land$\,\YE\,$\land$\,\JE\,$\land$\,\HE). Corresponding surface densities of \Euclid detectable AGN are presented as well as splits by obscured/unobscured classification and redshift range.}             
\label{tab:results_deep}      
\centering                          
\begin{tabular}{l c c c c c c}        
\hline            
Band & \head{2cm}{Detectable AGN} & \head{2cm}{Surface Density (deg$^{-2}$)} & \head{2cm}{Unobscured AGN} & \head{2cm}{Obscured AGN} & $0.1 \leq z \leq 4$ & $4 < z \leq 7$\\    
\hline                        
& & & & & & \\[-8pt]
    \IE & \NUMDetectableIeEDS & \SDDetectableIeEDS & $4.9\times10^{4}$ & $1.4\times10^{5}$ & $1.8\times10^{5}$ & $6.8\times10^{3}$ \\
    \YE & \NUMDetectableYeEDS & \SDDetectableYeEDS & $4.4\times10^{4}$ & $1.2\times10^{5}$ & $1.5\times10^{5}$ & $5.3\times10^{3}$ \\  
    \JE & \NUMDetectableJeEDS & \SDDetectableJeEDS & $4.5\times10^{4}$ & $1.5\times10^{5}$ & $1.9\times10^{5}$ & $6.3\times10^{3}$ \\  
    \HE & \NUMDetectableHeEDS & \SDDetectableHeEDS & $4.5\times10^{4}$ & $1.8\times10^{5}$ & $2.1\times10^{5}$ & $7.6\times10^{3}$ \\  
    (\IE\,|\,\YE\,|\,\JE\,|\,\HE) & \NUMDetectableOneFiltEDS & \SDDetectableOneFiltEDS & $4.9\times10^{4}$ & $1.9\times10^{5}$ & $2.3\times10^{5}$ & $8.7\times10^{3}$ \\
    (\IE\,$\land$\,\YE\,$\land$\,\JE\,$\land$\,\HE) & \NUMDetectableAllFiltEDS & \SDDetectableAllFiltEDS & $4.3\times10^{4}$ & $1.1\times10^{5}$ & $1.5\times10^{5}$ & $4.8\times10^{3}$ \\
\hline

\end{tabular}
\end{table*}

Our main results for the number and surface densities of AGN with $\geq$5$\sigma$ \Euclid photometric detections in the EWS are presented in Table \ref{tab:results_wide} and for the EDS in Table \ref{tab:results_deep}. We report our forecast for each of \Euclid's four photometric filters individually, for AGN detectable in at least one filter, and simultaneously in all filters. 

Of the AGN detectable in at least one \Euclid filter in the EWS, 3.9\,$\times$\,10$^{7}$ (98\%) lie in the redshift range $0.01 \leq z \leq 4$, where our XLF is well constrained by observations, and 1.4\,$\times$\,10$^{6}$ (2\%) at $4 < z \leq 7$. In the EDS 2.3\,$\times$\,10$^{5}$ detectable AGN (96\%) are in the redshift range $0.01 \leq z \leq 4$, while 8.7\,$\times$\,10$^{3}$ (4\%) are in the range $4 < z \leq 7$. In the case of both \Euclid surveys the vast majority of the detectable AGN are derived from the well constrained region of the input XLF. 

\begin{figure}
     \centering
     \begin{subfigure}[b]{\hsize}
         \centering
         \includegraphics[width=\hsize]{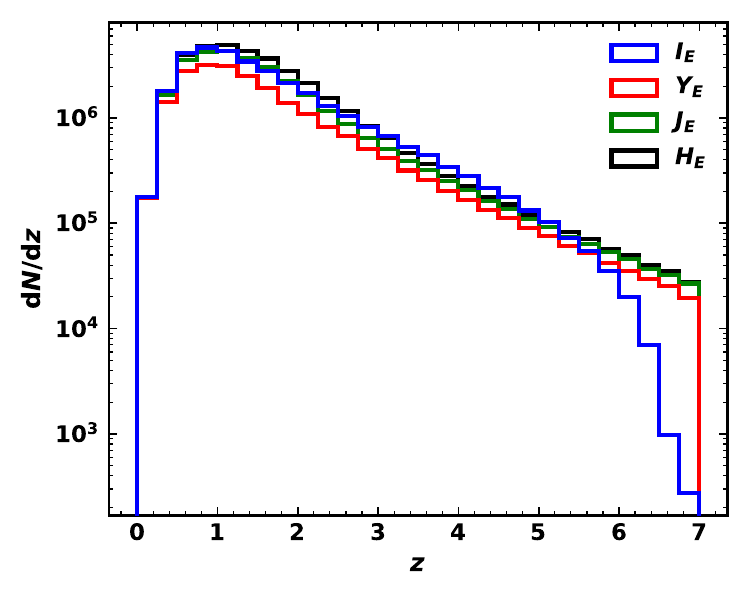}
         \caption{Euclid Wide Survey}
         \label{fig:ews_detection_zdist}
     \end{subfigure}
     \vfill
     \begin{subfigure}[b]{\hsize}
         \centering
         \includegraphics[width=\hsize]{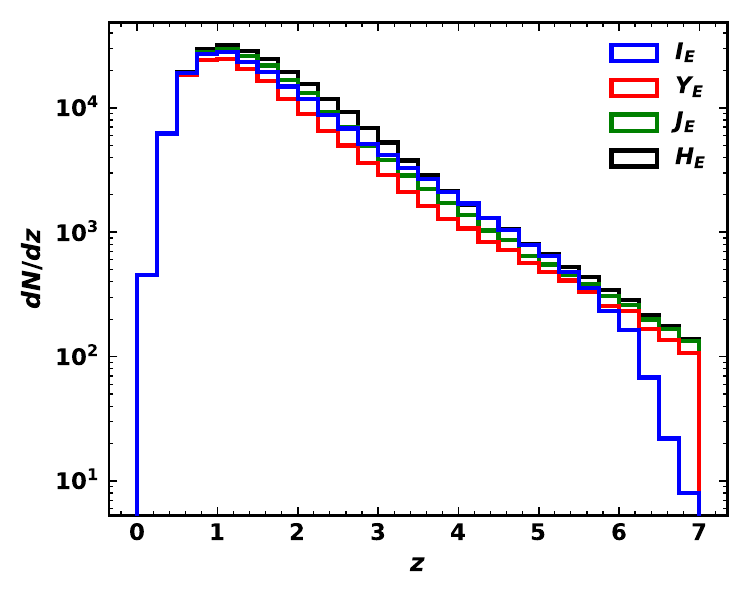}
         \caption{Euclid Deep Survey}
         \label{fig:eds_detection_zdist}
     \end{subfigure}
        \caption{Predicted redshift distributions of \Euclid detectable AGN in the redshift range $0.01 \leq z \leq 7$ in the EWS (top) and EDS (bottom). The data are binned in steps of $\delta z = 0.25$. Separate distributions are presented for AGN detectable in each of \Euclid's photometric filters; \IE{} (blue), \YE{} (red), \JE{} (green), and \HE{} (black).}
        \label{fig:detection_zdists}
\end{figure}

We present the distribution of our predicted AGN numbers in redshift bins for the EWS and EDS in Fig. \ref{fig:ews_detection_zdist} and Fig. \ref{fig:eds_detection_zdist}, respectively. Both distributions show a space-density peak of detectable AGN at $z\approx1$. This is consistent with the redshift at which the number density of moderate luminosity X-ray AGN peaks, as described by our XLF \citep{fotopoulou2016xlf}. From Fig. \ref{fig:ews_detection_zdist} we see that for $z \lesssim 5.5$ filters \IE{} (blue), \JE{} (green) and \HE{} (black) yield similar numbers of AGN in the EWS. For redshifts greater than this the yield of AGN detectable in the shortest wavelength \IE{} filter drops dramatically. This is caused by the portion of the AGN SED affected by IGM extinction shifting into the \IE{} pass-band. From $z \gtrsim 6.4$ AGN detections in the \IE{} band are limited to only the most luminous sources. This effect is also observed in the EDS redshift distributions (Fig. \ref{fig:eds_detection_zdist}), however with a marginally shallower fall-off at $z\approx6$ due to the deeper limiting magnitude of EDS allowing more AGN in this regime to be detected with the \IE{} filter.

\begin{figure*}
     \centering
     \begin{subfigure}[b]{0.49\textwidth}
         \centering
         \includegraphics[width=\textwidth]{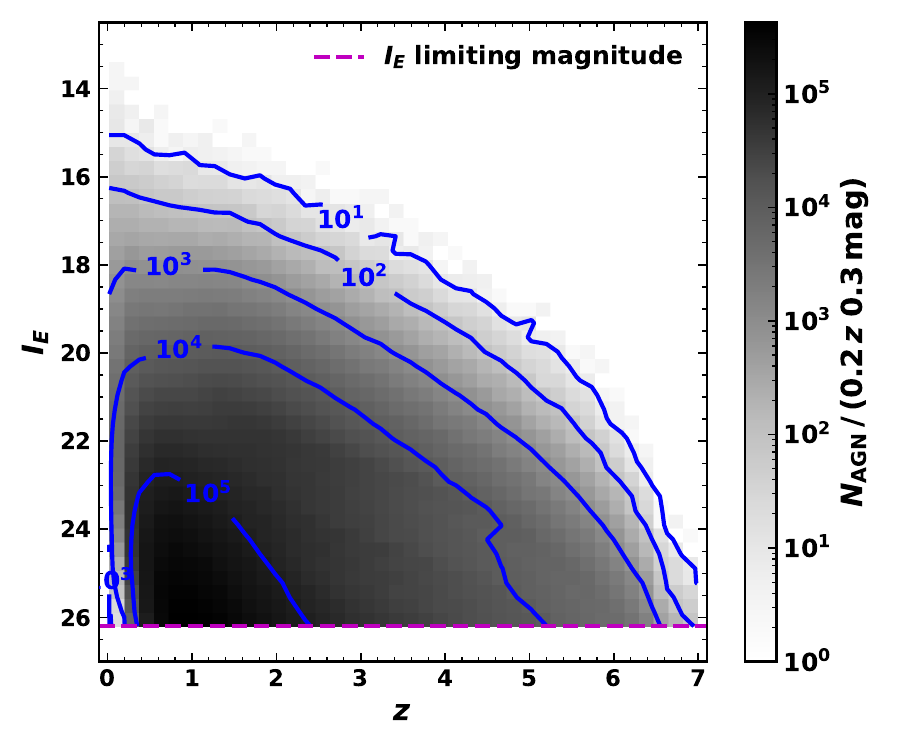}
         \label{fig:I_density_wide}
     \end{subfigure}
     \hfill
     \begin{subfigure}[b]{0.49\textwidth}
         \centering
         \includegraphics[width=\textwidth]{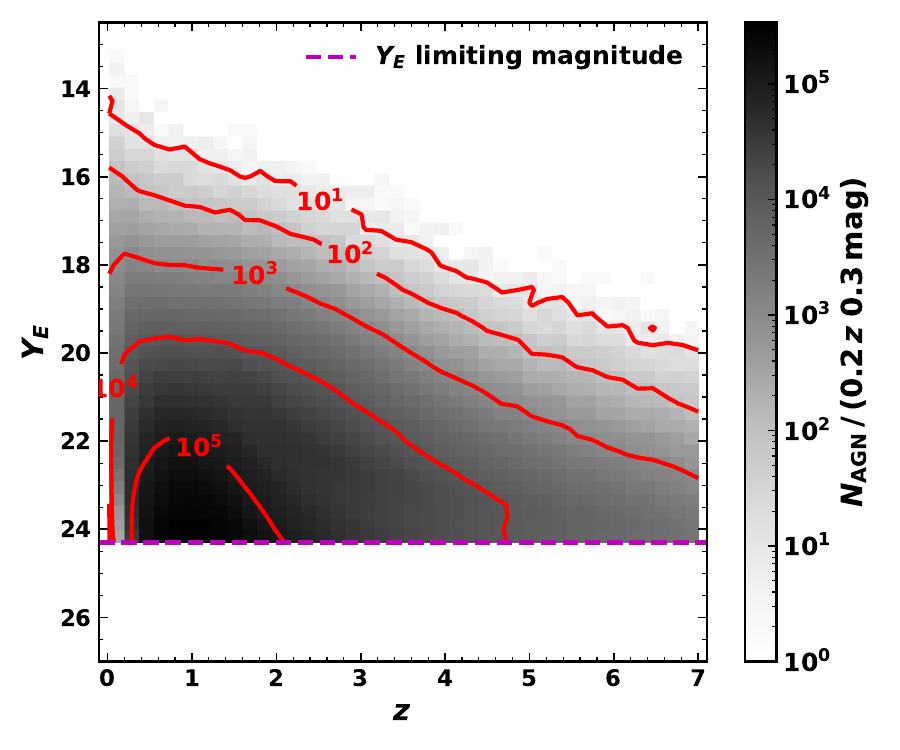}
         \label{fig:Y_density_wide}
     \end{subfigure}
     \vfill
     \begin{subfigure}[b]{0.49\textwidth}
         \centering
         \includegraphics[width=\textwidth]{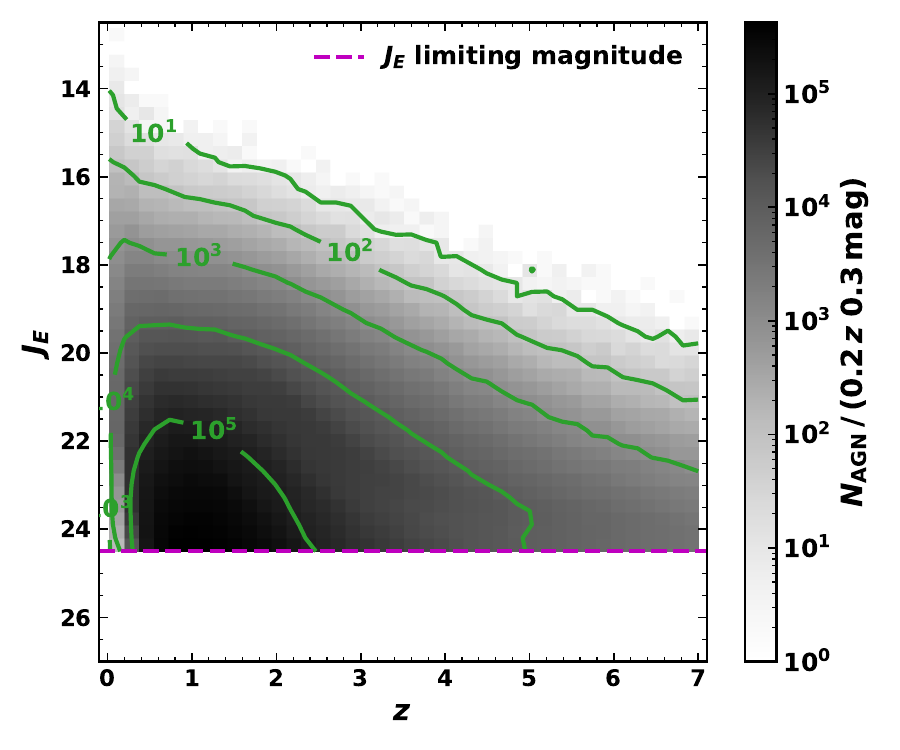}
         \label{fig:J_density_wide}
     \end{subfigure}
     \hfill
     \begin{subfigure}[b]{0.49\textwidth}
         \centering
         \includegraphics[width=\textwidth]{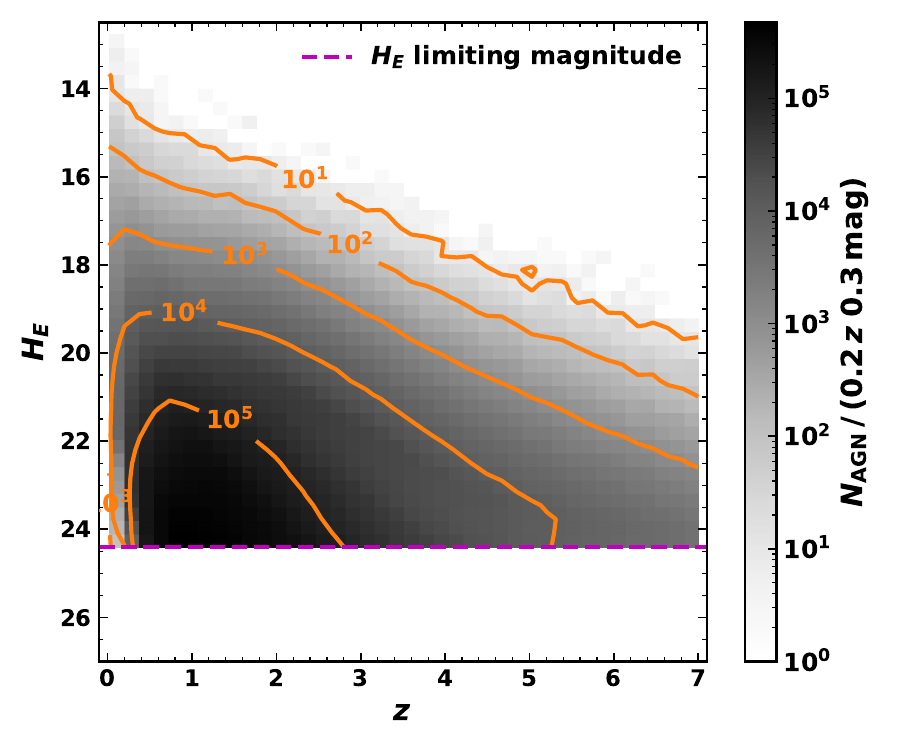}
         \label{fig:H_density_wide}
     \end{subfigure}
     
        \caption{Observed magnitude vs redshift density distributions for observable AGN in \Euclid's four photometric filters in the EWS. Two-dimensional histograms representing the density of observed AGN are plotted in grey-scale. Subplots correspond to \IE{} (top left), \YE{} (top right), \JE{} (bottom left), and \HE{} (bottom right). The data are binned with 40 bins in the $x$ and $y$ domains, giving two-dimensional AGN density in units of $N_{\rm AGN} / (0.2\,z,\,0.3\,{\rm mag})$. Contours of constant $\logten{N_{\rm AGN}}$ are plotted for $\logten{N_{\rm AGN}}$ = (1, 2, 3, 4, 5). Lines depicting the limiting magnitude in the EWS for each filter are plotted in dashed \textit{magenta} lines. }
        \label{fig:mag_density_wide}
\end{figure*}

\begin{figure*}
     \centering
     \begin{subfigure}[b]{0.49\textwidth}
         \centering
         \includegraphics[width=\textwidth]{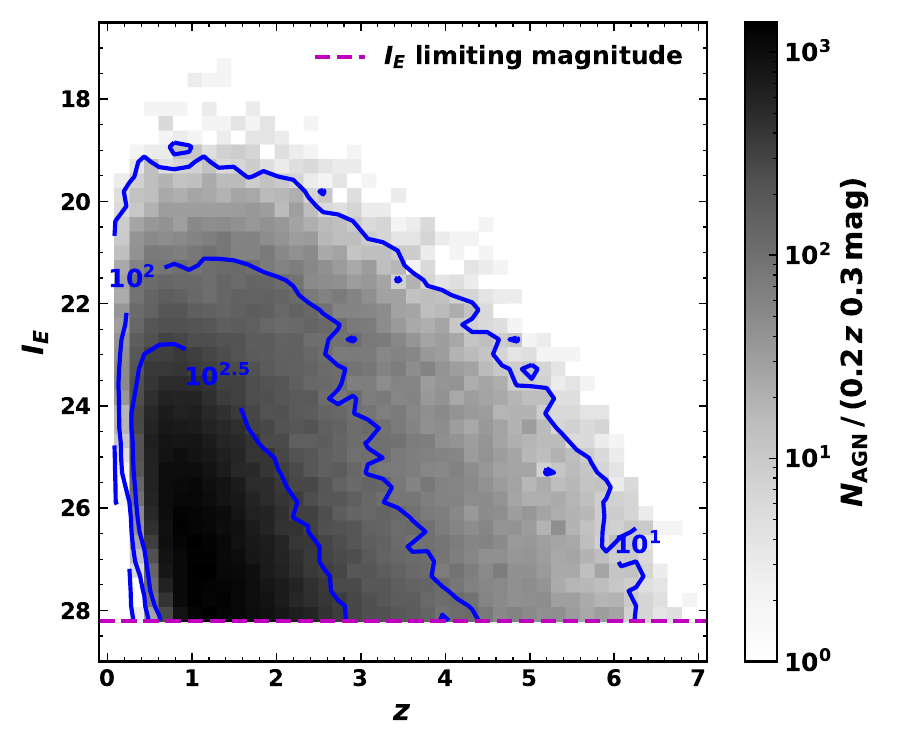}
         \label{fig:I_density_deep}
     \end{subfigure}
     \hfill
     \begin{subfigure}[b]{0.49\textwidth}
         \centering
         \includegraphics[width=\textwidth]{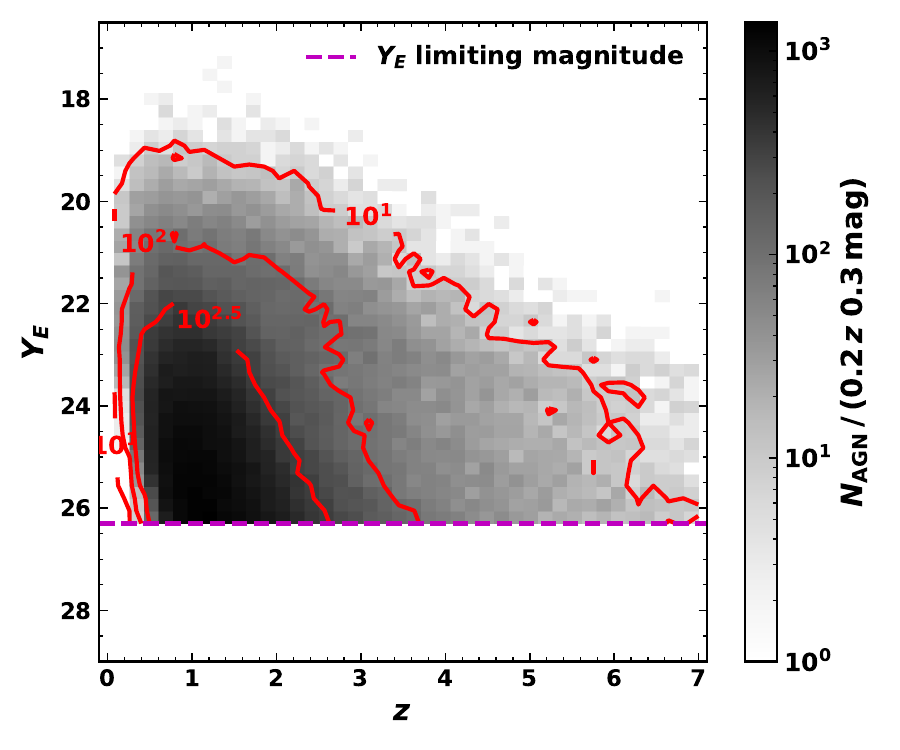}
         \label{fig:Y_density_deep}
     \end{subfigure}
     \vfill
     \begin{subfigure}[b]{0.49\textwidth}
         \centering
         \includegraphics[width=\textwidth]{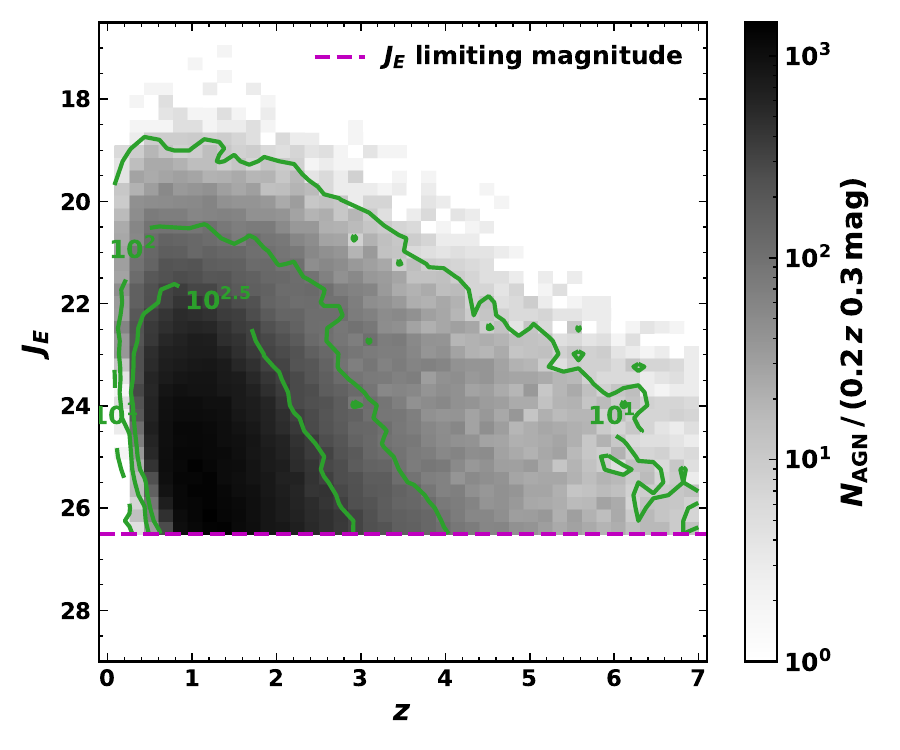}
         \label{fig:J_density_deep}
     \end{subfigure}
     \hfill
     \begin{subfigure}[b]{0.49\textwidth}
         \centering
         \includegraphics[width=\textwidth]{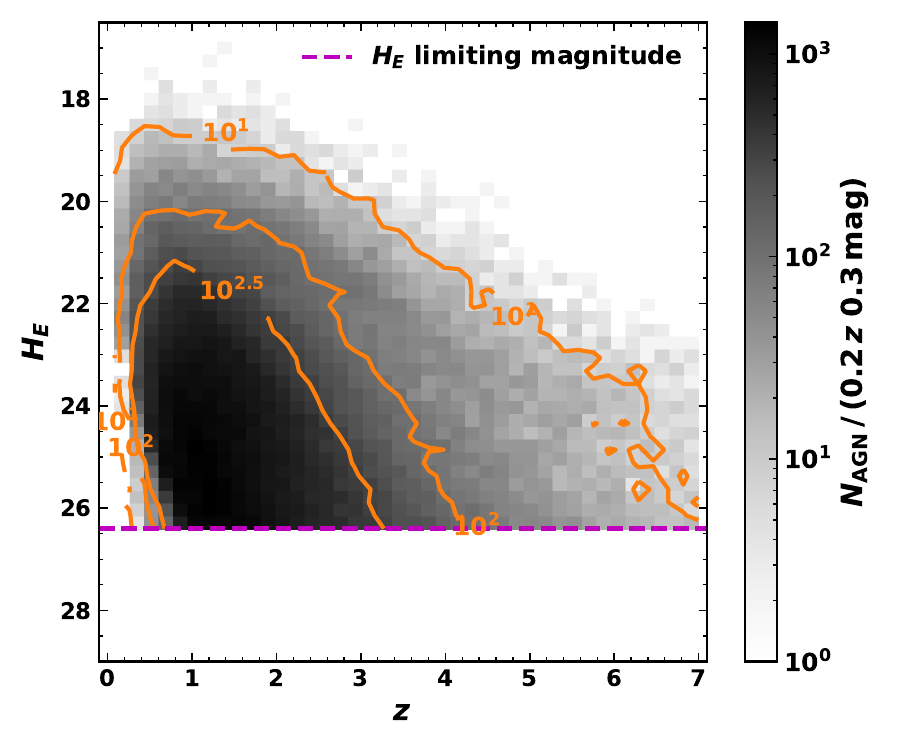}
         \label{fig:H_density_deep}
     \end{subfigure}
     
        \caption{Same as Fig. \ref{fig:mag_density_wide} for the EDS. Contours of constant $\logten{N_{\rm AGN}}$ are plotted for $\logten{N_{\rm AGN}}$ = (1, 2, 2.5).}
        \label{fig:mag_density_deep}
\end{figure*}

The predicted density distributions of \Euclid detectable AGN observed magnitudes as a function of redshift are depicted in Figs. \ref{fig:mag_density_wide} and \ref{fig:mag_density_deep} for EWS and EDS, respectively.
Comparing between the two surveys, we see that in each filter the deeper limiting magnitude of the EDS allows a high density of AGN close to the detection limit to be observed at $z=2$. Access to this parameter space in the EDS will therefore allow a greater surface density of AGN to be detected with \Euclid. The current deepest available NIR survey, ultra-VISTA \citep{mccracken2012}, reaches a similar depth to the EDS over 1.5\,deg$^{2}$. The EDS will match the depth of ultra-VISTA and surpass the area coverage by a factor of 30. 

\begin{figure}
\centering
\includegraphics[width=\hsize]{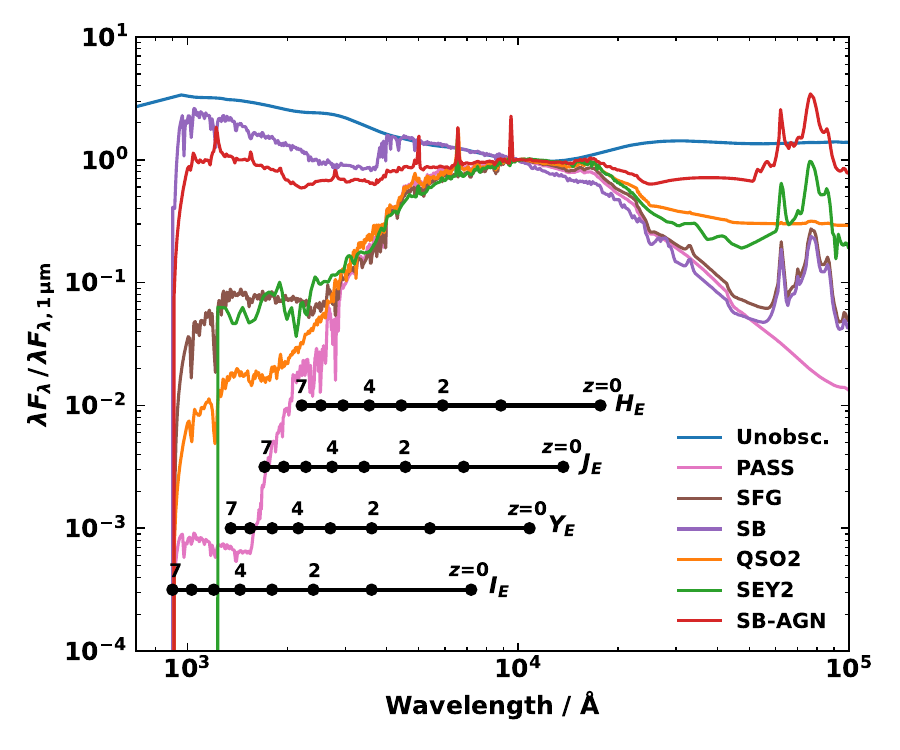}
  \caption{All SEDs (unobscured and obscured) assigned to AGN in this work, normalised at 1\,\micron. The redshift evolution of the effective wavelength for each \Euclid filter over the redshift range probed in this work are depicted below as black lines. The seven SED classes in this figure are: Unobscured AGN (`Unobsc.'; blue), Passive (`PASS'; pink), Star-forming (`SFG'; brown), Starburst (`SB'; purple), High-luminosity obscured AGN (`QSO2'; orange), Seyfert 2 (`SEY2'; green), and Starburst-AGN composite (`SB-AGN'; red).}
     \label{fig:filter_sed_zevo}
\end{figure}

For both the EWS and EDS the \YE{} band yields the smallest number of detectable AGN and \HE{} is capable of detecting the highest. There are a few effects driving this result. The \YE{} filter (along with \JE{}) is the narrowest \Euclid band. This means less photons will be captured by the bandpass. The \HE{} followed by \IE{} bands are the widest \Euclid bandpasses, generating the reciprocal of this effect. Alongside this consideration, \YE{} has the shallowest limiting magnitude of the \Euclid bands which will inhibit the detection of sources close to the detection limit compared to other, marginally deeper filters. The respective wavelength coverage plays an important role in the AGN yield of each band. Each un-reddened SED assigned in this work normalised at 1\,\micron{} is presented in Fig. \ref{fig:filter_sed_zevo}, with the effective wavelength position of each \Euclid filter plotted for the redshift range covered by our samples. It is clear that the longer wavelength bandpasses (\JE{}, \HE{}) will capture more flux from each SED compared to the shorter wavelength bands due to the shape of the SEDs where the flux diminishes at shorter wavelengths. This effect is apparent particularly at higher redshifts ($z \gtrsim 4$). With the addition of dust extinction to our SEDs the wavelength coverage driven performance difference is further accentuated for the longer wavelength \Euclid bands. We found specifically that more obscured AGN assigned the `QSO2' and `SEY2' SED classes are detected as the bandpass wavelength increases. It appears that the increased ability to detect obscured AGN to higher redshifts is the main driver for the differences in the number of AGN detected with each \Euclid filter.

The order of relative performance of each \Euclid band differs slightly between the EWS and EDS samples. For the EWS we see that in descending order the most AGN are detectable with $\HE>\IE>\JE>\YE$. In contrast, the results from the EDS go as $\HE>\JE>\IE>\YE$. The main difference between the two samples is that the EWS probes the bright-end slope of the XLF, detecting the brightest sources from across the extra-Galactic sky, whereas the EDS with its deeper limiting magnitudes and smaller area yields detectable AGN which lie on the faint-end slope of the XLF. Due to our optical type assignment model (see Sect. \ref{sec:opticaltype}) less luminous AGN are more likely to be assigned as obscured, with this probability increasing with redshift. We therefore found that more AGN are assigned as obscured in our EDS sample which probes the fainter AGN population. Our obscured AGN SED assignment model based on XXL SED fitting data (Sect. \ref{sec:type2agnSED}) allocates AGN2-like SEDs (classes `QSO2', `SEY2', `SB-AGN') with the highest probability to the low-luminosity group at higher redshifts. Pairing these factors together results in the EDS sample having a comparably higher number of AGN assigned AGN2-like SEDs compared to the EWS which contains more bright unobscured AGN. Referring to Fig. \ref{fig:filter_sed_zevo} and as discussed above, the longer wavelength bandpasses (\JE{}, \HE{}) perform better at detecting AGN with AGN2-like SEDs and therefore in the EDS the \JE{} filter detects more AGN than \IE{} due to the increased fraction of obscured AGN in the sample.

\section{Discussion}\label{sec:discussion}

\subsection{Uncertainties}\label{sec:uncertainties}

Throughout this work we make a number of key assumptions regarding models and empirical values adopted. Each of these choices introduces an element of uncertainty into our overall estimates of AGN numbers. In this section we work through our methodology and address five key assumptions we have made. We assess the impact each of these has on our predictions and attempt to quantify and understand the major drivers of the uncertainties within our framework. Many of the assumptions and choices we have made are due to inherent ambiguity in our current knowledge and understanding of astrophysical relations, particularly at higher redshifts. This work will therefore serve to point towards relations and models that can be extended and improved upon with data from future facilities, some of which may be addressed with \Euclid itself.

\subsubsection{Extrapolation of the XLF}
Through necessity, we extrapolated the XLF used in this work in redshift space. The data used to constrain the XLF lies in the range $0.01 < z < 4.0$. The lack of constraining observations at the low-luminosity end of the XLF, particularly at high redshift, leads to substantial uncertainty in the XLF in this regime. Our results for the EDS probe this area of parameter space and therefore will be most heavily affected by the uncertainty in this extrapolation. 

To understand the impact of extrapolating the XLF to higher redshifts we make use of parameter posterior distributions generated during the construction of the XLF in \citet{fotopoulou2016xlf}. For the nine XLF parameters (given in Table \ref{tab:XLFparams}) we obtained the final three hundred samples before final convergence, which we found effectively samples the 68\% credible interval (analogous to a $\pm\,1\sigma$ interval in frequentist statistics) of each parameter. From these parameter sets we generated three hundred corresponding realisations of the XLF. Using these realisations, we followed through our method (Sect. \ref{sec:method}) with each to assess the uncertainty on our numbers of \Euclid detectable AGN.

\begin{figure}
\centering
\includegraphics[width=\hsize]{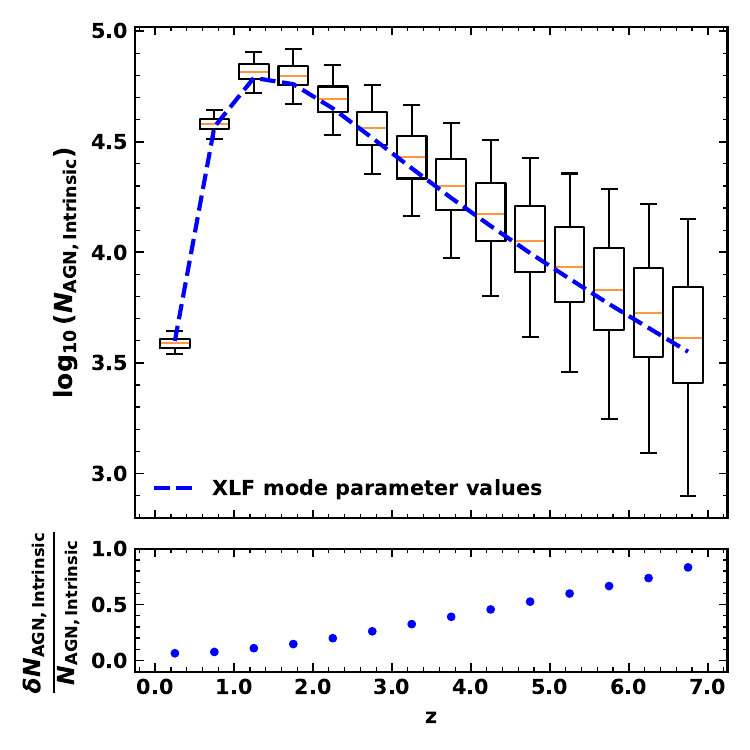}
  \caption{Distribution of the statistically expected number of AGN ($N_{\rm AGN,\,Intrinsic}$) in the EDS footprint in redshift bins ($\delta z = 0.5$) generated using three hundred realisations of the XLF used in this work. Each realisation samples XLF parameters visited in the final three hundred draws before convergence in the derivation of the XLF, probing the 1$\sigma$ XLF uncertainty. The limits of the box plots correspond to the first and third quartiles of the data, while the extent of the whiskers corresponds to $\pm\,1\sigma$. The orange lines in the boxes show the median number of intrinsic AGN in each redshift bin. The dashed blue line depicts the value attained when using the XLF constructed with the modal parameters presented in \citet{fotopoulou2016xlf}. The subplot depicts the relative 1$\sigma$ uncertainty of each redshift bin, when compared with the median value.}
     \label{fig:xlf_unc_intrinsic}
\end{figure}

In Fig. \ref{fig:xlf_unc_intrinsic} we show box plots depicting the distribution of the number of AGN in volume-limited samples derived from our XLF realisations for the EDS in redshift bins with $\delta z = 0.5$. There is a clearly increasing spread in the number of derived AGN with increasing redshift. This is a manifestation of the decreasing availability of constraining observations as redshift increases. While there is good agreement between each XLF realisation at low redshift, the $\pm1\sigma$ extent of the distributions extends to greater than 1\,dex in the $6.5 \leq z \leq 7$ bin. The subplot of Fig. \ref{fig:xlf_unc_intrinsic} depicts the relative $1\sigma$ uncertainty of each redshift bin, when compared with the median redshift bin value. We observe here that the relative uncertainty rises above 0.5 (50\%) at $z>4$.

The uncertainties exhibited by our XLF are largest on the low-luminosity end (see Fig. \ref{fig:lf_comparison}) and can be partly attributed to the propagation and consideration of photometric redshift uncertainties in the calculation of the overall XLF uncertainty. These uncertainties are seldom taken into consideration when constructing LFs. The uncertainty therefore appears large compared to other available AGN LFs, however may be a more truthful actualisation of the uncertainty. The XLF uncertainties observed at $z > 4$ are perpetuated as a consequence of the bias-variance trade-off philosophy adopted by its authors during computation. When there is little constraining observational data available in a region of parameter space, a decision must be made whether to allow for high variance in the resulting model or to attribute a greater significance to the data in the sparsely populated parameter space. The former will give rise to greater model uncertainties, however the truth is likely to be encapsulated within these. The latter should produce tighter model constraints that could perhaps bias the consequent model away from the ground truth. The authors of \citet{fotopoulou2016xlf} leaned towards a higher variance computation and hence when extrapolated we encounter large uncertainties. As discussed in Sect. \ref{sec:lumfuncs}, the XLF extrapolation in this work is in agreement with recent high-redshift constraints and the simulated unobscured AGN population is in agreement with UV/optical samples over $1 \leq z \leq 6$.

In our EDS analysis we found the median number of AGN observable in at least one \Euclid filter to be (2.4$\,\pm\,$0.3)\,$\times$\,10$^{5}$. The associated uncertainty corresponds to 12.5\% of the total number of \Euclid observable AGN. Considering the order of magnitude uncertainties observed in the number of intrinsic AGN this develops into a mild uncertainty in the number of \Euclid observable AGN. This is because AGN in the high-redshift regions, where the largest uncertainties for the intrinsic AGN are found, remain largely unobserved. Referring to Fig. \ref{fig:eds_detection_zdist} we see that the differential counts of the number of \Euclid observable AGN peaks between $z=1\text{--}2$ and drops below 10$^{3}$ at $z=4$. The majority of \Euclid-observable AGN (96\%) lie in the regions where the XLF is relatively well constrained and therefore the XLF uncertainty does not propagate into an extensive uncertainty when considering observed AGN.

We repeated the same analysis for the EWS, with reduced resolution in our method to save on computation. Specifically, for XLF integration bins with $\langle N \rangle \geq 100$, we re-sampled the data to treat one AGN in our data set to represent every 100 AGN. Any remainder AGN were treated as individual AGN with $\langle N \rangle = 1$, the same as for integration bins where $\langle N \rangle < 100$. We found that compared to our full resolution treatment of the EWS there is a difference of only 1\% in the resulting number of \Euclid-detectable AGN. The median number of EWS AGN observable in at least one \Euclid filter was (4.1$\,\pm\,$0.2)\,$\times$\,10$^{7}$. The associated uncertainty corresponds to 6.7\% of the total number of \Euclid observable AGN in the EWS. This marks an approximately halved relative uncertainty derived from the XLF between the EDS and EWS. In turn, this is a consequence of the EWS observing AGN that occupy the well constrained bright end of the XLF, whilst the EDS will probe AGN which occupy the faint end of the XLF, where the greatest uncertainties are present. 

We note that the median value derived from our XLF realisations are slightly larger than the number found for our sample derived with the XLF mode parameter values. As we can see in Fig. \ref{fig:xlf_unc_intrinsic} the XLF mode parameter value realisation (blue dashed line) used for the bulk of this work lies slightly below the median value in the distributions (orange lines) of the sampled XLF realisations at all redshifts. This is due to the shape of the posteriors that we have sampled parameter values from systematically producing more realisations with a marginally higher space density of AGN than the mode values. 

For comparison, we ran our EDS and XLF uncertainty analysis utilizing the hard XLF of \citet{ueda2014}. To obtain an uncertainty using the \citet{ueda2014} XLF we generated three hundred realisations using three hundred sets of Gaussian sampled parameters considering their published 1$\sigma$ parameter uncertainties. Following our method as above, we found a total of (2.6$\,\pm\,$0.8)\,$\times$\,10$^{5}$ AGN detectable in at least one \Euclid band. This is consistent with our results adopting the \citet{fotopoulou2016xlf} XLF within our derived XLF uncertainty and makes an almost indistinguishable difference to our final result. Analysing the redshift-dependent uncertainties, we found that within the well-constrained $z \leq 4$ regime the number of \Euclid detectable AGN derived with the \citet{fotopoulou2016xlf} and \citet{ueda2014} XLFs agree within 1$\sigma$ in all redshift bins. In the extrapolated $z > 4$ region, the \citet{ueda2014} XLF predicts a steeply decreasing yield of detectable AGN with increasing redshift compared to the \citet{fotopoulou2016xlf} XLF. This is consistent with the space-density decline observed in Fig. \ref{fig:lf_comparison}. The disparity between results with the two XLFs reaches 1\,dex in the $5.5 \leq z \leq 6.0$ bin. The relative uncertainty of the \citet{ueda2014} XLF predictions increases significantly with increasing redshift, inflated by the lower numbers of detectable AGN, up to 100\% in the $5.5 \leq z \leq 6.0$ bin. A larger overall relative uncertainty of 30\% is found with the \citet{ueda2014} XLF, where the relative uncertainty is larger than obtained with the \citet{fotopoulou2016xlf} XLF at all redshifts. The larger uncertainty in the low-redshift regime has a greater impact when propagated through our method to \Euclid detectable AGN as there are more observable AGN in this domain. In Sect. \ref{sec:euclidselection} we demonstrate that \Euclid is expected to identify $\sim 4\times10^{4}$ AGN in the EWS at $z > 4$ using colour cuts alone. These data will serve to point towards which extrapolation of the LF is correct.

\subsubsection{Extrapolation of the optically obscured AGN fraction}
The extrapolation of the \citet{merloni2014} optically obscured AGN fraction evolution led to some modification at lower luminosities where the fraction would exceed 1.0 beyond $z = 3$. We also found that when extrapolated beyond $z\approx6.5$ the obscured fraction of the highest luminosity AGN becomes greater than that of the medium luminosity sources (Fig. \ref{fig:absorption_evo}). This is touched upon in \citet{merloni2014}, where it is described that the incidence of obscuration shows significant redshift evolution only for the most luminous AGN, which appear to be more commonly obscured at higher redshifts.
\citet{treister&urry2006} also reported the relative optically obscured fraction of X-ray selected AGN increases with redshift in the range $0 \leq z \leq 4$, when corrected for optical counterpart selection effects.

In regard to X-ray obscuration, a range of works present evidence that the absorbed but Compton-thin fraction of X-ray AGN increases with redshift at fixed luminosity \citep[e.g.,][]{lafranca2005, hasinger2008, triester&2009, aird2015}. \citet{aird2015} reported with low significance that the absorbed fraction of the highest luminosity AGN may increase more substantially with redshift than lower luminosity AGN, remarkably similarly to our extrapolation of the \citet{merloni2014} model, albeit over a smaller redshift range (see \citealt{aird2015}, Fig. 15).
The heavily obscured ($\logten [N_{\rm H} / {\rm cm}^{-2}] > 23$), high-redshift ($3\leq z\leq 6$) fraction of AGN across a range of luminosities has been suggested to be $\sim \text{0.6--0.8}$ \citep[e.g.,][]{vito2018,signorini2023,pouliasis2024}. No dependence on X-ray luminosity or redshift was found for these values, however the derived fractions are higher than the local Universe obscured fraction for the same column density \citep[43\%;][]{burlon2011}.

In the high X-ray luminosity regime, \citet{vijarnwannaluk2022} reported that the obscured ($\logten [N_{\rm H} / {\rm cm}^{-2}] > 22$) fraction of X-ray selected AGN with $\logten (L_{\rm X} / {\rm erg}\, {\rm s}^{-1}) > 44.5$ lies around 76\% at $z=2$. This again marks an increase compared to obscured fractions derived in the local Universe.
\citet{buchner2015} similarly reported that averaged over cosmic time, obscured AGN with $\logten (N_{\rm H} / {\rm cm}^{-2}) > 22$ account for 77\% of the number and luminosity density of the AGN population with $\logten (L_{\rm X}  / {\rm erg}\, {\rm s}^{-1}) > 43$. They also described evidence of a redshift evolution, with the obscured fraction of Compton-thin AGN increasing towards $z=3$ where it is 25\% higher than the local Universe value. 
\citet{malizia2012} performed a study on the X-ray and optical obscuration properties of a sample of INTEGRAL/IBIS AGN spanning the redshift range $0.0014 \leq z \leq 3.7$. They presented evidence that the X-ray absorbed ($\logten [N_{\rm H} / {\rm cm}^{-2}] > 22$) fraction of AGN is a function of both X-ray luminosity and redshift, with the absorbed fraction increasing with increasing redshift but decreasing with increasing luminosity. They also determined that when corrected for bias, X-ray absorbed AGN account for up to 80\% of the population, in agreement with the results discussed above. Furthermore, they determined that the optically obscured and X-ray obscured classifications agree in 88\% of AGN.
Assuming the agreement of X-ray and optical obscuration in the majority of objects \citep[e.g.,][]{malizia2012,merloni2014,burtscher2016,fotopoulou2016}, we expect the fraction of optically obscured AGN to correlate with these general trends.

Obscuration in AGN is driven by attenuating gas and dust from both the AGN torus and ISM of the host galaxy in the line-of-sight to the observer. The contribution from the AGN torus is suggested to decrease with increasing AGN luminosity, a trend described by the receding torus model \citep[e.g.,][]{lawrence1991, simpson2005, assef2013}. A similar idea is proposed in the AGN evolution scheme of \citet{hopkins2008}, in which obscured AGN precede an unobscured AGN phase after blowing out gas and dust accumulated from galaxy mergers. The latter model provides a reasoning for the earlier peak in space density of obscured AGN compared to unobscured \citep[e.g.,][]{lacy2015}.
\citet{gilli2022} suggested that host galaxy ISM plays a larger role with increasing redshift, with $z \gtrsim 6$ AGN expected to be primarily obscured by the ISM of their hosts.

Considering the results above, for low X-ray luminosity AGN extrapolated to higher redshifts our model may over-predict the obscured fraction of AGN by 20--40\%. In this regime, the overall trend of the optically obscured fraction of AGN increasing from local values is preserved and in agreement with the literature. For medium and high X-ray luminosities our extrapolated model agrees with observational trends of optically obscured fractions of AGN increasing with redshift and the highest luminosity AGN having a more significant evolution compared to medium luminosity counterparts. Overall, the extrapolated model used in our method is consistent with the available literature, however at high redshifts it is not feasible to discern the true evolution of the optically obscured fraction due to lack of observable objects with the current technological limitations. 

\subsubsection{AGN bolometric correction dispersion}

Intrinsic scatter in bolometric corrections is to be expected as AGN spectra are diverse and exhibit different scaling from object to object with the population following a correlated trend rather than an exact underlying rule. We decided to conduct our analysis using a nominal bolometric correction conversion as a population average. To test the impact of this assumption on our overall estimates of \Euclid observable AGN we re-ran our analysis for the EDS, separately enabling scatter for each bolometric correction, for three hundred runs each. 

We explored the dispersion for the employed bolometric corrections in \citet{duras2020} and \citet{runnoe2012b} through Gaussian sampling of each parameterising variable considering the 1$\sigma$ uncertainties reported in their respective publications. For the \citet{duras2020} X-ray bolometric correction we sampled an additional factor derived from the reported intrinsic spread of 0.27\,dex. New parameter and intrinsic spread values were drawn for each obscured AGN we consider.

Introducing scatter in the X-ray bolometric correction we found a median number AGN detectable in at least one \Euclid band of (2.150$\,\pm\,$0.001)\,$\times$\,10$^{5}$. The 1$\sigma$ uncertainty of our analysis corresponds to 0.05\% of the total number of \Euclid observable AGN. This difference is marginal to the overall results of our analysis. 

Completing the analysis with scatter in the 2\,\micron{} bolometric correction we report a median number of AGN detectable in at least one \Euclid band of (2.156$\,\pm\,$0.001)\,$\times$\,10$^{5}$. Identical to our X-ray bolometric correction uncertainty analysis, the uncertainty associated with the 2\,\micron{} bolometric correction corresponds to 0.05\% of the total number of \Euclid observable AGN. 


\subsubsection{Unobscured AGN emission lines}

The \citet{shen2020} mean quasar SED used to model unobscured AGN in this work lacks AGN emission lines. Strong AGN emission lines, such as the Ly$\alpha$ and H$\alpha$ hydrogen recombination lines, have been demonstrated to cause significant deviations to photometric colours. These changes are a strong function of redshift as the emission lines are shifted in and out of different photometric filters \citep{temple2021}. We focus on Ly$\alpha$ and H$\alpha$ in the present analysis because these emission lines are the most prominent emission line features in many AGN. The findings of this section therefore provide a lower limit to the overall effect the inclusion of emission lines has on our results.

The H$\alpha$$+$[\ion{N}{ii}] emission line complex is composed of the broad and narrow component H$\alpha$ emission line and narrow [{\sc{N ii}}]$\lambda$6549 and [\ion{N}{ii}]$\lambda$6583 lines. H$\alpha$ has a rest-frame wavelength of 6563\,\AA{} and is hence redshifted beyond the extent of the \HE{} band at $z=2.05$. The analysis of H$\alpha$ in this section is therefore limited to the redshift range $0.01 \leq z \leq 2.05$. Ly$\alpha$ is also comprised of a broad and narrow component and has rest-frame wavelength 1216\,\AA{}. Ly$\alpha$ therefore enters the \IE{} band at $z=3.4$. Accordingly, the analysis of the impact of including Ly$\alpha$ in our unobscured AGN template is constrained to $3.4 \leq z \leq 7.0$.

To attain realistic Ly$\alpha$ and H$\alpha$$+$[\ion{N}{ii}] flux excess values we exploited measurements of stacked quasar templates produced in \citet{lusso2023}. We refer the reader to the aforementioned work for full details of the spectral construction and analysis, however we give a brief overview here. Nine empirical SDSS quasar stacked spectra were generated from a parent sample of 91\,579 SDSS-DR7 quasars following the method of \citet{lusso2015}. Each empirical stack was binned in H$\beta$ equivalent width (EW) and FWHM ranges. The nine templates were modified using a grid of redshifts, \ebv, and bolometric luminosities. The ranges of these parameters were constructed to probe the expected observed parameter space of unobscured AGN in \Euclid. The spectra were scaled to the given luminosity, optically reddened with the assigned \ebv{} value and the \citet{prevot1984} SMC law and finally redshifted with IGM extinction applied via the curve of \cite{prochaska2014}. Across the parameter grid a random sub-sample of 1248 of these spectra that satisfied the \Euclid depths of VIS and NISP (5$\sigma$) were selected. 

The emission line fluxes for each of the empirical unobscured AGN spectra included in the final sample were measured using the Quasar Spectral Fitting library \citep[\qsfit;][]{calderone2017,selwood2023} AGN spectral fitting package. We leveraged the \qsfit{} spectral fitting results to obtain the H$\alpha$$+$[\ion{N}{ii}] complex integrated luminosity excess for each of the nine unobscured AGN templates. For each empirical template the median luminosity and its median absolute deviation ($\sigma_{\rm MAD}$) was determined. We then assigned each unobscured AGN in our EWS and EDS samples a random unobscured AGN template class and sampled a corresponding H$\alpha$$+$[\ion{N}{ii}] luminosity based on the derived median and $\sigma_{\rm MAD}$ values determined for the corresponding template class. Transforming to a H$\alpha$$+$[\ion{N}{ii}] complex flux excess, we added the flux perturbation to the measured incident flux in the \Euclid photometric band associated with the H$\alpha$ observed-frame centroid. We then re-calculated the apparent magnitude considering the emission line perturbation. Adopting this procedure we assessed the number of AGN that would become observable in any \Euclid band with the addition of the H$\alpha$$+$[\ion{N}{ii}] emission line complex in our unobscured AGN SED. We followed the same procedure for Ly$\alpha$ measurements, however due to there only being $\sim 300$ spectra with $z > 3.4$ in the \citet{lusso2023} sample we do not separate by template class. Instead, we take the median Ly$\alpha$ luminosity and $\sigma_{\rm MAD}$ of the full $z > 3.4$ sub-sample.

In our EWS sample, there were 1.0\,$\times$\,10$^{7}$ candidate unobscured AGN in the H$\alpha$$+$[\ion{N}{ii}] range $0.01 \leq z \leq 2.05$. We found 4644 (0.05\%) of these AGN became newly observable with the inclusion of H$\alpha$$+$[\ion{N}{ii}]. There were 1.9\,$\times$\,10$^{6}$ unobscured AGN at $3.4 \leq z \leq 7.0$ of which 119 (0.006\%) became observable with the inclusion of Ly$\alpha$. In total a lower limit of 4763 out of 1.2\,$\times$\,10$^{7}$ (0.04\%) unobscured AGN could have been observed with the inclusion of broad emission lines in our unobscured AGN template.

For the EDS sample the inclusion of H$\alpha$$+$[\ion{N}{ii}] resulted in an additional 60 detectable unobscured AGN out of 3.3\,$\times$\,10$^{4}$ at $0.01 \leq z \leq 2.05$, a gain of 0.2\%. There were 5.9\,$\times$\,10$^{3}$ unobscured AGN at $z > 3.4$ of which only two (0.03\%) became observable by \Euclid with Ly$\alpha$ considered. In total 62 of 3.9\,$\times$\,10$^{4}$ (0.16\%) unobscured AGN could be observed with the inclusion of broad emission lines in our unobscured AGN template.

The enhanced proportion of newly detected AGN from H$\alpha$$+$[\ion{N}{ii}] in the EDS is due to the deeper apparent magnitudes (i.e., fluxes) probed in the survey. As the 5$\sigma$ flux threshold is deeper, a small change in the source flux provided by emission lines is more likely to allow an AGN to breach this threshold and become detectable with \Euclid. In both surveys the inclusion of Ly$\alpha$ made a smaller difference to detectability compared to H$\alpha$$+$[\ion{N}{ii}]. This is due to the higher redshift range where Ly$\alpha$ occupies the \Euclid bands. The additive Ly$\alpha$ flux excess in this regime is comparatively smaller and so has a lower impact on detectability than with H$\alpha$$+$[\ion{N}{ii}]. We observe from Figs. \ref{fig:mag_density_wide} and \ref{fig:mag_density_deep} that whilst there is a high density of AGN near the detection threshold at $0.01 \leq z \leq 2.05$, the parameter space is more sparse at $z > 3.4$. This means there are less AGN that could become observable due to a flux perturbation from emission lines at higher redshifts.

Despite these numbers amounting to a lower limit, the derived number of newly observed AGN are a negligible percentage of the total observable sources and fall well within the Poisson error ($\sigma = \sqrt{N}$) of our data sets. We therefore conclude that the omission of AGN emission lines contributes a minimal effect and uncertainty on our total AGN number estimates in the \Euclid photometric surveys. 

\subsubsection{Uncertainty comparison}

\begin{figure}
\centering
\includegraphics[width=\hsize]{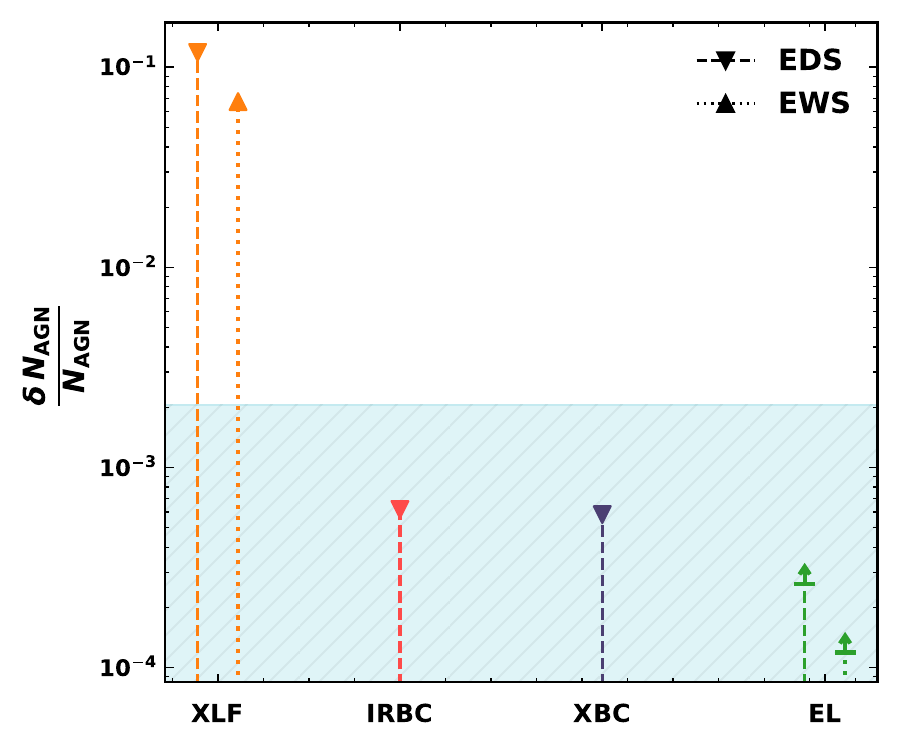}
  \caption{Relative 1$\sigma$ uncertainty of the number of AGN detectable in at least one \Euclid band ($N_{\rm AGN}$) for each of the quantified sources of uncertainty introduced in this work. Displayed in this plot are uncertainties introduced by the X-ray luminosity function (XLF), infrared bolometric correction dispersion (IRBC), X-ray bolometric correction dispersion (XBC) and unobscured AGN emission lines (EL) for which a lower limit was quantified using the Ly$\alpha$ emission line and H$\alpha$$+$[\ion{N}{ii}] complex. The expected relative Poisson noise for the EDS, calculated as $\sigma = \sqrt{N_{\rm AGN}}$, is indicated by the shaded blue region.}
     \label{fig:uncertainty}
\end{figure}

Here we compare the uncertainties determined for different models and assumptions in this work. In Fig. \ref{fig:uncertainty} we present a collation of the relative uncertainties quantified in this section on the number of AGN with a detection in at least one \Euclid band. The blue shaded region signifies the expected relative Poisson noise ($\sigma = \sqrt{N}$) for our EDS detectable AGN. 

The dominant source of uncertainty in our analysis is the uncertainty in the XLF. The XLF relative uncertainty is several orders of magnitude greater than any other source of uncertainty quantified here and is the only source that generates an uncertainty greater than the derived Poisson noise for our EDS sample. Amounting to 12.5\% for the EDS and 6.7\% for the EWS, we may consider this the overall uncertainty of our detectable AGN estimates. 
Well within the EDS Poisson noise, the second largest uncertainty is driven by the IR and X-ray bolometric correction dispersion. These relative uncertainties are already negligible compared to that of the XLF. 

We were unable to quantify the impact that extrapolating the \citet{merloni2014} optically obscured AGN fraction model has on our results. Although we have established our extrapolation agrees with the general trends of literature results, the ground truth of the redshift evolution and X-ray luminosity dependence of the optically obscured AGN fraction remains unresolved. Despite the complexity in constraining such dependencies we hope that the extension of such models in the redshift and luminosity domain can be an area of investigation with next-generation facilities. 

In a scenario where our predictions in this analysis are proven to be in tension with \Euclid observations, we can back-propagate through our assumed models and framework to identify where disparities lie between our current understanding and the ground truth. This will in turn serve to inform the community of areas where there are gaps and inconsistencies in our current modelling of AGN properties and demographics. 



\subsection{Photometric AGN selection in \Euclid}\label{sec:euclidselection}

Optical to MIR colour-colour cuts are often invoked to select samples of AGN from photometric data sets \citep[e.g.][]{stern2005, donley2012, assef2018}. The colours used in AGN selection exploit spectral features unique to AGN, such as the MIR excess, to separate AGN from inactive galaxies and stars in the colour-colour parameter space. Because they can be quickly and cheaply applied to large data sets, simple colour selection criteria are used in many cases as an initial selection for candidates of a population before applying more complex identification methods. This will likely be the use-case for colour selection of AGN in \Euclid. 

Only a fraction of the AGN detected by \Euclid, i.e., present in at least one image, will actually be selected as AGN based on \Euclid photometry alone. Indeed, half of the obscured AGN SED classes we assign in Sect. \ref{sec:type2agnSED} (PASS, SFG, SB) are entirely free from AGN spectral features in the \Euclid wavelength range. This means that without external data these AGN will appear indistinguishable from non-active galaxies. Therefore, it will not be possible to identify all of our \Euclid detectable AGN using photometric criteria, with \Euclid data alone or otherwise. 

Colour selection criteria for AGN using \Euclid photometry has been explored in depth in Bisigello et al. (in prep). Optimal selection criteria were derived for the EWS and EDS based on \Euclid photometry exclusively and with the inclusion of ancillary photometric observations, such as with \textit{Spitzer}/IRAC \citep{fazio2004} and Rubin/LSST. Each selection was defined to target either unobscured AGN or all AGN including obscured and composite sources, with optimal criteria derived to maximise the F1-score, i.e. the harmonic mean of sample completeness ($C$; fraction of selected AGN with respect to the total AGN population) and purity ($P$; fraction of selected sample that are AGN, not contaminants). Each criterion exhibits a selection function with a unique redshift dependence. In the following, we have applied each of the optimal colour selection criteria to our samples of simulated AGN to assess how many \Euclid detectable AGN we can expect to select using \Euclid colours. We verified that the AGN colours derived with our models are consistent with those derived in Bisigello et al. (in prep) over the relevant ranges for AGN selection. 

We consider only AGN in our samples with 5$\sigma$ detections in the constituent bands for each colour criterion. For AGN selection schemes incorporating Rubin/LSST bands we consider the 5$\sigma$ point-source final co-added depths reported in \citet{ivezic2019}; (\textit{u, g, r, i, z, Y}) = (25.6, 26.9, 26.9, 26.4, 25.6, 24.8). Detailed breakdowns of the photometric selection criteria and the results of applying each to our data are provided in Appendix \ref{app:colourcriteria}. We additionally present density plots depicting our AGN sample with each colour selection criterion considered in this section (Figs. \ref{fig:ews_selection}, \ref{fig:eds_selection}).  

\begin{table*}
\caption{Performance of \Euclid photometry AGN selection criteria (Bisigello et al., in prep.) applied to our samples of EWS and EDS AGN. We consider only AGN that are detected above the 5$\sigma$ limiting magnitude for the four constituent filters in each colour criterion. For each criterion we report the total number of selected AGN, the surface density of selected AGN in deg$^{-2}$, the completeness ($C$) at the 5$\sigma$ level in the relevant filters, and the expected purity ($P$) for the criterion, as estimated in Bisigello et al. (in prep.). The final row for each survey denotes the union ($\cup$) of all considered selection criteria.}             
\label{tab:selection}      
\centering                          
\begin{tabular}{l c c c c c c c c}        
\hline            
Survey & Target Class & Photometry & \head{2cm}{Total Selected AGN} & \head{2cm}{Surface Density (deg$^{-2}$)} & \head{2cm}{Unobscured AGN} & \head{2cm}{Obscured AGN} & $C$ & $P$\\    
\hline                        
& & & & & & & & \\[-8pt]
    EWS & Unobscured AGN & \Euclid & \NUMSelectedEuclidOnlyTypeOneEWS & \SDSelectedEuclidOnlyTypeOneEWS & 4.4\,$\times$\,10$^{6}$ & 3.7\,$\times$\,10$^{5}$ & \CSelectedEuclidOnlyTypeOneEWS & 0.17 \\
    EWS & Unobscured AGN & \Euclid, Rubin/LSST & \NUMSelectedEuclidLSSTTypeOneEWS & \SDSelectedEuclidLSSTTypeOneEWS & 5.5\,$\times$\,10$^{6}$ & 1.8\,$\times$\,10$^{5}$ & \CSelectedEuclidLSSTTypeOneEWS & 0.92 \\
    EWS & All AGN & \Euclid, Rubin/LSST & \NUMSelectedEuclidLSSTAllEWS & \SDSelectedEuclidLSSTAllEWS & 4.3\,$\times$\,10$^{6}$ & 1.7\,$\times$\,10$^{6}$ & \CSelectedEuclidLSSTAllEWS & 0.19 \\
    EWS & $\cup$ & \Euclid, Rubin/LSST & 8.1\,$\times$\,10$^{6}$ & 556 & 6.0\,$\times$\,10$^{6}$ & 2.1\,$\times$\,10$^{6}$ & 0.37 & - \\
    \hline
& & & & & & & & \\[-8pt]
    
    EDS & Unobscured AGN & \Euclid & \NUMSelectedEuclidOnlyTypeOneEDS & \SDSelectedEuclidOnlyTypeOneEDS & 1.6\,$\times$\,10$^{4}$ & 400 & \CSelectedEuclidOnlyTypeOneEDS & 0.23 \\
    EDS & Unobscured AGN & \Euclid, Rubin/LSST & \NUMSelectedEuclidLSSTTypeOneEDS & \SDSelectedEuclidLSSTTypeOneEDS & 1.9\,$\times$\,10$^{4}$ & 390 & \CSelectedEuclidLSSTTypeOneEDS & 0.92 \\
    EDS & All AGN & \Euclid, Rubin/LSST & \NUMSelectedEuclidLSSTAllEDS & \SDSelectedEuclidLSSTAllEDS & 2.0\,$\times$\,10$^{4}$ & 9.5\,$\times$\,10$^{3}$ & \CSelectedEuclidLSSTAllEDS & 0.58 \\
    EDS & $\cup$ & \Euclid, Rubin/LSST & 3.5\,$\times$\,10$^{4}$ & 692 & 2.5\,$\times$\,10$^{4}$ & 1.0\,$\times$\,10$^{4}$ & 0.22 & - \\
\hline

\end{tabular}
\end{table*}

Table \ref{tab:selection} provides a summary of the performance of each photometric selection criterion applied to our AGN sample. Absolute expected numbers of selected AGN and surface densities are presented as well as an assessment of the sample completeness. In all circumstances the quoted completeness refers to the completeness of selected AGN relative to 5$\sigma$ detected AGN available in the same bands. Due to our sample, by construction, containing only the expected colours of AGN and not other astrophysical sources we are unable to estimate the purity of our selected samples, however we note the expected purity of each selection criterion from Bisigello et al. (in prep.). 

Adopting the \Euclid-only unobscured AGN criterion, we expect a completeness $C=\CSelectedEuclidOnlyTypeOneEWS$ in the EWS and $C=\CSelectedEuclidOnlyTypeOneEDS$ in the EDS. When considering completeness with respect to available unobscured AGN, the target class of the selection, the respective EWS and EDS completeness values rise to $C=\CTypeOneSelectedEuclidOnlyTypeOneEWS$ and $C=\CTypeOneSelectedEuclidOnlyTypeOneEDS$. AGN selected using colour cuts with \Euclid photometry alone represent 40\% and 35\% of the unobscured AGN detectable in at least one \Euclid band in the EWS and EDS, respectively. The same samples correspond to 12\% and 8\% of the total AGN population detectable in at least one \Euclid band. The surface densities resulting from these selections have a difference of only 15\,deg$^{-2}$ between the EWS and EDS, despite the EDS probing two magnitudes deeper. The disparity stems from the low overall completeness for the EDS selection. Although there is around twice the surface density of detected AGN in the EDS, the \Euclid-only unobscured AGN selection has around half the completeness. The inefficiency of the EDS selection is noted in Bisigello et al. (in prep.), emerging from the fact that most EDS AGN are expected to be faint, lying on the detection boundary of each band.

The addition of Rubin/LSST bands, when available, are expected to supplement the purity, completeness, and size of AGN samples selected with \Euclid data. Crucially, the inclusion of ancillary data will allow the selection of obscured and composite AGN as well as unobscured. 
To determine the total sample sizes of selectable AGN with \Euclid, we took the union of all selected samples resulting from the criteria discussed in this section, leveraging both \Euclid and Rubin/LSST photometry. The total selected samples in the EWS and EDS have overall completeness $C=0.37$ and $C=0.22$. Considering selected and available unobscured AGN, the total samples have completeness $C=0.66$ and $C=0.58$ in the EWS and EDS. The completeness is $C=0.15$ and $C=0.09$ with respect to obscured AGN. These total samples equate to 21\% and 15\% of the AGN found to obtain a 5$\sigma$ detection in at least one \Euclid photometric band in the EWS and EDS, respectively.

\begin{figure}
\centering
\includegraphics[width=0.49\textwidth]{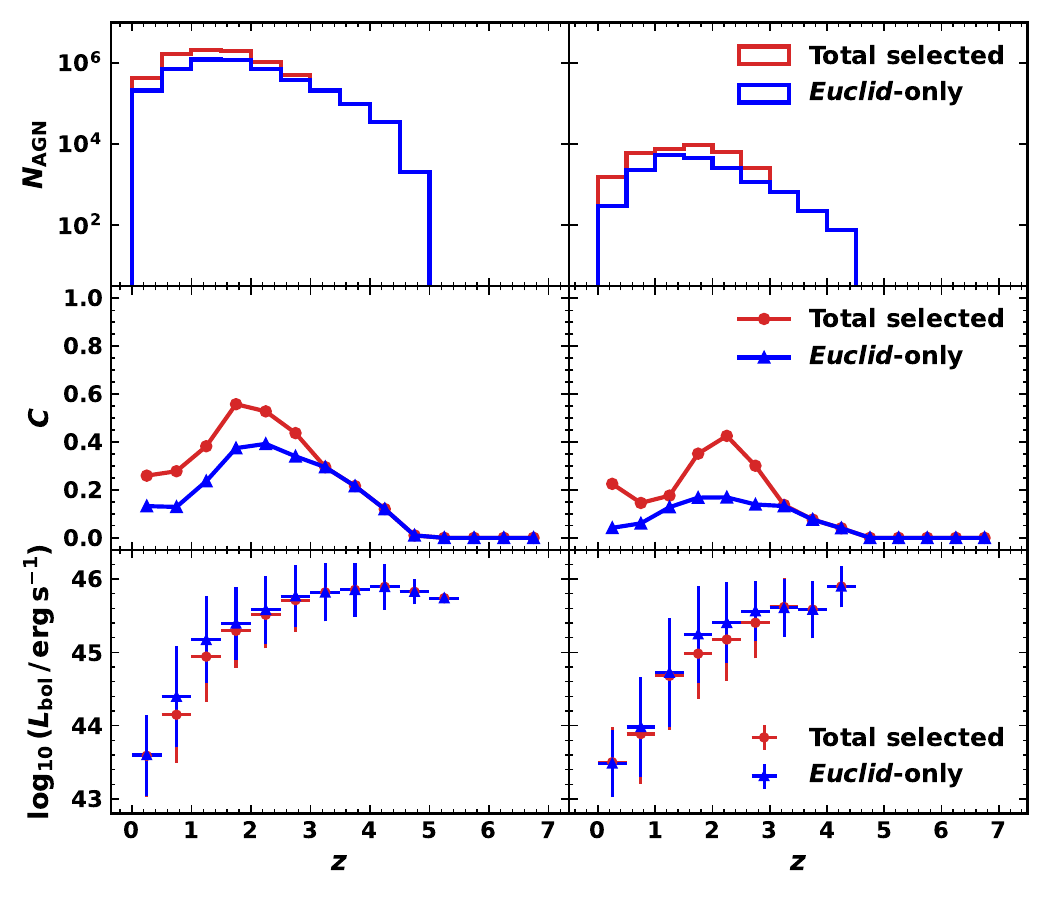}
  \caption{Redshift distributions (top), redshift-dependent completeness (middle), and $L_{\rm bol}$-$z$ planes (bottom) for the selected AGN in the EWS (left) and EDS (right). AGN selected with \Euclid-only photometric criteria (blue) and the total selected sample defined as the union ($\cup$) of all \Euclid and ancillary $ugrz$ photometric criteria discussed in Sect. \ref{sec:euclidselection} are plotted for each panel. In all plots redshift is binned with width $\delta z = 0.5$. In the $L_{\rm bol}$-$z$ plane points represent the median, vertical lines represent $1\sigma$ standard deviation and horizontal lines represent the width of the redshift bin.}
     \label{fig:selectioninfo}
\end{figure}

Figure \ref{fig:selectioninfo} presents redshift distributions, redshift-dependent completeness, and $L_{\rm bol}$-$z$ planes of the \Euclid-only selection and union of all selections for the EWS and EDS.
The middle panels of Fig. \ref{fig:selectioninfo} show the completeness is highest for all selected samples around cosmic noon ($z \sim 2$), matching the expectations of Bisigello et al. (in prep.). This marks the redshift at which the 4000\,\AA{} break enters the \YE{} band, providing one of the cruxes the photometric selection criteria exploits to separate AGN from galaxies. 
An elevated completeness is observed around this redshift for the total selected sample compared to the \Euclid-only selected sample in both surveys and is particularly accentuated for the EDS selections. The shorter wavelength optical bands help to further distinguish AGN colours from the generally redder galaxy colours.

The derived 5$\sigma$ level completeness for each of the applied selection criteria may appear as a point of concern for our samples. Many analyses however do not have the benefit of first assessing the underlying population of available detected sources. Those that do perform such an exploration report similar or diminished levels of completeness. For example, \cite{assef2018} estimate a completeness of 0.17 and 0.28 for their colour-colour selections of AGN from the AllWISE survey with a reliability of 90\% and 75\%, respectively \citep[see also][]{stern2012, assef2013}. 

The redshift distributions (top panels of Fig. \ref{fig:selectioninfo}) show the highest number of AGN are selected around around cosmic noon in both surveys. This is congruent with the peak in the number of detected AGN (Fig. \ref{fig:detection_zdists}) as well as the highest selection completeness. AGN selected at $z > 3$ are dominated by those identified by the \Euclid-only unobscured AGN criterion. The addition of ancillary photometry will help to distinguish AGN at low and intermediate redshifts, whilst \Euclid colours alone can identify the luminous unobscured AGN at $z > 3$. We project $\sim 4\times10^{4}$ AGN at $z > 4$ will be selected in the EWS, with $\sim 2000$ selected AGN at $z > 4.5$.

Unobscured AGN are selected efficiently in all cases, with around half of the available AGN of this class identified using \Euclid photometry alone, and up to two thirds when employing ancillary optical colours. Obscured AGN are less efficiently identified with only a tenth of those available in the EDS selected and a fifth in the EWS. Indeed, across all selections for both surveys only $\sim 15$\% of selected AGN have $E(B-V) \geq 0.05$. In comparison, $\sim 50$\% of the detected AGN available in the selection bands have $E(B-V) \geq 0.05$. The majority of the colour selections used here were fine-tuned to identify the unobscured population, hence it is no surprise the reddened population is missed. It is clear however that there is an abundance of AGN with higher \ebv{} available to be extracted from the data through alternative means.

The obscured AGN which are identified in either survey have the SB-AGN, SEY2, or SB SED classes. The SB-AGN and SB classes have optical--NIR colours similar to unobscured AGN (see Fig. \ref{fig:colourvalidation_logan2020}). As such these SED classes are often captured in the same selection cuts identifying unobscured AGN. No obscured AGN with the PASS, SFG or QSO2 SED classes were selected. These are all galaxy-dominated SEDs (see Fig. \ref{fig:filter_sed_zevo}) with optical--NIR colours that are hard to distinguish from inactive galaxies. As such, we would not expect these classes to be efficiently selected via \Euclid photometry.

AGN will be selected over the bolometric luminosity range $43 \leq \log_{10} (L_{\rm bol}\,/\,{\rm erg\,s}^{-1}) \leq 47$, a significant portion of the range used as input for our simulation. 
The $L_{\rm bol}$-$z$ planes in the bottom panel of Fig. \ref{fig:selectioninfo} show that ancillary photometry allows us to select lower-luminosity AGN in both the EWS and EDS. This is apparent compared to the \Euclid-only unobscured AGN selection at $0.5 \leq z \leq 3$ for both surveys. The range of the $1\sigma$ vertical bars suggest AGN selected around $0.5 \leq z \leq 3$ in the EDS are relatively lower-luminosity than those in the EWS. This is due to the two magnitude deeper observations in the EDS reaching a fainter population of AGN. The highest redshift selected AGN from our simulation is at $z = 5.16$ and selected from the EWS. Due to the much larger area probed by the EWS it is more likely that exceptionally luminous high-redshift AGN will be observed and selected. We discuss predictions of high-redshift ($z \sim 7$) AGN detections in more detail compared with the results of \citet{barnett2019} in Appendix \ref{app:barnett}.

The large disparity between the projected number of AGN we will detect with \Euclid versus those we can identify as AGN should not be neglected. Our findings suggest work should be undertaken to improve upon and devise new AGN selection methods in order to maximize the AGN yield with \Euclid as well as other facilities. This is particularly crucial for intermediate and high-redshift AGN selection. We show in this work we expect to detect an abundance of AGN in this regime with \Euclid, however we lack a means to efficiently identify such AGN from the data. Of course, this is partly due to \Euclid probing wavelengths where it is difficult to distinguish AGN emission from other populations of galaxies. A multiwavelength approach where ancillary observations allow, or explore the viability of machine learning driven techniques (Bisigello et al., in sub., Signor et al., in prep.) is necessary.

Before Rubin/LSST bands become available, it is possible to exploit readily accessible ancillary optical data to perform AGN selection in conjunction with \Euclid bands. The Dark Energy Survey \citep[DES;][]{abbott2018}, covering a footprint of $\sim$5000\,deg$^{2}$ in the southern hemisphere, provides optical to NIR photometry in the \textit{g, r, i, z, Y} bands to 5$\sigma$ photometric depths of 25.0, 24.5, 23.7, 22.6 and 21.3, respectively. Unfortunately, a one-to-one mapping is not possible with DES for the photometric selection criteria defined in Bisigello et al. (in prep.) as DES lacks observations in a \textit{u} band. The Ultraviolet Near Infrared Optical Northern Survey (UNIONS) however will observe $\sim$5000\,deg$^{2}$ of the northern hemisphere in the \textit{gri} bands and $\sim$10\,000\,deg$^{2}$ in the crucial \textit{u} band \citep{ibata2017}. 
We assessed the number of AGN in the EWS with a 5$\sigma$ detection in each DES band to be 1.2\,$\times$\,10$^{7}$ in $g$, 1.3\,$\times$\,10$^{7}$ in $r$, 1.1\,$\times$\,10$^{7}$ in $i$, 6.1\,$\times$\,10$^{6}$ in $z$ and 2.1\,$\times$\,10$^{6}$ in $Y$. For AGN in the EDS we predict 4.2\,$\times$\,10$^{5}$ in $g$, 4.3\,$\times$\,10$^{5}$ in $r$, 3.4\,$\times$\,10$^{5}$ in $i$, 1.9\,$\times$\,10$^{5}$ in $z$ and 5.9\,$\times$\,10$^{4}$ in $Y$.


\subsection{Comparison to other surveys}\label{sec:surveycomparison}


\begin{table*}
\caption{Characteristics and AGN identification statistics for a selection of medium area and wide field surveys across different wavebands, which we use for comparison with our \Euclid yields. The quoted area relates to the area considered in the survey AGN selection procedure as described in the corresponding reference(s). \Euclid survey detected and selected yields as plotted in Fig. \ref{fig:survey_comparison} are provided. Detected samples refer to the number of AGN with a 5$\sigma$ detection in at least one \Euclid filter. Selected samples correspond to the union of all selection criteria considered in Sect. \ref{sec:euclidselection}.}             
\label{tab:surveycomp}      
\begin{threeparttable}
\centering                          
\begin{tabular}{l c c c c c}        
\hline            
Survey & Band & Area (deg$^{2}$) & \head{2cm}{Number of AGN} & \head{2cm}{AGN Surface Density (deg$^{-2}$)} & Reference(s) \\    
\hline                        
& & & & & \\[-8pt]

   EWS (detected) & optical -- NIR & 14\,500 & 4.0\,$\times$\,10$^{7}$ & 2800 & This work \\
   EWS (selected) & optical -- NIR & 14\,500 & 8.1\,$\times$\,10$^{6}$ & 556 & \textquotedbl \\
   EDS (detected) & optical -- NIR & 50 & 2.4\,$\times$\,10$^{5}$ & 4700 & \textquotedbl \\
   EDS (selected) & optical -- NIR & 50 & 3.5\,$\times$\,10$^{4}$ & 692 & \textquotedbl \\
   \hline
& & & & & \\[-8pt]
   
   XMM-SERVS & X-ray & 13.1 & 10\,300 & 786 & \citet{chen2018,qingling2021} \\
   XXL-3XLSS & X-ray & 50 & 26\,056 & 521 & \citet{chiappetti2018} \\
   eFEDS & X-ray & 142 & 22\,079 & 155 & \citet{liu2022,salvato2022} \\
   SDSS & optical & 9376 & 750\,414 & 80\tnote{a} & \citet{lyke2020} \\
   DES & optical -- NIR & 4913 & 945\,860 & 193 & \citet{yang2023} \\
   \textit{Spitzer} cryogenic\tnote{b} & NIR -- MIR & 55 & $\sim 20\,000$ & 364 & \citet{lacy2020} \\
   AllWISE (R90) & MIR & 30\,093 & 4.54\,$\times$\,10$^{6}$ & 151 & \citet{assef2018} \\
   AllWISE (C75) & MIR & 30\,093 & 2.09\,$\times$\,10$^{7}$ & 695 & \citet{assef2018} \\

\hline

\end{tabular}
\begin{tablenotes}
   \item[a] Lower limit as the sample includes only spectroscopically confirmed unobscured quasars
   \item[b] Collation of AGN identified in \Spitzer cryogenic surveys: The Spitzer Wide-Area Infrared Extragalactic Survey \citep[SWIRE;][]{lonsdale2003}, The AGN and Galaxy Evolution Survey \citep[AGES;][]{kochanek2012}, The Spitzer First Look survey \citep[][]{lacy2005,fadda2006}, and the Spitzer COSMOS survey \citep[S-COSMOS;][]{sanders2007}
\end{tablenotes}
\end{threeparttable}
\end{table*}

\begin{figure*}
\centering
\includegraphics[width=\textwidth]{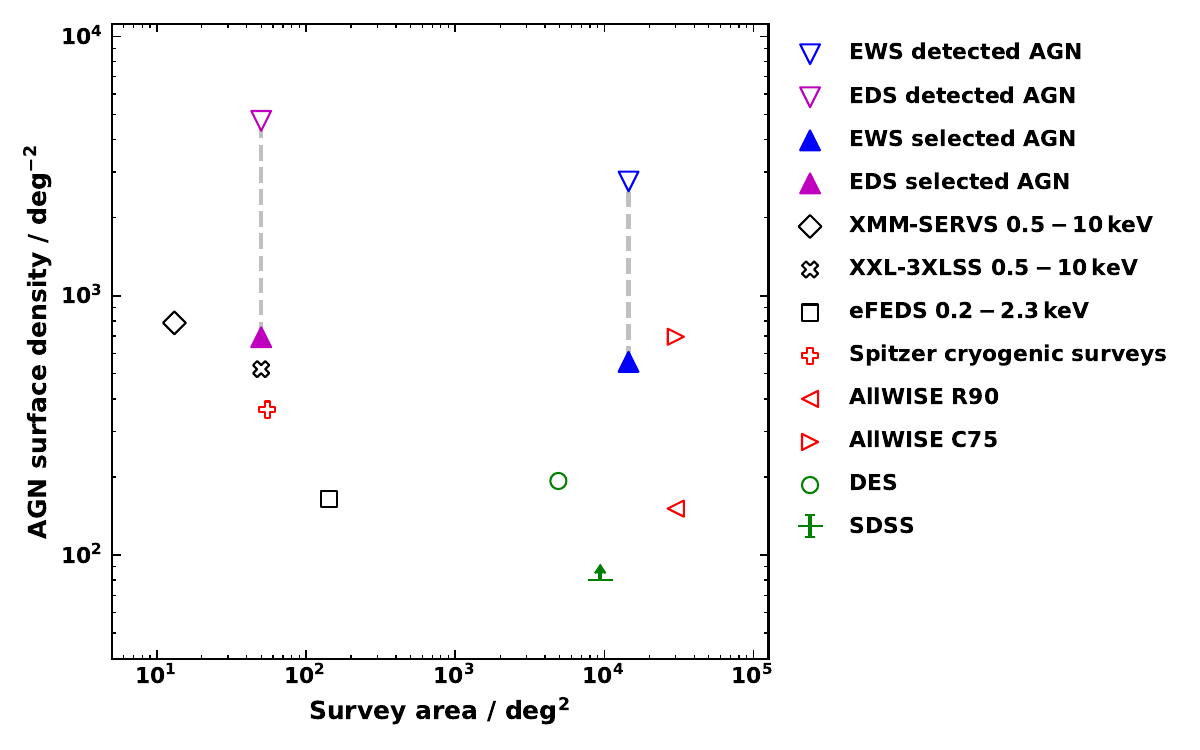}
  \caption{AGN surface density versus survey area comparison for a number of wide field and medium area surveys in different wavebands. Included AGN surface densities are: EWS detectable AGN (blue downwards triangle, unfilled), EWS selected AGN (blue upwards triangle, filled), EDS detectable AGN (magenta downwards triangle, unfilled), EDS selected AGN (magenta upwards triangle, filled), DES (green circle, unfilled), SDSS spectroscopic lower limit (green lower limit), \Spitzer combined cryogenic surveys (red plus, unfilled), AllWISE R90 selection (red left-facing triangle, unfilled), AllWISE C75 selection (red right-facing triangle, unfilled), eFEDS (black square, unfilled), XXL-3XLSS (black cross, unfilled), and XMM-SERVS (black diamond, unfilled). The considered \Euclid detectable AGN have $\geq5\sigma$ detection in at least one \Euclid band. The plotted values and references for each survey are given in Table \ref{tab:surveycomp}.}
     \label{fig:survey_comparison}
\end{figure*}

We compared the surface density of AGN in a mixture of wide field and medium area surveys described in Table \ref{tab:surveycomp} to those predicted for \Euclid in this work in Fig. \ref{fig:survey_comparison}. In terms of detectable AGN, \Euclid will detect far more AGN in the EWS and EDS than are identified in other surveys that cover similar areas. Of course in this case we are comparing detectable AGN in the \Euclid photometric surveys to AGN that have been selected through various means in different surveys. 

Considering AGN that are selected with \Euclid-only photometric criteria (see Sect. \ref{sec:euclidselection}), we expect AGN surface densities that are on par with surveys of similar area.
In the EDS AGN selected with \Euclid-only unobscured AGN criteria yields an AGN surface density of \SDSelectedEuclidOnlyTypeOneEDS\,deg$^{-2}$, marginally lower than that of the collated cryogenic \textit{Spitzer} surveys and the XXL-3XLSS survey. The XXL-3XLSS is the deepest X-ray survey we have compared with in our analysis and is likely to recover AGN with colours that cannot be selected using our photometric criteria as some may appear indistinguishable from non-active galaxies in the optical--NIR.
The AGN surface density of \SDSelectedEuclidOnlyTypeOneEWS\,deg$^{-2}$ derived for the EWS using \Euclid-only unobscured AGN colour selection is roughly double the AGN surface density found for DES and AllWISE, using their 90\% reliability (R90) criterion. Working with \Euclid photometry alone then, we predict that the surface density of selected AGN in the EWS will be greater than ground-based optical and space-based MIR peers.   

Considering the total selected AGN sample (union of all criteria discussed in Sect. \ref{sec:euclidselection}), we forecast the EDS will yield a greater AGN surface density than the XMM-3XLSS survey. The total sample surface density is marginally below that of XMM-SERVS, which marks the highest AGN surface density from a survey considered in this analysis. 
The total selected AGN surface density for the EWS will be greater than optical surveys and comparable to the reported AGN surface density for the AllWISE 75\% completeness (C75) selection. Again, this emphasises that we expect \Euclid to perform to a similar degree as a space-based MIR counterpart in regard to AGN selection. 

Towards the end of \Euclid survey operations both the EWS and EDS are expected to yield surface densities of selected AGN that are similar or in excess of their wide and medium-field counterparts across a range of wavebands. If AGN selection in \Euclid is improved upon by combining different techniques and fine tuning the criteria considered here, \Euclid may well surpass the AGN selection performance of any prior surveys. This represents a promising outcome considering that \Euclid was not designed for the facilitation of AGN identification.

\subsubsection{Expected X-ray counterparts}
\label{sec:xraycounterparts}

\begin{figure}
\centering
\includegraphics[width=\hsize]{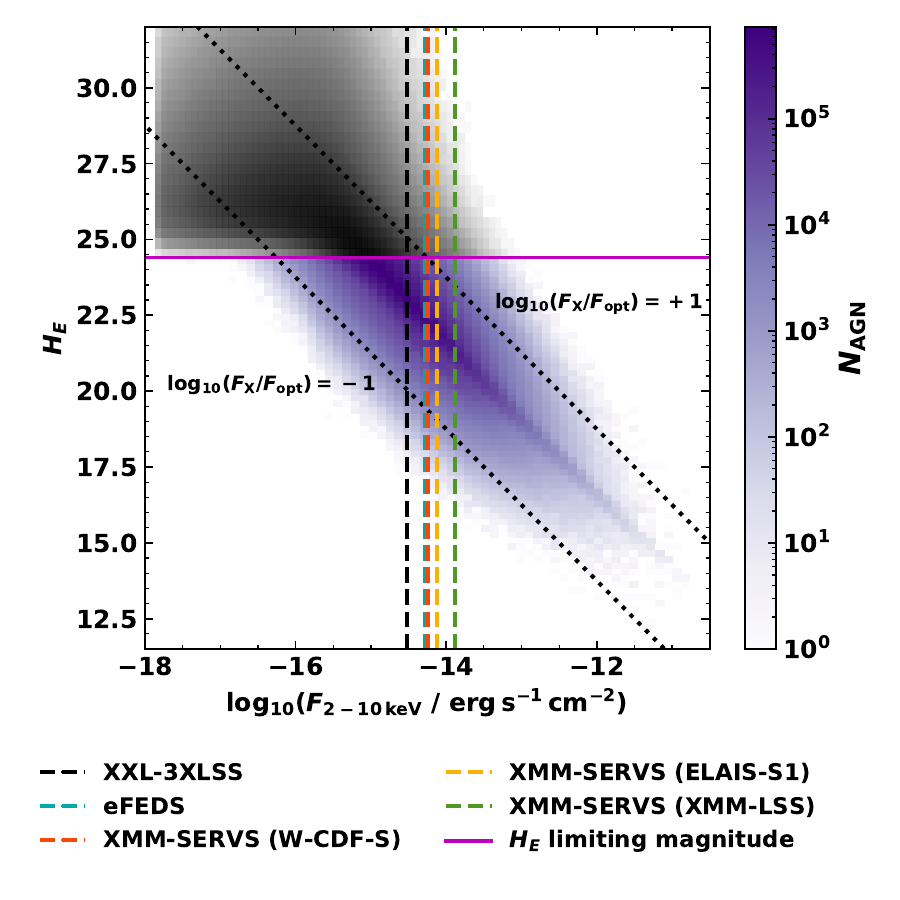}
  \caption{Density of EWS AGN in the 2--10\,keV X-ray flux vs \HE{} observed magnitude parameter space. The \Euclid detectable AGN are shown in purple, whilst undetectable AGN are represented in grey. The AGN population detectable with \Euclid photometry will, in part, reside beyond the X-ray flux limits of current X-ray surveys. There is also a portion of parameter space where a population of AGN detectable in modern X-ray surveys remain undetectable in the \Euclid \HE{} band. X-ray flux limits converted to the 2--10\,keV band, where needed, of the XXL-3XLSS (black), eFEDS (red) and XMM-SERVS surveys are plotted for comparison. We plot the XMM-SERVS fields separately as they have differing depths; XMM-LSS (orange), W-CDF-S (cyan), ELAIS-S1 (green). The dotted black lines represent $\logten(F_{\rm X}/F_{\rm opt}) = \pm 1$. The majority of AGN should lie between these lines.}
     \label{fig:xrayflux_vs_euclidmag}
\end{figure}

We examined the range of 2--10\,keV X-ray fluxes probed by \Euclid detectable AGN in our samples. Figure \ref{fig:xrayflux_vs_euclidmag} shows the two-dimensional density distribution of our EWS AGN sample on a 2--10\,keV X-ray flux vs \Euclid \HE{} apparent magnitude plane.
\HE{} was selected for this visualisation as we found the greatest yield of \Euclid detectable AGN in this band.
The 2--10\,keV flux limits of the X-ray surveys considered in this section are plotted for reference: XXL-3XLSS (limiting flux), eFEDs (eROSITA final depth; limiting flux), and the three XMM-SERVS fields (90\% sky coverage): XMM-LSS, W-CDF-S, and ELAIS-S1. For surveys not observing directly in the 2--10\,keV band we converted their flux limits from the survey native X-ray band assuming an X-ray power law with photon index $\Gamma = 1.9$. The over-density of sources on an approximately linear trend in the figure is populated by our AGN SEDs and is a consequence of using the nominal X-ray and 2\,\micron{} bolometric corrections without considering their dispersion. 

AGN detectable in the \HE{} band can exhibit 2--10\,keV X-ray fluxes ranging from 8.8\,$\times$\,10$^{-12}$ to 2.5\,$\times$\,10$^{-17}$\,erg\,s$^{-1}$\,cm$^{-2}$. Therefore, the AGN population detectable with \Euclid photometry will, in part, reside beyond the X-ray flux limits of all the X-ray surveys considered here. Equally, there is also a portion of parameter space where a population of AGN detectable in modern X-ray surveys remain undetectable in the \Euclid \HE{} band. We therefore expect \Euclid to probe a different population of AGN to those selected purely from X-ray surveys, similarly to what is observed with MIR AGN selection \citep[e.g.,][]{eckart2010}. It is likely that AGN detected by \Euclid but not in X-ray surveys are either low-luminosity ($\logten [L_{\rm \text{2--10}\,keV} / {\rm erg}\, {\rm s}^{-1}] \sim 42\text{--}43$) AGN at intermediate ($z \sim 1$) to high redshifts ($z > 3$), or AGN that are absorbed in the X-rays. The smaller converse population observable in X-ray surveys but not with \Euclid are likely to be largely made up of AGN that are under-luminous compared to their host galaxies \citep[e.g.,][]{mendez2013}.

To the flux limit of the deepest X-ray survey considered here (XXL-3XLSS; $F_{\rm \text{2--10}\,keV}\ $= 3\,$\times$\,10$^{-15}$\,erg\,s$^{-1}$\,cm$^{-2}$), we assessed the number of possible X-ray counterparts for our \Euclid observable AGN. 
We forecast 1.8\,$\times$\,10$^{7}$ (5.9\,$\times$\,10$^{4}$) AGN, corresponding to 45\% (25\%) of the total detectable population in the EWS (EDS) will exhibit X-ray fluxes that could be detected in the XXL-3XLSS survey. Of these AGN, 6.3\,$\times$\,10$^{6}$ (2.1\,$\times$\,10$^{4}$) are unobscured and 1.3\,$\times$\,10$^{7}$ (3.8\,$\times$\,10$^{4}$) are obscured in the EWS (EDS).
This again highlights the difference in AGN population that will be probed by \Euclid compared to those selected in the X-ray regime as up to 55\% (75\%) of EWS (EDS) \Euclid-detected AGN would not be detectable in the deepest modern medium area X-ray surveys. The lower X-ray detection rate of the EDS sample stems from the fainter NIR magnitudes probed corresponding to lower luminosity AGN at the faintest X-ray fluxes.

\section{Summary}

In this work we made forecasts of the observational expectations for $z<7$ AGN in the EWS and EDS. 
Starting from the \citet{fotopoulou2016xlf} 5--10\,keV XLF we generated volume-limited samples of the statistically expected AGN in the \Euclid survey footprints (Sect. \ref{sec:samplegeneration}). Our samples cover the redshift range $0.01 \leq z \leq 7$ and bolometric luminosity range $43 \leq \logten (L_{\rm bol} / {\rm erg}\, {\rm s}^{-1}) \leq 47$, corresponding to $41.8 \leq \logten (L_{\rm \text{2--10}\,keV} / {\rm erg}\, {\rm s}^{-1}) \leq 46.3$, or $-29.0 \leq M_{1450} \leq -17.2$.

Each AGN in our sample was assigned an SED based on its X-ray luminosity and redshift (Sect. \ref{sec:sed_models}). 
As the observed 5--10\,keV XLF considers obscured and unobscured AGN in an unbiased fashion up to $N_{\rm H} \sim$ 10$^{23}$\,cm$^{-2}$, we employed the optically obscured AGN fraction evolution model of \citet{merloni2014} to assign each AGN as optically obscured or unobscured. 
Unobscured AGN were assigned the mean quasar SED collated in \citet{shen2020} with $\alpha_{\rm ox}$ values sampled from the empirical distribution determined in \citet{lusso2010}. 
For obscured AGN we leveraged XXL AGN SED fitting results \citep{fotopoulou2016} to probabilistically allocate an empirical SED template class based on X-ray luminosity and redshift (Sect. \ref{sec:type2agnSED}). Finally, we applied dust extinction to each AGN SED, sampling from obscured and unobscured AGN \ebv{} distributions derived from XXL survey AGN \citep{fotopoulou2016}, as well as IGM extinction \citep[][Sect. \ref{sec:xrayopt_extinction}]{madau1995}.  
Through the probabilistic assignment of optical obscuration class, \ebv{} values, $\alpha_{\rm ox}$ in unobscured AGN, and SED template class in obscured AGN, we ensured empirically driven diversity in the resultant photometry of our AGN, even at similar luminosities and redshifts. 

Once assigned and scaled we performed mock observations of each AGN SED in our sample, convolving with the \Euclid bands and an assortment of ancillary photometric bands (Sect. \ref{sec:detectabilityassessment}). We utilized the resulting photometric catalog to investigate the observable population of $z < 7$ AGN in the \Euclid surveys.


Our main findings are summarised as follows.
   \begin{enumerate}
      \item We estimated \NUMDetectableOneFiltEWS AGN will have a $\geq 5\sigma$ detection in at least one \Euclid band in the EWS. Of these AGN 31\% were unobscured and 69\% were obscured. Our predicted yield corresponds to a detectable AGN surface density in the EWS of \SDDetectableOneFiltEWS\,deg$^{-2}$. In the EDS we expect a $\geq 5\sigma$ detection in at least one \Euclid band for \NUMDetectableOneFiltEDS AGN, of which 21\% are unobscured AGN and 79\% are obscured AGN. A detectable AGN surface density of \SDDetectableOneFiltEDS\,deg$^{-2}$ was found for the EDS. Full four-band $\geq 5\sigma$ \Euclid photometry coverage will be available for \NUMDetectableAllFiltEWS AGN in the EWS and \NUMDetectableAllFiltEDS AGN in the EDS. This is equivalent to AGN surface densities of \SDDetectableAllFiltEWS\,deg$^{-2}$ and \SDDetectableAllFiltEDS\,deg$^{-2}$ for the EWS and EDS, respectively.\\

      \item The dominant source of uncertainty in our analysis is from uncertainties in the XLF. We quantified the relative uncertainty on our numbers of \Euclid-detectable AGN to correspond to 6.7\% for the EWS and 12.5\% for the EDS. The disparity in relative uncertainties for the two surveys is due to the EWS probing the more tightly constrained bright end of the XLF, while the EDS is able to probe the lesser constrained faint end of the XLF. \\

      \item Using the \Euclid colour selection criteria derived in Bisigello et al. (in prep.), we obtained expectations on the number of AGN we will \textit{select} with \Euclid data. Employing \Euclid bands only we selected \NUMSelectedEuclidOnlyTypeOneEWS (\SDSelectedEuclidOnlyTypeOneEWS\,deg$^{-2}$) AGN in the EWS, comprised of 92\% unobscured AGN and 8\% obscured AGN. In the EDS we selected a sample of \NUMSelectedEuclidOnlyTypeOneEDS (\SDSelectedEuclidOnlyTypeOneEDS\,deg$^{-2}$) AGN, which consisted of 97\% unobscured AGN and 3\% obscured AGN. \\

      \item Ancillary $ugrz$ photometric bands from Rubin/LSST improved the completeness, purity, and size of colour-selected AGN samples. Including these selection criteria we selected a total of $8.1\times10^{6}$ (556\,deg$^{-2}$) and $3.5\times10^{4}$ (692\,deg$^{-2}$) AGN in the EWS and EDS, respectively. The EWS total selected AGN sample consisted of 75\% unobscured AGN and 25\% obscured AGN, while the EDS total selected AGN sample consisted of 71\% unobscured AGN and 29\% obscured AGN. These samples represent a yield of 20\% and 15\% of the EWS and EDS samples of AGN with a $\geq 5\sigma$ detection in at least one \Euclid band. The total expected sample of colour-selected AGN across both \Euclid surveys contains 6.0\,$\times$\,10$^{6}$ (74\%) unobscured AGN and 2.1\,$\times$\,10$^{6}$ (26\%) obscured AGN, covering $0.02 \leq z \lesssim 5.2$ and $43 \leq \log_{10} (L_{\rm bol} / {\rm erg}\, {\rm s}^{-1}) \leq 47$. \\
      
      \item The predicted surface densities of \Euclid selected AGN are comparable to those derived from other modern wide-field and medium-area surveys, across a range of wavebands. Our EWS yield is most comparable to the \textit{WISE} C75 AGN selection, with a slightly lower surface density. Our EDS selected surface density is marginally greater than that of the XXL-3XLSS survey, which is of similar area. \\ 
      
      \item We project that 1.8\,$\times$\,10$^{7}$ (5.9\,$\times$\,10$^{4}$) AGN, corresponding to 45\% (25\%) of the total \Euclid detectable population in the EWS (EDS) will exhibit X-ray fluxes that could be detected in the XXL-3XLSS survey. Therefore, we assess that up to 55\% (75\%) of EWS (EDS) \Euclid-detected AGN would not be detectable in the deepest modern medium area X-ray surveys.
      
   \end{enumerate}
   
We expect \Euclid to yield a sizeable statistical sample of AGN, in the order of tens of millions \textit{detected} AGN and millions of \textit{selected} AGN. The deep optical to NIR magnitudes, high spatial resolution and large areas interrogated in the \Euclid surveys will facilitate a diverse range of scientific studies with AGN. 

\begin{acknowledgements}
We thank J.~T.~Schindler and R.~Gilli for their helpful comments and discussion on the manuscript.
\AckEC
This work is supported by the UKRI AIMLAC CDT, funded by grant EP/S023992/1. 
V.~A. acknowledges support from INAF-PRIN 1.05.01.85.08. 
A.~F. acknowledges the support from project "VLT-MOONS" CRAM 1.05.03.07, INAF Large Grant 2022 "The metal circle: a new sharp view of the baryon cycle up to Cosmic Dawn with the latest generation IFU facilities" and INAF Large Grant 2022 "Dual and binary SMBH in the multi-messenger era".
F.~R. and F.~L.~F. acknowledge support from PRIN 2017 "Black hole winds and the baryon life cycle of galaxies: the stone-guest at the galaxy evolution supper" contract \#2017PH3WAT.
\end{acknowledgements}

%
%

\bibliography{references} 

\begin{appendix}

\section{Obscured AGN SED class distributions}\label{app:sedclassdists}
\begin{figure}
\centering
\includegraphics[width=\hsize]{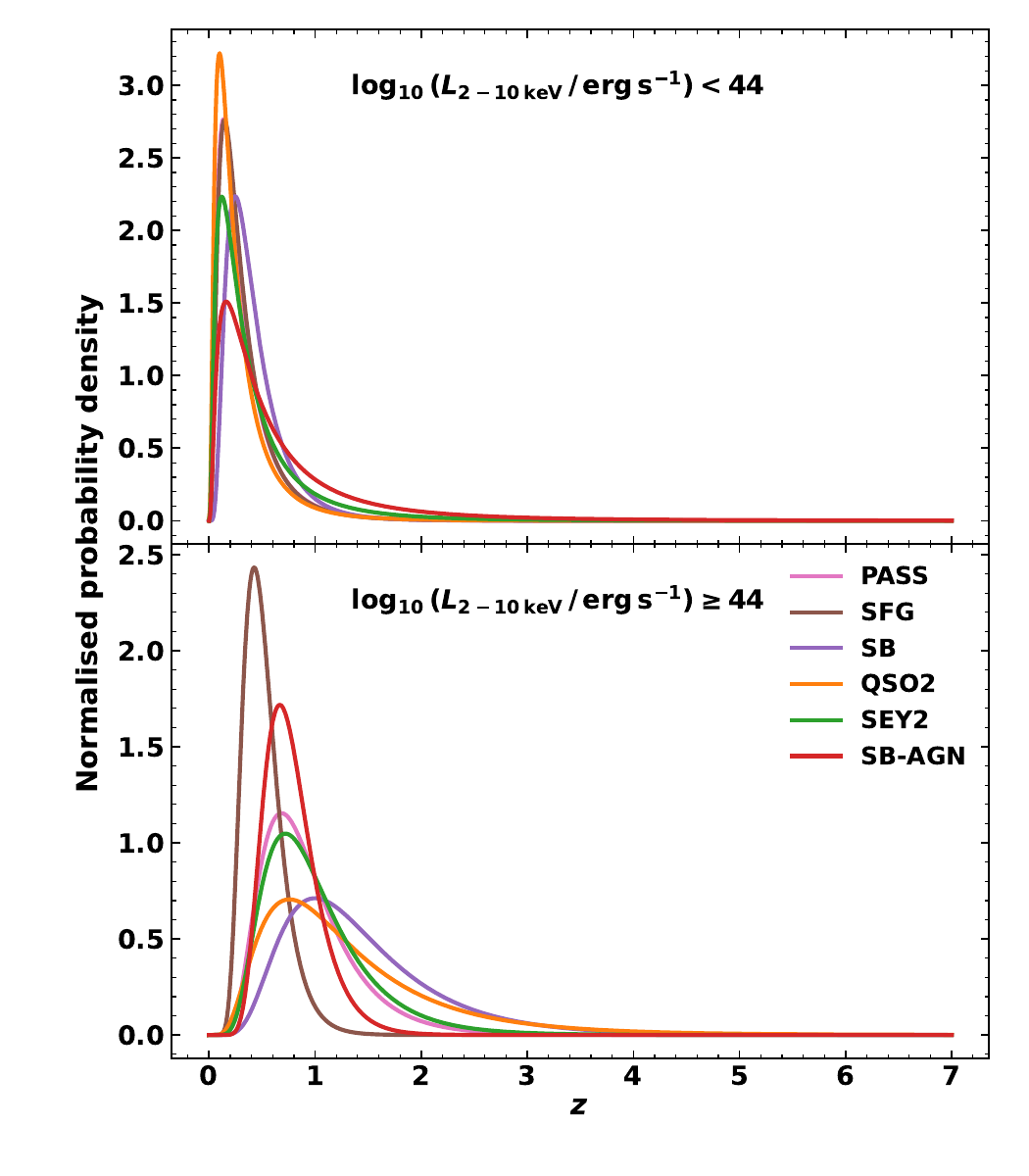}
  \caption{Normalised and extrapolated (to $z = 7$) redshift-space probability distributions for each obscured AGN SED class assigned in this work. Distributions are presented separately for the low (top) and high (bottom) X-ray luminosity groups.}
     \label{fig:sedclassdists}
\end{figure}

In Fig. \ref{fig:sedclassdists} we present the normalised and extrapolated (to $z = 7$) redshift-space probability distribution for each obscured AGN SED class assigned in this work (Sect. \ref{sec:type2agnSED}). We separately present distributions for the high ($\logten [L_{\rm \text{2--10}\,keV} / {\rm erg}\, {\rm s}^{-1}] \geq 44$) and low ($\logten [L_{\rm \text{2--10}\,keV} / {\rm erg}\, {\rm s}^{-1}] < 44$) X-ray luminosity groups. 

\section{Ancillary photometry}
\label{app:photometry}
In addition to the \Euclid filter set presented in Table \ref{tab:euc_filters}, we also perform synthetic photometric observations with bands from complimentary surveys covering UV--MIR. The addition of these bands aid in our discussion and assessment of the observational expectations for AGN with \Euclid compared to existing and upcoming surveys. The characteristics of each of the ancillary bands utilized in our work are given in Table \ref{tab:filters}.

\begin{table}
\caption{Characteristics of ancillary filters included in our catalog, used for synthetic photometric observations of AGN in this work.}             
\label{tab:filters}      
\centering                          
\begin{tabular}{l c c c c}        
\hline                 
Survey & Filter & $\lambda_{\rm eff}$ (\micron) & Reference\\    
\hline                        
& & & \\[-8pt]

   2MASS & $J$ & 1.66 & \citet{skrutskie2006} \\
   2MASS & $H$ & 1.24 & \textquotedbl \\
   2MASS & $Ks$ & 2.15 & \textquotedbl \\
   DES & $g$ & 0.473 & \citet{morganson2018} \\
   DES & $r$ & 0.642 & \textquotedbl \\
   DES & $i$ & 0.784 & \textquotedbl \\
   DES & $z$ & 0.926 & \textquotedbl \\
   DES & $Y$ & 1.01 & \textquotedbl \\
   \textit{GALEX} & FUV & 0.154 & \citet{morrissey2007} \\
   \textit{GALEX} & NUV & 0.230 & \textquotedbl \\
   Rubin/LSST & $u$ & 0.368 & \citet{ivezic2019} \\
   Rubin/LSST & $g$ & 0.478 & \textquotedbl \\
   Rubin/LSST & $r$ & 0.622 & \textquotedbl \\
   Rubin/LSST & $i$ & 0.753 & \textquotedbl \\
   Rubin/LSST & $z$ & 0.869 & \textquotedbl \\
   Rubin/LSST & $Y$ & 0.973 & \textquotedbl \\
   \textit{Spitzer} & IRAC1 & 3.53 & \citet{fazio2004} \\
   \textit{Spitzer} & IRAC2 & 4.48 & \textquotedbl \\
   \textit{Spitzer} & IRAC3 & 5.70 & \textquotedbl \\
   \textit{Spitzer} & IRAC4 & 7.80 & \textquotedbl \\
   VISTA & $J$ & 1.25 & \citet{sutherland2015} \\
   VISTA & $H$ & 1.65 & \textquotedbl \\
   VISTA & $Ks$ & 2.15 & \textquotedbl \\
   \textit{WISE} & $W1$ & 3.35 & \citet{wright2010} \\
   \textit{WISE} & $W2$ & 4.60 & \textquotedbl \\
   \textit{WISE} & $W3$ & 11.6 & \textquotedbl \\
   \textit{WISE} & $W4$ & 22.1 & \textquotedbl \\
\hline                                   
\end{tabular}
\end{table}

\section{Derived AGN colours}
\label{app:colourvalidation}

We validate the colours of AGN generated in this work by showing that our derived colours are consistent with AGN colour-colour diagrams from the literature \citep[e.g.,][]{stern2005, mateos2012, fotopoulou&paltani2018}. In each case we test that our AGN occupy the expected positions on the diagrams using our EDS data where we require a detection with S/N > 5 in at least one \Euclid band.

\begin{figure}
\centering
\includegraphics[width=0.49\textwidth]{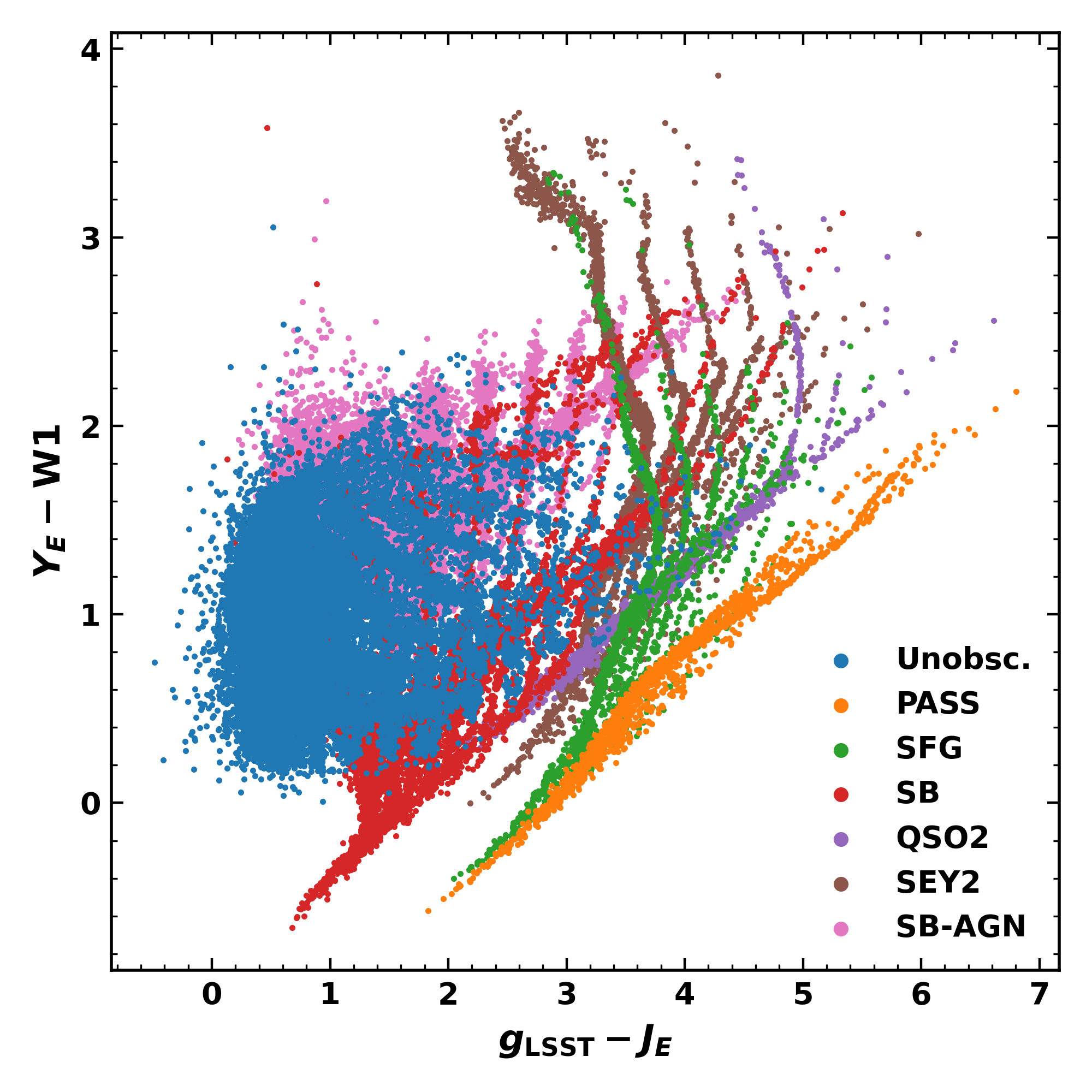}
  \caption{EDS AGN simulated in this work plotted on the optical-NIR-MIR colour space used in \citet{fotopoulou&paltani2018}. We used the \Euclid \YE{} and \JE{} bands, the Rubin/LSST $g$ band ($g\sfont{LSST}$), and the \textit{WISE} 3.4\,\micron{} (W1) band. Only AGN with S/N > 5 in at least one \Euclid band are plotted. Each of our AGN SED classes are plotted individually: Unobscured AGN (`Unobsc.'; blue), Passive (`PASS'; orange), Star-forming (`SFG'; green), Starburst (`SB'; red), High-luminosity obscured AGN (`QSO2'; purple), Seyfert 2 (`SEY2'; brown), and Starburst-AGN composite (`SB-AGN'; pink).}
     \label{fig:colourvalidation_logan2020}
\end{figure}

In Fig. \ref{fig:colourvalidation_logan2020} we plot our AGN on an adapted optical--NIR--MIR colour-colour space used in \citet{fotopoulou&paltani2018} and \citet{logan2020}. We substituted the \Euclid \YE{} and \JE{} bands in place of the VISTA $Y$ and $J$ bands, and the Rubin/LSST $g$ band ($g\sfont{LSST}$) in place of the SDSS $g$ band. This colour-colour space is known to separate well population loci of stars, galaxies and unobscured AGN. The unobscured AGN are expected to fall in a locus on the blue side of the diagram and galaxies (i.e., the obscured AGN classes in this work) in a more extended locus on the red side of unobscured AGN \citep[see Fig. 4 in][]{fotopoulou&paltani2018}. For clarity in the locations of different SED shapes, we plot each SED class with separate colours. We see that the different SED classes assigned to AGN in this work fall into the expected positions in the colour space. The unobscured AGN locus is contaminated by the SB-AGN and SB SED classes, specifically AGN with $z > 2$ for the latter. This is no surprise given the similarity of these SEDs at optical wavelengths (see Fig. \ref{fig:filter_sed_zevo}). We note that a portion of the unobscured AGN stray into the galaxy locus of the diagram. Akin to what is reported in \citet{logan2020}, these interlopers are affected by extinction, either from intrinsic dust absorption with $E(B-V) \gtrsim 0.2$ or from IGM absorption affecting $g\sfont{LSST}$ at $z \gtrsim 3.4$.

\begin{figure}
\centering
\includegraphics[width=\hsize]{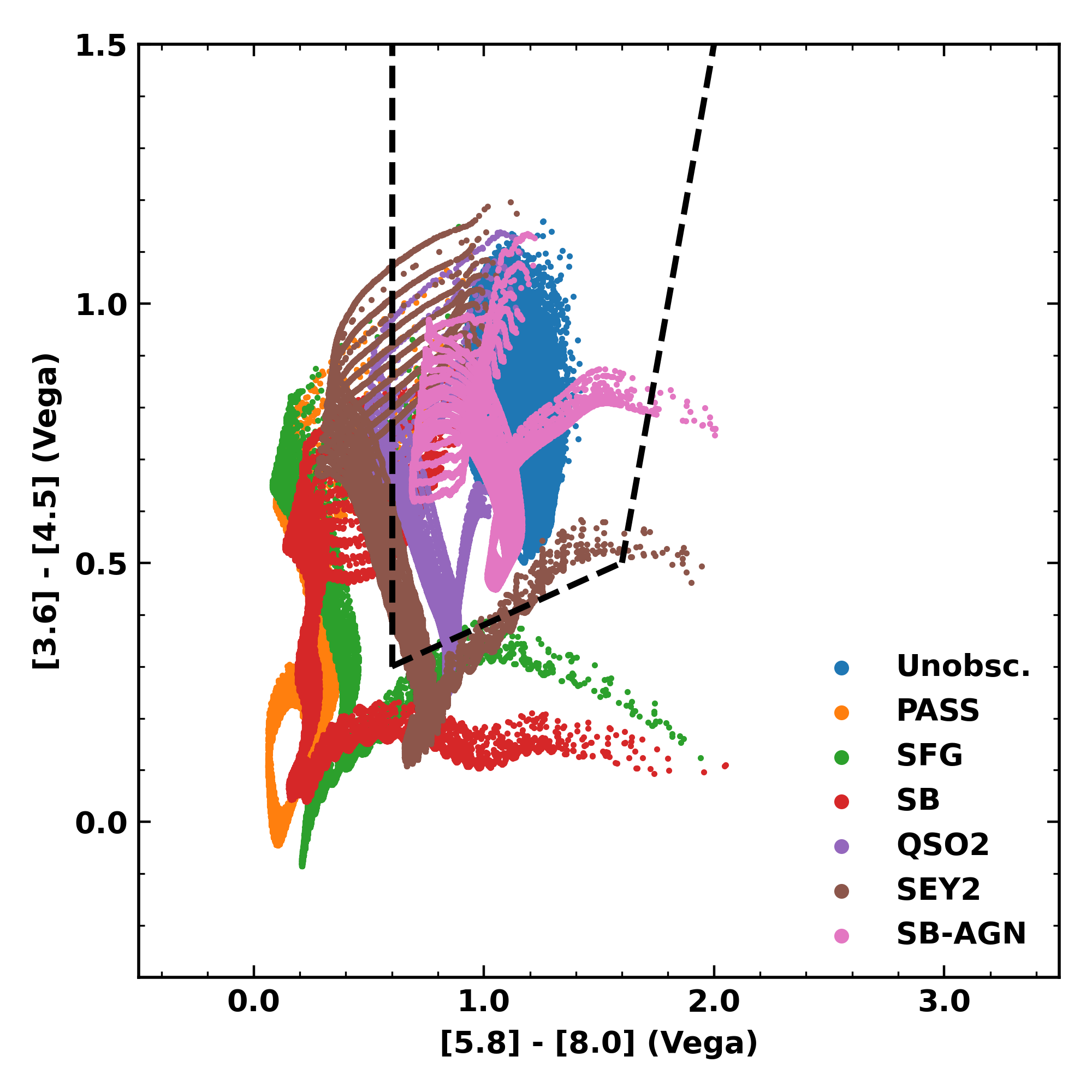}
  \caption{\textit{Spitzer}/IRAC AGN selection of \citet{stern2005} applied to our EDS AGN. We plot only AGN with S/N > 5 in at least one \Euclid band with $z < 4$. Each of our AGN SED classes are plotted individually with corresponding colours identical to Fig. \ref{fig:colourvalidation_logan2020}.}
     \label{fig:colourvalidation_stern2005}
\end{figure}

We plot the \textit{Spitzer}/IRAC [3.6]-[4.5] and [5.8]-[8.0] Vega system colours of our EDS AGN sample with $z < 4$ in Fig. \ref{fig:colourvalidation_stern2005}. The black dashed lines in this plot denote the AGN selection criterion identified by \citet{stern2005}, valid for redshifts $z < 4$. We found that 58.2\% of our total available AGN were identified using this selection. By class we select 100\% of the unobscured AGN and 47.5\% of the obscured AGN. This is expected as we see that the majority of non-selected obscured AGN belong to SED classes where the representative SED does not include a strong AGN component at these wavelengths (i.e., PASS, SFG, SB). Furthermore, it is reassuring that 100\% of unobscured AGN assigned with our mean quasar SED are selected using this method. Such AGN identification methods should in theory be well optimised towards including the `average' quasar in the resulting selected sample. 

\begin{figure}
\centering
\includegraphics[width=\hsize]{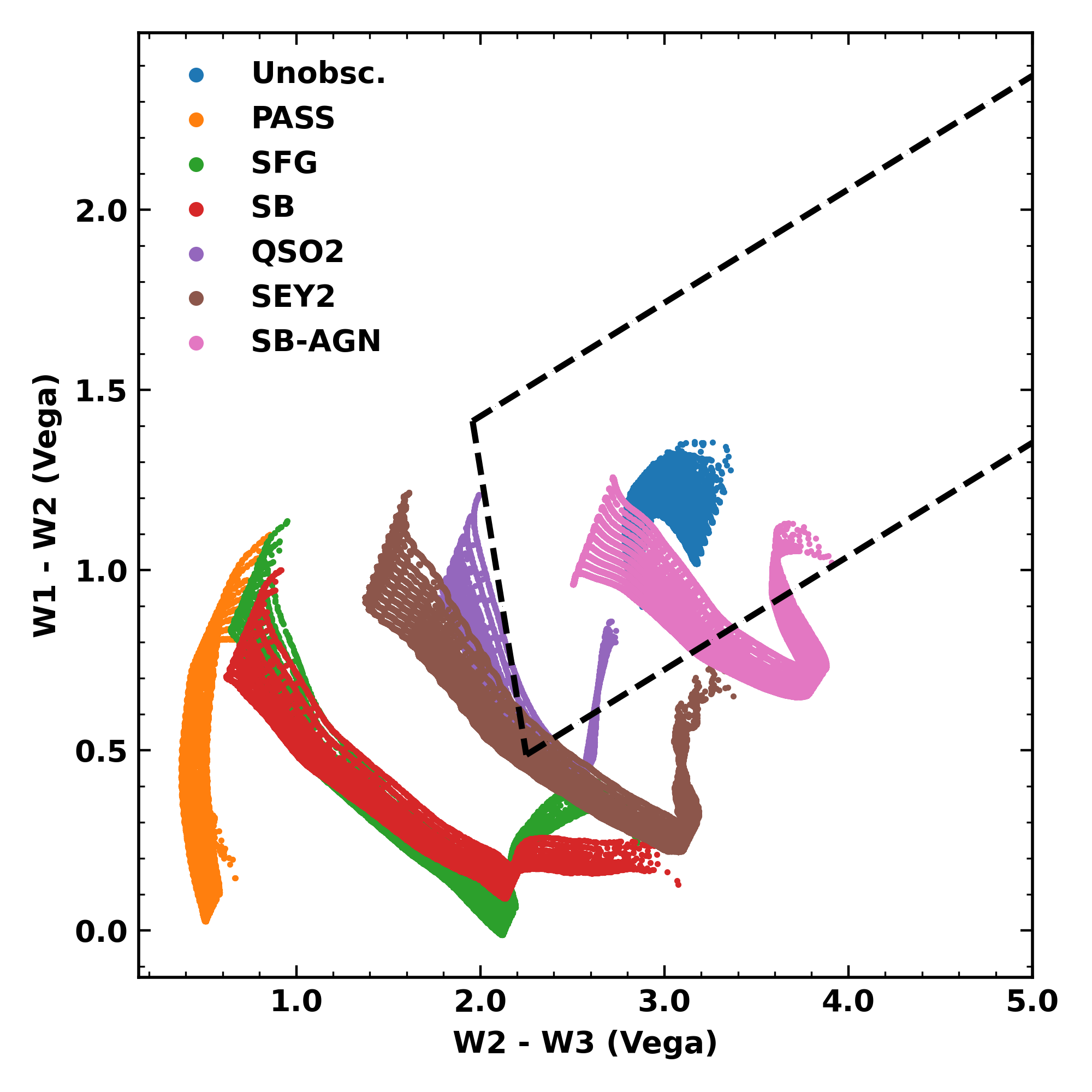}
  \caption{\textit{WISE} three-colour selection of \citet{mateos2012} applied to our EDS AGN. This selection uses the \textit{WISE} 3.4, 4.6, and 12\,\micron{} bands (W1, W2, and W3 respectively). We plot only AGN with S/N > 5 in at least one \Euclid band and with $z < 2$. Each of our AGN SED classes are plotted individually with corresponding colours identical to Fig. \ref{fig:colourvalidation_logan2020}.}
     \label{fig:colourvalidation_mateos2012}
\end{figure}

The three-band AGN colour selection criterion of \citet{mateos2012} utilizes the \textit{WISE} 3.4, 4.6 and 12\,\micron{} bands (W1, W2, and W3, respectively). In Fig. \ref{fig:colourvalidation_mateos2012} we show the W1-W2 and W2-W3 Vega system colours along with the \citet{mateos2012} three-band AGN colour selection criterion in black dashed lines. Colours are plotted for our EDS AGN sample with $z < 2$, the redshift range for which this selection is valid. We found 37.1\% of the total available AGN are identified in this case. Again, 100\% of the unobscured AGN are selected, an outcome validating our mean quasar colours as discussed above. With this selection criterion however, only 22.9\% of our obscured AGN are identified. The low identification rate of obscured AGN in this case can be attributed to the MIR selection scheme being optimised for high X-ray luminosity AGN that have a power-law spectral shape in the MIR. The SED templates that are well identified have the corresponding MIR spectral shape. As we explore in Sect. \ref{sec:surveycomparison}, our \Euclid detected AGN probe to faint X-ray fluxes which will lead to incompleteness in this selection. The overall positions of our different SED classes in the colour space agree well with the redshift evolution tracks for corresponding populations presented throughout \citet{mateos2012}.

We conclude that the colours derived for AGN in this work are consistent with observations.


\section{High-redshift predictions}\label{app:barnett}

The predicted yield of $7 \leq z \leq 9$ quasars in the EWS was explored in \citet{barnett2019}. Quasars were incorporated in their work utilizing the \citet{jiang2016} high-redshift quasar LF with two different assumed rates of decline (modest and steep) in space density at $z\geq6$. The decline in quasar space density is parameterised as $\phi \propto 10^{k(z-6)}$, where $\phi$ is the quasar LF and $k$ takes the values $k=-0.72$ or $k=-0.92$ for the modest and steep rate of decline, respectively. Contaminant populations such as M-type stars, L and T-type dwarfs and compact early-type galaxies were additionally modelled \citep{hewett2006}. 
Quasar selection functions were integrated over the sample of quasars and contaminants to determine the predicted yield of quasars with $7 \leq z \leq 9$ in the EWS. Over the full considered redshift range quasars were successfully selected to the effective depth $\JE \sim 22$ when using only \Euclid bands. 
Selection with a modest\,(steep) quasar LF decline predicted 87\,(51) quasars in the redshift range $7\leq z \leq 7.5$. This corresponds to a quasar surface density of 6.0\,$\times$\,10$^{-3}$\,deg$^{-2}$\,(3.5\,$\times$\,10$^{-3}$\,deg$^{-2}$). 

For comparison with the predicted EWS quasar yields of \citet{barnett2019}, we extended our simulation to $z = 7.5$. To approximate the available quasar candidates, we imposed conditions on our EWS sample for a detection with $\JE < 22$, an unobscured optical classification, and occupation of the redshift range $7.0\leq z \leq 7.5$. Applying these constraints we found 1053 unobscured AGN, equal to a surface density of 0.07\,deg$^{-2}$, a significantly higher yield.
We consider that we have made only a magnitude and redshift cut for our estimates, therefore recovering the detectable quasars in the target parameter space, but not incorporating the complex selection function constructed in \citet{barnett2019}. The selection function is likely to reject a number of the unobscured AGN included in our approximated sample.

The AGN selected to define the \citet{jiang2016} LF fulfilled two colour cuts primarily so that contaminants were avoided in the final quasar sample. SDSS main survey quasars required no detection in the $ugr$ bands and obeyed
\begin{equation}
    i - z > 2.2,
\end{equation}
\noindent where $i$ and $z$ refer to the SDSS bands. These criteria select $i$-band dropout objects and separate quasars (and cool brown dwarfs) from the majority of stellar objects \citep{fan1999, strauss1999}. Final quasar candidates also satisfied the criterion
\begin{equation}
    z - J < 0.5 (i - z) + 0.5,
\end{equation}
\noindent where $J$ refers to the UKIRT Infrared Deep Sky Survey \citep[UKIDSS;][]{warren2007} band. This colour-space cut separates quasars from the cool brown dwarf population.
We applied these criteria to our $7.0\leq z \leq 7.5$ unobscured AGN sample with $\JE < 22$, substituting the the UKIDSS $J$ band for the \Euclid \JE{} band. With these additional constraints we recover 668 detectable quasars, which is a factor of eight higher than forecast in \citet{barnett2019}. 
 
High-redshift quasar yield predictions are especially sensitive to the assumed LF shape and redshift evolution \citep{tee2023}. The XLF employed in this work, when extrapolated, does not exhibit the steep and accelerating space density decline at $z \gtrsim 6$ seen in UV/optical and other X-ray determinations of the AGN LF \citep[e.g.,][see Fig. \ref{fig:lf_comparison}]{ueda2014,jiang2016,wang2019,matsuoka2023}. 
At $z > 6$ our simulation predicts an unobscured AGN space density excess of up to 1\,dex at $M_{1450} \sim -24$, when compared with empirical UV/optical quasar observations. This ultimately is the driver of the apparent excess of quasars we predict at $7\leq z \leq 7.5$ compared to the \citet{barnett2019} analysis.

Equally as impactful as the assumed LF are the integration range and selection parameters of high-redshift quasar yield predictions. \citet{schindler2023} make $z > 7$ EWS quasar predictions also utilizing the $k=-0.7$ \citet{jiang2016} quasar LF. Integrating to $M_{1450} \lesssim -22.4$ and using $\HE < 24$ as the effective survey depth, they predicted 809 detectable quasars in the EWS at $7 \leq z \leq 8$. This sample is of comparable size to our $7 \leq z \leq 7.5$ simulated detectable sample of 1053, which incorporates unobscured AGN to $M_{1450} \sim -22.6$, albeit over a larger redshift slice.

At present there are only 532 quasars confirmed at $z \geq 5.3$ and 275 quasars known at $z \geq 6$ \citep[][and references therein]{fan2023}. Our predictions suggest there are still many quasars at these redshifts to be uncovered, with \Euclid capable of detecting many of them. We found that $\sim 600$ obscured AGN will have a $\JE < 22$ detection in the redshift range $7.0\leq z \leq 7.5$. For the reasons discussed above this number is uncertain, however suggests that \Euclid may uncover a population of high-redshift obscured AGN that are half as numerous in the data as unobscured AGN.


\section{Comparison to \spr{} FIR predictions}\label{app:FIRcomparison}

A complementary analysis to our work was carried out utilizing the \emph{s}pectro-\emph{p}hotometric \emph{r}ealisations of \emph{i}nfrared-selected \emph{t}argets at all-\emph{z} \citep[\spr{} {\sc{v\,1.13}};][]{bisigello2021,bisigello2022} simulation. The simulation is based on observed galaxy and AGN IR LFs and agrees with a wide set of observables, including the total galaxy stellar mass function, LFs at several wavelengths and number counts (from radio to X-rays), and relation between the SFR and the stellar mass. Relevant to this work, \spr{} incorporates two populations of AGN. The first population contains systems that have their SEDs dominated by radiation produced from AGN activity. These objects are split into AGN1 (unobscured) and AGN2 (obscured) sub-populations based on their optical extinction. These AGN are incorporated in the simulation via an IR LF obtained with a simultaneous fit of the \textit{Herschel} FIR \citep{gruppioni2013} and UV observed LFs \citep{croom2009, mcgreer2013, ross2013, akiyama2018, schindler2019}, derived in \cite{bisigello2021}. The second population, composite systems, have energetics dominated by star-formation activity but have a sub-dominant AGN component. These objects are again split into two sub-populations: star-forming AGN with an intrinsically faint AGN contribution and starburst AGN that have a bright yet heavily obscured ($\logten [N_{\rm H} / {\rm cm}^{-2}] = \text{23.5--24.5}$) AGN component. Composite systems are included in the simulation based on the \textit{Herschel} FIR observed LFs of \citet{gruppioni2013} with some updates reported in Bisigello et al. (in prep.). 

The \spr{} simulated EWS and EDS catalogs include sources with signal-to-noise ratio (S/N) > 3 in at least one \Euclid band, considering the depths given in Table \ref{tab:euc_filters}. For the EWS the \spr{} simulated catalog predicts 2.1\,$\times$\,10$^{8}$ AGN in the redshift range $0.01 \leq z \leq 7$, encompassing 2.9\,$\times$\,10$^{7}$ AGN-dominated objects and 1.8\,$\times$\,10$^{8}$ composite sources. Over the same redshift range, the \spr{} EDS catalog yields 1.1\,$\times$\,10$^{6}$ AGN, comprised of 1.7\,$\times$\,10$^{5}$ AGN emission dominated objects and 9.2\,$\times$\,10$^{5}$ composite objects. 

When applying identical S/N criteria to our data we found 4.5\,$\times$\,10$^{7}$ AGN in the EWS and 2.6\,$\times$\,10$^{5}$ AGN in the EDS. We observe in both surveys that the total number of AGN predicted using \spr{} are around five times greater than those derived using our methodology. Considering separated populations, our XLF-based analysis predicts numbers of AGN that are slightly greater than those derived for AGN-dominated objects in the IR LF rooted \spr{} analysis.
One reason for this could be that our XLF-based method allows consistent analysis of the AGN-dominated systems and the more luminous star-forming AGN composite objects in the \spr{} simulation. The heavily obscured starburst AGN composite systems however, have not been considered in our analysis as they appear absorbed to Compton-thick in the X-ray sky. Indeed, the entire population of starburst AGN composites in \spr{} have $\logten (N_{\rm H} / {\rm cm}^{-2}) > 23$. From cosmic X-ray background considerations \citet{gilli2007} predict that Compton-thick ($\logten [N_{\rm H} / {\rm cm}^{-2}] \geq 24$) AGN are at least as numerous as Compton-thin AGN, and up to four times as abundant as unobscured AGN with $\logten (L_{\rm X} / {\rm erg}\, {\rm s}^{-1}) < 43.5$. More extreme estimates postulate that Compton-thick AGN could be up to ten times more abundant than unobscured AGN \citep[][]{akylas2012}. Compton-thick AGN unobserved in the hard X-rays could therefore explain much of the deficit between the two analyses.

The above argument is likely the case for some AGN in the \spr{} samples, however this is based on the assumption that the XLF and IR LFs probe the same underlying population of AGN with selection effects. There is strong evidence however that X-ray selected AGN and IR selected AGN probe separate, incomplete samples of AGN, albeit with some overlap between the two \citep[e.g.,][]{mendez2013,ji2022,lyu2022}. X-ray selection can identify AGN with lower luminosities and accretion rates when compared to samples of IR selected AGN. This allows X-ray selected samples to contain AGN where host galaxy light dominates the MIR \citep[e.g.,][]{mendez2013,chung2014,menzel2016}. Consistent analysis of the same fields have highlighted differences in the derived IR and X-ray LFs \citep{runburg2022}. It is therefore not surprising that different numbers of AGN are derived using LFs from different bands in these similar analyses because each offers a different, incomplete window into the total census of AGN. 

On top of these differences in observed AGN populations between the XLF and IR LFs we must also consider that substantial uncertainties are introduced during the redshift extrapolation of the \textit{Herschel} FIR LF. As we explored in Sect. \ref{sec:uncertainties}, the largest uncertainty introduced in our work stems from the poor high-redshift constraints on the XLF. The lack of observable AGN at the faint-end slope of the LF in the high-redshift regime is not a challenge unique to X-ray frequencies. Indeed, the \citet{gruppioni2013} \textit{Herschel} FIR LF utilizes relatively shallow FIR observations in its construction and is well constrained up to $z\approx3$. The \spr{} implementation of this LF extrapolates beyond $z=3$ with some modifications described in \citet{bisigello2021}. During extrapolation the break luminosity is held constant and the galaxy number density is decreased as $\phi^{*} \propto (1+z)^{k_{\phi}}$ where $k_{\phi}\in[-4,-1]$. The extrapolated populations generated with \spr{} are shown to be consistent with a range of LFs in alternate wavebands up to $z=6$. The XLFs predicted by \spr{} however are in excess of observed XLFs when considering only Compton-thin AGN \citep{bisigello2021}. The uncertainties associated with both the \spr{} extrapolations and observed LFs at larger redshifts are however considerable due to the lack of constraining observations, just as in this work. Therefore, as we showed with our XLF, the uncertainty on the derived number of AGN is extensive, particularly at high redshift. 

The main difference between these analyses is that the IR AGN LF probes an overlapping but differing sample of AGN to the XLF. The nature of IR AGN selection also allows more heavily obscured AGN to be considered that are Compton-thick in the X-ray regime and therefore are not probed by the XLF employed in this work. We assess that the numbers of detectable AGN presented in this work may be a conservative estimate due to the selection effects biased against heavily obscured and Compton-thick AGN in the X-ray regime.

\section{AGN colour selection criteria}
\label{app:colourcriteria}

Bisigello et al. (in prep.) derives a number of different photometric selections for AGN in \Euclid using \Euclid, Rubin/LSST and \Spitzer/IRAC bands. In this section we applied each criterion to our AGN sample and report the results. Throughout this section we use the nomenclature of {\tt AND}, {\tt OR} corresponding to the logical AND, OR operators.

\begin{figure*}
     \centering
     \begin{subfigure}[b]{0.49\textwidth}
         \centering
         \includegraphics[width=\textwidth]{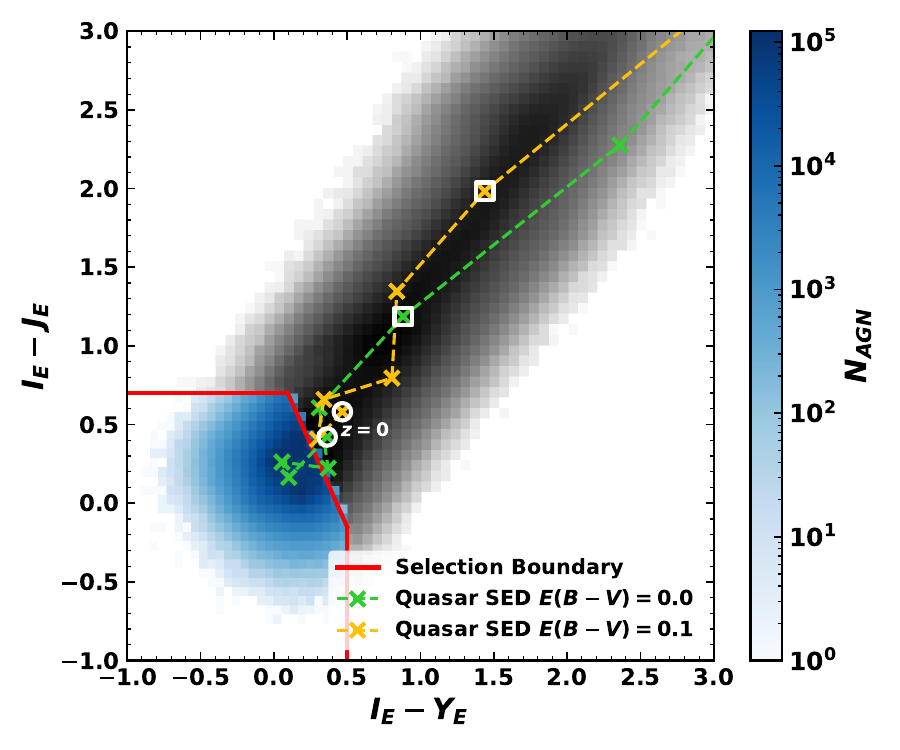}
         \caption{EWS unobscured AGN \Euclid Selection}
         \label{fig:ews_t1selection}
     \end{subfigure}
     \hfill
     \begin{subfigure}[b]{0.49\textwidth}
         \centering
         \includegraphics[width=\textwidth]{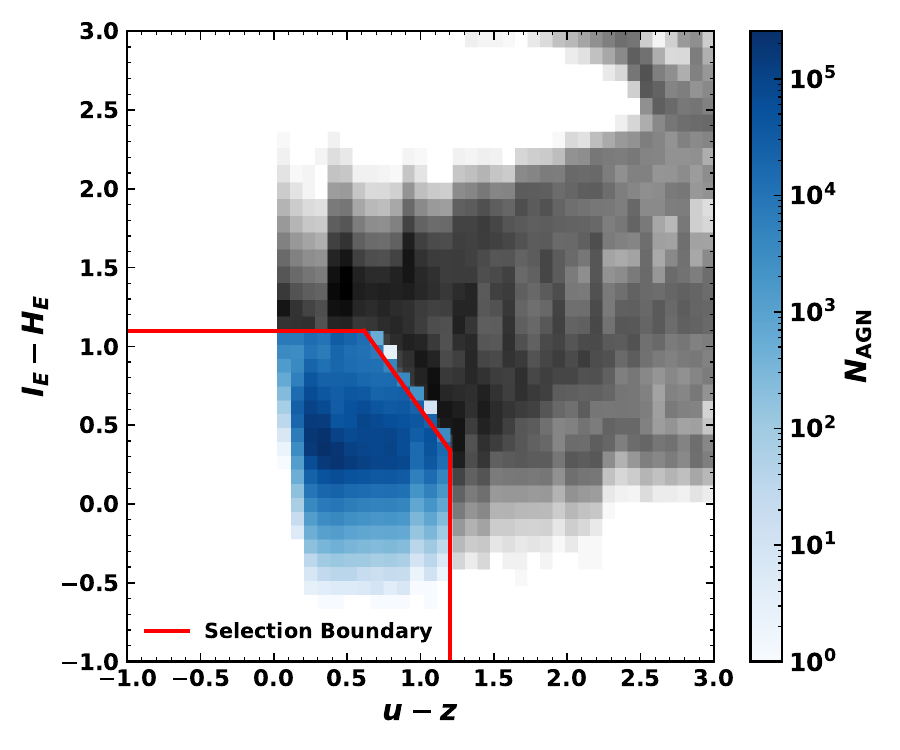}
         \caption{EWS unobscured AGN \Euclid + Rubin/LSST Selection}
         \label{fig:ews_t1selection_lsst}
     \end{subfigure}
     \vfill
     \centering
     \begin{subfigure}[b]{0.49\textwidth}
         \centering
         \includegraphics[width=\textwidth]{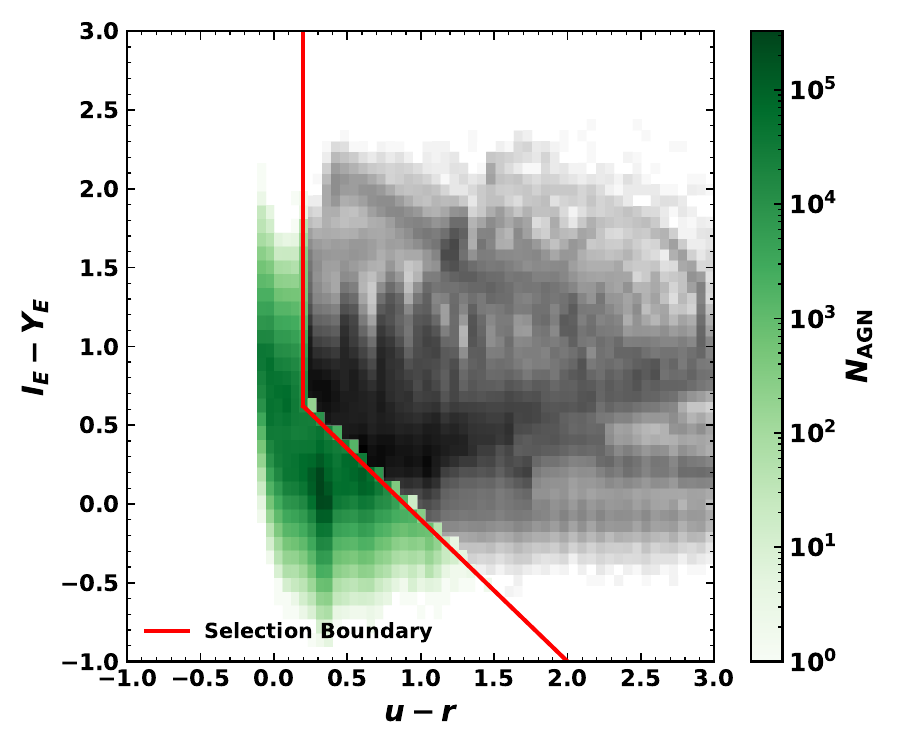}
         \caption{EWS all AGN \Euclid + Rubin/LSST Selection}
         \label{fig:ews_allagnselection}
     \end{subfigure}
        \caption{Optimal \Euclid photometry AGN selection criteria derived in Bisigello et al. (in prep.) for the EWS applied to our data. In each case we show two-dimensional density plots for AGN that are detected above the 5$\sigma$ point-source depths in all four relevant bands. The grey points show the distribution of AGN that are not selected, while the coloured regions show the selected AGN. The sub-figures correspond to optimal selection criteria for: (\textit{a}) Unobscured AGN with \Euclid photometry only, (\textit{b}) Unobscured AGN with \Euclid and Rubin/LSST photometry and (\textit{c}) All AGN with \Euclid and Rubin/LSST photometry. In panel (a) we plot the colour-space redshift evolution of our unobscured AGN template with $E(B-V) = 0$ (green) and $E(B-V) = 0.1$ (orange) for $z \in [0, 7]$ in steps of $\delta z = 1$. White circles show the $z=0$ points and white squares denote the $z=5$ points. The $E(B-V) = 0$ green crosses in the selection region of (a) correspond to $z = 1$ and $z = 2$.}
        \label{fig:ews_selection}
\end{figure*}

\begin{figure*}
     \centering
     \begin{subfigure}[b]{0.49\textwidth}
         \centering
         \includegraphics[width=\textwidth]{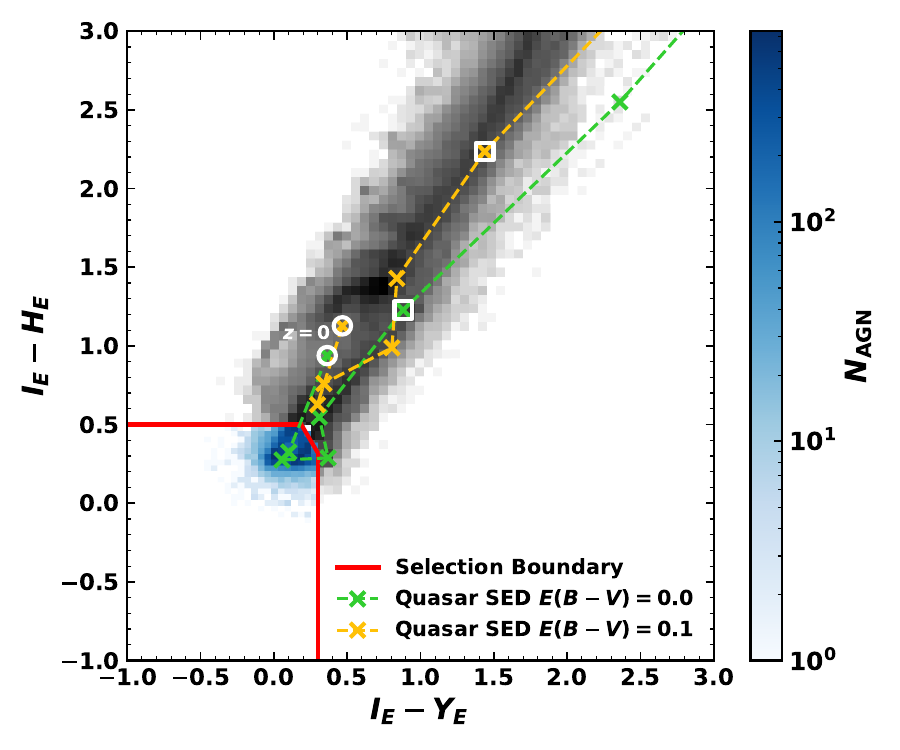}
         \caption{EDS unobscured AGN \Euclid Selection}
         \label{fig:eds_t1selection}
     \end{subfigure}
     \hfill
     \begin{subfigure}[b]{0.49\textwidth}
         \centering
         \includegraphics[width=\textwidth]{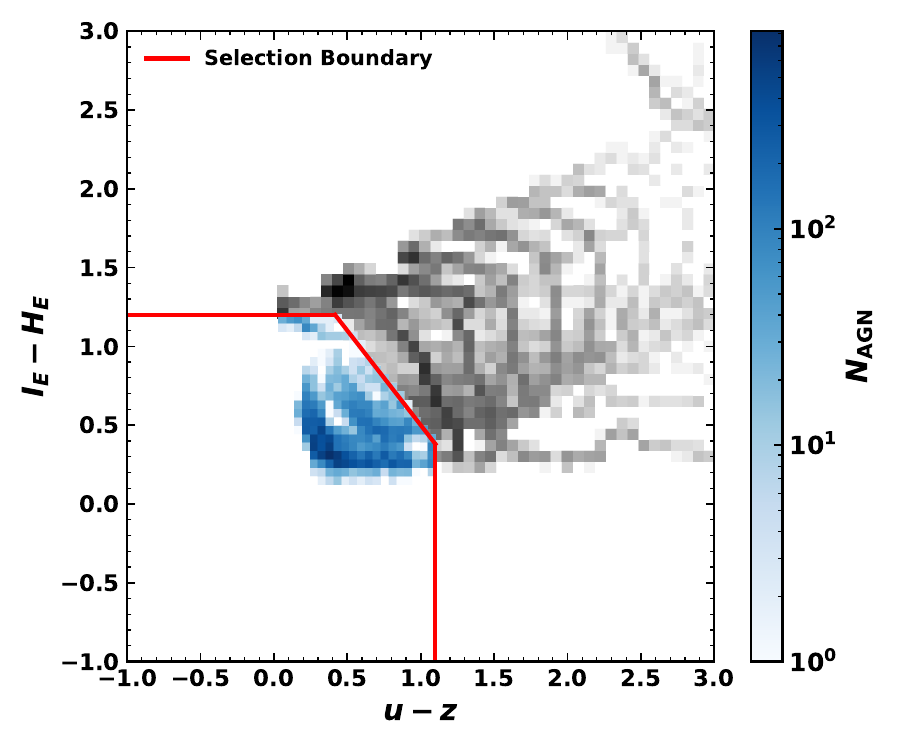}
         \caption{EDS unobscured AGN \Euclid + Rubin/LSST Selection}
         \label{fig:eds_t1selection_lsst}
     \end{subfigure}
     \vfill
     \begin{subfigure}[b]{0.49\textwidth}
         \centering
         \includegraphics[width=\textwidth]{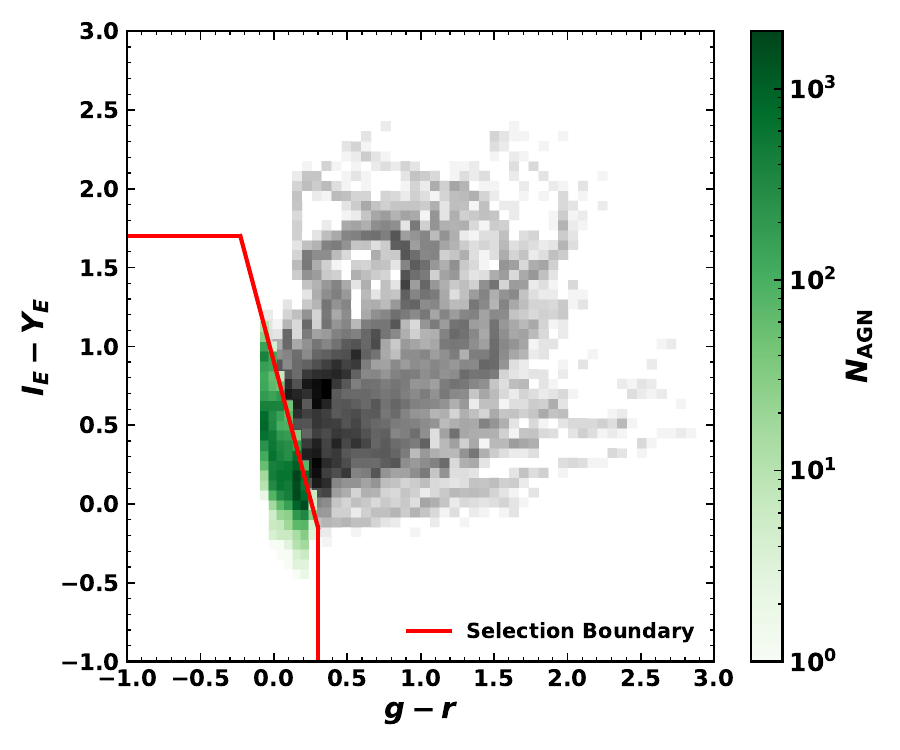}
         \caption{EDS all AGN \Euclid + Rubin/LSST Selection}
         \label{fig:eds_allagnselection_lsst}
     \end{subfigure}
        \caption{Same as Fig. \ref{fig:ews_selection} for the EDS.}
        \label{fig:eds_selection}
\end{figure*}

Density plots showing the \Euclid colour AGN selection criteria discussed in Sect. \ref{sec:euclidselection} applied to our samples of EWS and EDS AGN are shown in Figs. \ref{fig:ews_selection} and \ref{fig:eds_selection}, respectively. In Fig. \ref{fig:ews_t1selection} and Fig. \ref{fig:eds_t1selection} we additionally plot trails of the colour-space redshift evolution of our unobscured AGN template with $E(B - V) = 0$ and $E(B - V) = 0.1$ for $z\in[0, 7]$ in steps of $\delta z = 1$. These lines illustrate how the position of our AGN on the considered selection plots are influenced by SED and source parameters. We observe that much of the spread of our population in the top right portion of the diagram is occupied by high-redshift sources on the right and more heavily optically obscured sources on the left. Interestingly, we observe that the lowest redshift unobscured AGN and mildly dust reddened unobscured AGN would not be selected using these \Euclid photometric criteria.

\begin{table*}
\caption{Redshift-dependent performance of \Euclid photometry AGN selection criteria (Bisigello et al., in prep.) applied to our samples of EWS and EDS AGN. For the \Euclid-only unobscured AGN criterion and union ($\cup$) of all \Euclid and Rubin/LSST selection criteria we report the number of selected AGN, completeness ($C$) at the 5$\sigma$ level in the relevant filters, and the median bolometric luminosity ($\pm 1\sigma$) in each redshift bin. These data are visualised in Fig. \ref{fig:selectioninfo}.}             
\label{tab:z_selection}      
\centering                          
\begin{tabular}{l c c c c c c}        
\hline            
Survey & Target Class & Photometry & Redshift & Selected AGN & $C$ & $\log_{10} (L_{\rm bol} / {\rm erg\,s}^{-1})$ \\    
\hline                        
& & & & & & \\[-8pt]

EWS & Unobscured AGN & \Euclid & $0.0 \leq z < 0.5$ & $2.1\times10^{5}$ & 0.13 & $43.6 \pm 0.6$ \\
&  &  & $0.5 \leq z < 1.0$ & $7.3\times10^{5}$ & 0.13 & $44.4 \pm 0.7$ \\
&  &  & $1.0 \leq z < 1.5$ & $1.2\times10^{6}$ & 0.24 & $45.2 \pm 0.6$ \\
&  &  & $1.5 \leq z < 2.0$ & $1.2\times10^{6}$ & 0.37 & $45.4 \pm 0.5$ \\
&  &  & $2.0 \leq z < 2.5$ & $7.2\times10^{5}$ & 0.39 & $45.6 \pm 0.5$ \\
&  &  & $2.5 \leq z < 3.0$ & $3.9\times10^{5}$ & 0.34 & $45.8 \pm 0.4$ \\
&  &  & $3.0 \leq z < 3.5$ & $2.1\times10^{5}$ & 0.29 & $45.8 \pm 0.4$ \\
&  &  & $3.5 \leq z < 4.0$ & $9.9\times10^{4}$ & 0.22 & $45.9 \pm 0.4$ \\
&  &  & $4.0 \leq z < 4.5$ & $3.6\times10^{4}$ & 0.12 & $45.9 \pm 0.3$ \\
&  &  & $4.5 \leq z < 5.0$ & $2.0\times10^{3}$ & 0.01 & $45.8 \pm 0.2$ \\
&  &  & $5.0 \leq z < 5.5$ & 3 & 0.00 & $45.7 \pm 0.1$ \\
\hline
& & & & & & \\[-8pt]

EWS & $\cup$ & \Euclid, Rubin/LSST & $0.0 \leq z < 0.5$ & $4.3\times10^{5}$ & 0.26 & $43.6 \pm 0.6$ \\
&  &  & $0.5 \leq z < 1.0$ & $1.7\times10^{6}$ & 0.28 & $44.1 \pm 0.7$ \\
&  &  & $1.0 \leq z < 1.5$ & $2.1\times10^{6}$ & 0.38 & $44.9 \pm 0.6$ \\
&  &  & $1.5 \leq z < 2.0$ & $2.0\times10^{6}$ & 0.56 & $45.3 \pm 0.5$ \\
&  &  & $2.0 \leq z < 2.5$ & $1.1\times10^{6}$ & 0.53 & $45.5 \pm 0.5$ \\
&  &  & $2.5 \leq z < 3.0$ & $5.2\times10^{5}$ & 0.44 & $45.7 \pm 0.4$ \\
&  &  & $3.0 \leq z < 3.5$ & $2.1\times10^{5}$ & 0.29 & $45.8 \pm 0.4$ \\
&  &  & $3.5 \leq z < 4.0$ & $9.9\times10^{4}$ & 0.22 & $45.9 \pm 0.4$ \\
&  &  & $4.0 \leq z < 4.5$ & $3.6\times10^{4}$ & 0.12 & $45.9 \pm 0.3$ \\
&  &  & $4.5 \leq z < 5.0$ & $2.0\times10^{3}$ & 0.01 & $45.8 \pm 0.2$ \\
&  &  & $5.0 \leq z < 5.5$ & 3 & 0.00 & $45.7 \pm 0.1$ \\
\hline
& & & & & & \\[-8pt]

EDS & Unobscured AGN & \Euclid & $0.0 \leq z < 0.5$ & $3.0\times10^{2}$ & 0.04 & $43.5 \pm 0.5$ \\
&  &  & $0.5 \leq z < 1.0$ & $2.4\times10^{3}$ & 0.06 & $44.0 \pm 0.7$ \\
&  &  & $1.0 \leq z < 1.5$ & $5.5\times10^{3}$ & 0.13 & $44.7 \pm 0.7$ \\
&  &  & $1.5 \leq z < 2.0$ & $4.5\times10^{3}$ & 0.17 & $45.2 \pm 0.7$ \\
&  &  & $2.0 \leq z < 2.5$ & $2.5\times10^{3}$ & 0.17 & $45.4 \pm 0.6$ \\
&  &  & $2.5 \leq z < 3.0$ & $1.2\times10^{3}$ & 0.14 & $45.6 \pm 0.4$ \\
&  &  & $3.0 \leq z < 3.5$ & $6.5\times10^{2}$ & 0.13 & $45.6 \pm 0.4$ \\
&  &  & $3.5 \leq z < 4.0$ & $2.2\times10^{2}$ & 0.08 & $45.6 \pm 0.4$ \\
&  &  & $4.0 \leq z < 4.5$ & 78 & 0.04 & $45.9 \pm 0.3$ \\
\hline
& & & & & & \\[-8pt]

EDS & $\cup$ & \Euclid, Rubin/LSST & $0.0 \leq z < 0.5$ & $1.5\times10^{3}$ & 0.23 & $43.5 \pm 0.5$ \\
&  &  & $0.5 \leq z < 1.0$ & $6.2\times10^{3}$ & 0.15 & $43.9 \pm 0.7$ \\
&  &  & $1.0 \leq z < 1.5$ & $7.5\times10^{3}$ & 0.18 & $44.7 \pm 0.7$ \\
&  &  & $1.5 \leq z < 2.0$ & $9.4\times10^{3}$ & 0.35 & $45.0 \pm 0.6$ \\
&  &  & $2.0 \leq z < 2.5$ & $6.4\times10^{3}$ & 0.43 & $45.2 \pm 0.6$ \\
&  &  & $2.5 \leq z < 3.0$ & $2.6\times10^{3}$ & 0.30 & $45.4 \pm 0.5$ \\
&  &  & $3.0 \leq z < 3.5$ & $6.8\times10^{2}$ & 0.14 & $45.6 \pm 0.4$ \\
&  &  & $3.5 \leq z < 4.0$ & $2.2\times10^{2}$ & 0.08 & $45.6 \pm 0.4$ \\
&  &  & $4.0 \leq z < 4.5$ & 78 & 0.04 & $45.9 \pm 0.3$ \\
\hline

\end{tabular}
\end{table*}

In the EWS the optimal selection criteria for unobscured AGN exploiting only \Euclid photometry obeys 
\begin{equation}
  \begin{aligned}
    &\IE - \YE < 0.5\ \\
    &{\tt AND}\ \IE - \JE < 0.7\ \\
    &{\tt AND}\ \IE - \JE < -2.1\,(\IE - \YE) + 0.9,
  \end{aligned}
\end{equation}
\noindent and is expected to achieve F1\,$\sim$\,0.2 with $P=0.2$. Applied to our \Euclid detectable AGN we selected \NUMSelectedEuclidOnlyTypeOneEWS AGN, corresponding to a selected AGN surface density of \SDSelectedEuclidOnlyTypeOneEWS\,deg$^{-2}$. We select 92\% unobscured and 8\% obscured AGN SEDs with this selection. An overall completeness $C=\CSelectedEuclidOnlyTypeOneEWS$ results for our full candidate sample. Considering only unobscured AGN, the target class of the selection, the completeness rises to $C=\CTypeOneSelectedEuclidOnlyTypeOneEWS$.

The colour selection performance of unobscured AGN in the EWS, and indeed for all AGN colour selections in the EWS and EDS, is improved with the addition of Rubin/LSST optical bands. The optimal selection criteria for unobscured AGN in the EWS including Rubin/LSST is given by 
\begin{equation}
    \begin{aligned}
        &\IE - \HE < 1.2\ \\
        &{\tt AND}\ u - z < 1.1\ \\
        &{\tt AND}\ \IE - \HE < -1.3\,(u - z) + 1.9,
    \end{aligned}
\end{equation}
\noindent where $u$ and $z$ correspond to the Rubin/LSST optical filters. The criterion is expected to achieve F1 = 0.9 with $P=0.9$. Implemented on our sample this selection yields \NUMSelectedEuclidLSSTTypeOneEWS selected AGN, which corresponds to a surface density of \SDSelectedEuclidLSSTTypeOneEWS\,deg$^{-2}$. With this criterion our resulting selected sample contained 97\% unobscured SEDs and 3\% obscured AGN SEDs. We derived an overall completeness of $C=\CSelectedEuclidLSSTTypeOneEWS$, with a completeness $C=\CTypeOneSelectedEuclidLSSTTypeOneEWS$ when considering unobscured AGN only.

In the EWS the selection of all AGN, i.e. obscured, unobscured and composite systems utilizing \Euclid photometry only is disregarded due to the difficulty of separating AGN powered sources from contaminants with the limited optical and NIR filters available. This selection task is improved with the addition of optical Rubin/LSST bands. The optimal criterion with such external data achieves F1 = 0.3 with $P=0.2$ and is formulated as 
\begin{equation}
    \begin{aligned}
        &u - r < 0.2\ \\
        &{\tt OR}\ \IE - \YE < -0.9\,(u - r) + 0.8,
    \end{aligned}
\end{equation}
\noindent where $u$ and $r$ are the Rubin/LSST optical filters. We note that this selection is for objects that fall on the outside of the defined boundary in colour space (referred to as ``type-B'' in Bisigello et al., in-prep). Applying this colour selection to our data we selected a sample of \NUMSelectedEuclidLSSTAllEWS AGN, corresponding to a surface density of \SDSelectedEuclidLSSTAllEWS deg$^{-2}$. Our selected sample with this criterion comprised 72\% unobscured and 28\% obscured AGN SEDs. As raised in Bisigello et al. (in prep.), we note that this colour selection has a low sample purity and therefore is expected to select a large fraction of contaminant inactive galaxies. We found the overall completeness of this selected sample is $C=\CSelectedEuclidLSSTAllEWS$. For unobscured (obscured) AGN populations the completeness achieved is $C=\CTypeOneSelectedEuclidLSSTAllEWS\;(\CTypeTwoSelectedEuclidLSSTAllEWS)$. 

The optimal colour criteria for AGN selection presented in Bisigello et al. (in prep.) differ between the EWS and EDS. Different criteria are required given that the two surveys probe different parts of the AGN LF. For unobscured AGN in the EDS the best selection criteria using \Euclid photometry alone, with F1 = 0.2 and $P=0.2$, is given by 
\begin{equation}
    \begin{aligned}
        &\IE - \YE < 0.3\ \\
        &{\tt AND}\ \IE - \HE < 0.5\ \\
        &{\tt AND}\ \IE - \HE < -1.6\,(\IE - \YE) + 0.8.
    \end{aligned}
\end{equation}
This selection criterion yields \NUMSelectedEuclidOnlyTypeOneEDS AGN at a surface density of \SDSelectedEuclidOnlyTypeOneEDS deg$^{-2}$ when applied to our sample. Our selected sample contained 97\% unobscured SEDs and 3\% obscured AGN SEDs. The completeness derived for this selection is $C=\CSelectedEuclidOnlyTypeOneEDS$ for the overall population and $C=\CTypeOneSelectedEuclidOnlyTypeOneEDS$ considering the target class of unobscured AGN only. We note that this selection is expected to be particularly contaminated by dwarf irregular galaxies in practice. 

Adding Rubin/LSST bands raises the quality of unobscured AGN selection in the EDS to an expected F1 = 0.8 with $P=0.9$. The criterion in this case is 
\begin{equation}
    \begin{aligned}
        &\IE - \HE < 1.1\ \\
        &{\tt AND}\ u - z < 1.2\ \\
        &{\tt AND}\ \IE - \HE < -1.2\,(u - z) + 1.7,
    \end{aligned} 
\end{equation}
\noindent where $u$ and $z$ are the Rubin/LSST optical filters. Using this selection on our \Euclid detectable sample we selected \NUMSelectedEuclidLSSTTypeOneEDS AGN, with a corresponding surface density of \SDSelectedEuclidLSSTTypeOneEDS deg$^{-2}$. Our selected sample in this case contained 98\% unobscured and 2\% obscured AGN SEDs. The completeness achieved with this selection is $C=\CSelectedEuclidLSSTTypeOneEDS$ overall and $C=\CTypeOneSelectedEuclidLSSTTypeOneEDS$ when only unobscured AGN are considered.

As for the all AGN selection with only \Euclid photometry in the EWS, the EDS counterpart is also disregarded due to the difficulty of separating a general active galaxy population from inactive galaxies. Adding ancillary Rubin/LSST bands for the selection of all AGN in the EDS improves the performance of the optimal colour selection to have an expectation of F1 = 0.3 and $P=0.6$. The optimal criterion in this case follows 
\begin{equation}
    \begin{aligned}
        &\IE - \YE < 1.7\ \\
        &{\tt AND}\ g - r < 0.3\ \\
        &{\tt AND}\ \IE - \YE < -3.5\,(g - r) + 0.9,
    \end{aligned}
\end{equation}
\noindent where $g$ and $r$ are the Rubin/LSST optical filters. Applied to our sample this criterion selects \NUMSelectedEuclidLSSTAllEDS of our AGN, providing a surface density of \SDSelectedEuclidLSSTAllEDS deg$^{-2}$. Our resulting selected sample was comprised of 67\% unobscured SEDs and 33\% obscured AGN SEDs. We derived an overall completeness $C=\CSelectedEuclidLSSTAllEDS$ for this selection. Considering separated AGN classes we found a completeness of $C=\CTypeOneSelectedEuclidLSSTAllEDS$ for unobscured and $C=\CTypeTwoSelectedEuclidLSSTAllEDS$ for obscured AGN.

\end{appendix}
\label{LastPage}
\end{document}